\documentclass[reprint,amsmath,amssymb,aps,prl,twocolumn,superscriptaddress]{revtex4-2}
\usepackage{amsmath}
\usepackage{graphicx}
\usepackage{dcolumn}
\usepackage{bm}
\usepackage{braket}
\usepackage{hyperref}
\usepackage{tabularx}
\usepackage{threeparttable}
\usepackage[normalem]{ulem}

\hypersetup{colorlinks=true,citecolor=magenta,urlcolor=magenta,linkcolor=magenta}
\allowdisplaybreaks[4]

\begin{document}

\preprint{APS/123-QED}

\title{Hybrid Acousto-Optical Double Dressing of a Two-Level System}
\author{Yuan Zhan}
\affiliation{JILA, University of Colorado, Boulder, Colorado 80309, USA}
\affiliation{Department of Physics, University of Colorado, Boulder, Colorado 80309, USA}

\author{Zixuan Wang}
\affiliation{Department of Physics, University of Colorado, Boulder, Colorado 80309, USA}
\affiliation{National Institute of Standards and Technology, Boulder, Colorado 80305, USA}

\author{Richard P. Mirin}
\affiliation{National Institute of Standards and Technology, Boulder, Colorado 80305, USA}

\author{Kevin L. Silverman}
\affiliation{National Institute of Standards and Technology, Boulder, Colorado 80305, USA}

\author{Shuo Sun}
\email{shuosun@colorado.edu}
\affiliation{JILA, University of Colorado, Boulder, Colorado 80309, USA}
\affiliation{Department of Physics, University of Colorado, Boulder, Colorado 80309, USA}

\date{\today}

\begin{abstract}
We experimentally investigate resonance fluorescence from a two-level system in a novel configuration where a strong laser drives an optical Rabi oscillation while an acoustic field parametrically modulates the frequency of the two-level system. We observe emission spectra that deviate markedly from the standard Mollow triplet, including dynamical cancellation of the central peak. A doubly dressed state model incorporating hybridization among the emitter, optical field, and acoustic field captures these features. Guided by this model, we experimentally validate the condition for optimal cooling of acoustic phonons in an emitter-optomechanical system. These results reveal new regimes of strongly driven quantum nonlinear interactions.
\end{abstract}

\maketitle

Resonance fluorescence from a single two-level system has been a central topic in quantum optics since its inception~\cite{Scully:1997aa}. A hallmark of its spectrum is the Mollow triplet, a trio of emission lines that emerges when the atom is strongly driven by a laser~\cite{Mollow:1969aa}. This behavior can be intuitively understood through the concept of atom-photon dressed states, with each emission peak corresponding to a transition between dressed states of different energy ladders~\cite{Cohen-Tannoudji:1977aa}. The Mollow triplet not only highlights the quantum nature of light-matter interactions but also enables various quantum optical applications, including the generation of single and heralded photons~\cite{Aspect:1980aa,Ates:2009aa,Ulhaq:2012aa}, entangled photon pairs~\cite{Peiris:2017aa,Masters:2023aa,Lopez-Carreno:2024aa}, $N$-photon bundles~\cite{Munoz:2014aa}, atom-photon entangled states~\cite{Nick-Vamivakas:2009aa}, and single-atom lasing~\cite{Zakrzewski:1991aa,Quang:1993aa}.

Another remarkable feature of the Mollow physics is that it reveals additional resonances and transitions within what is fundamentally a simple two-level system. Probing these transitions with a second drive uncovers rich quantum interference phenomena and multi‑photon dynamics. For example, when a second laser is tuned into resonance with two selected dressed states of different energy ladders, quantum interference between competing transition channels suppresses specific spectral lines of spontaneous emission, a phenomenon predicted theoretically~\cite{Ficek:1999aa} and observed experimentally~\cite{He:2015aa,Gustin:2021aa}. Meanwhile, parametric drives can facilitate Floquet engineering of an emitter's emission properties~\cite{Lukin:2020aa} and induce rich yet distinct quantum interference phenomena~\cite{Groll:2026aa}. Furthermore, they can directly dress the atom-light states within the same energy ladder to bridge frequencies differing by several orders of magnitude, with applications in microwave control of resonance fluorescence~\cite{Anton:2017aa,Lukin:2020aa}, optical readout of thermal motions~\cite{Munsch:2017aa,Spinnler:2024aa}, optical cooling of acoustic phonons~\cite{Wilson-Rae:2004aa,Rabl:2010aa}, and quantum transduction~\cite{Schuetz:2015aa}. However, despite many theoretical investigations~\cite{Yan:2016aa,Yan:2019aa,Groll:2026aa}, experimental studies of emitters driven simultaneously by strong transverse and longitudinal fields have remained elusive.

\begin{figure*}[t]
\centering
\includegraphics[width=2.0\columnwidth]{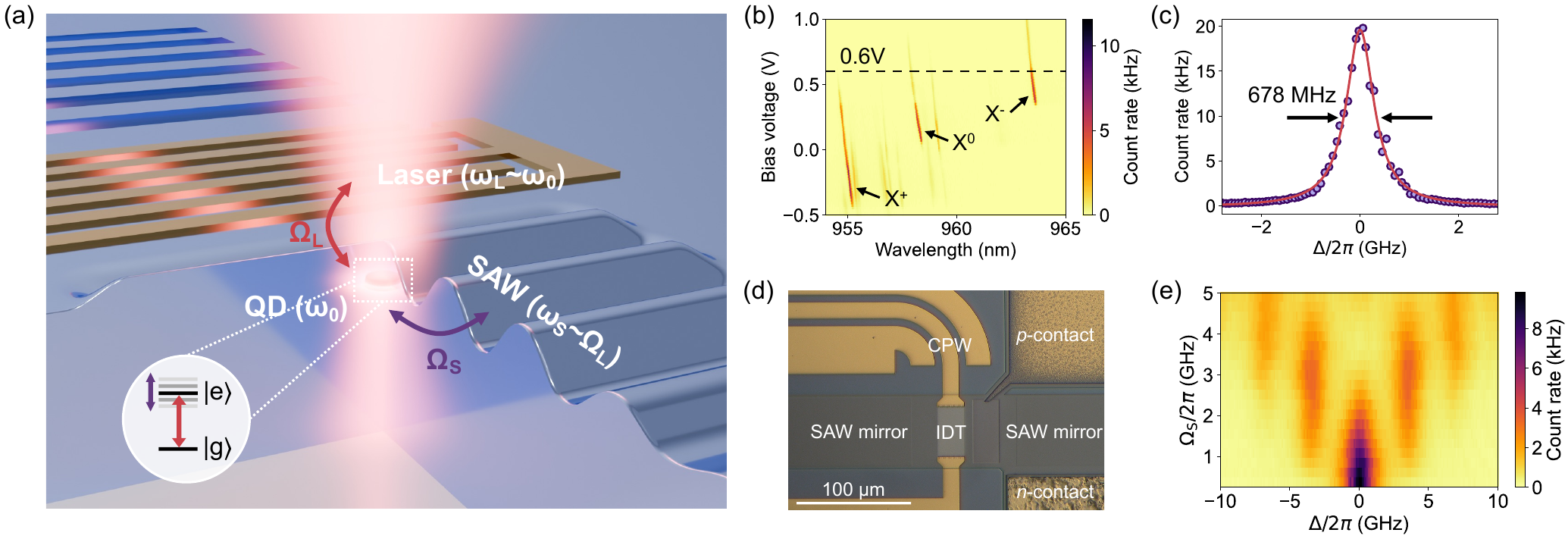}
\caption{(a) Schematic of a laser-dressed quantum dot that is simultaneously driven parametrically by a surface acoustic wave. (b) Photoluminescence spectra of the quantum dot as a function of the bias voltage. We operate our experiments at a bias voltage of 0.6 V (black dashed line), where the quantum dot is charged with a single electron. (c) Intensity of the quantum dot resonance fluorescence as a function of the frequency of a weak excitation laser (i.e. absorption spectrum), showing a transition linewidth of 678 MHz. The excitation power is 25\% of the quantum dot saturation power. (d) Optical microscope image of the device used for launching and confining the surface acoustic wave. SAW: surface acoustic wave. IDT: interdigital transducer. CPW: coplanar waveguide. \textit{p}- and \textit{n}-contacts: ohmic contacts to the \textit{p}- and \textit{n}-doped layers. (e) Absorption spectra of the quantum dot as a function of the acoustic driving strength $\Omega_\text{S}$.}
\label{fig1}
\end{figure*}

In this Letter, we experimentally demonstrate strong parametric drive of a laser-dressed two-level system via a surface acoustic field [Fig.~\ref{fig1}(a)]. We observe emission spectra significantly altered from the standard Mollow triplet, including the dynamical cancellation of resonance fluorescence at the central emission frequency, a feature previously seen only with bichromatic optical excitation~\cite{He:2015aa,Gustin:2021aa}. The observed spectra are well explained by a theoretical model that incorporates the hybridization of the two-level system, the optical field, and the acoustic field. Beyond its significance in fundamental quantum optics, the spectrally resolved resonance fluorescence observed in our experiment enables direct measurement of the optimal condition of optical cooling of acoustic phonons mediated by a single two-level system. Altogether, these results deepen our understanding of quantum interactions among single two‐level systems, optical fields, and acoustic fields in the strong driving regime, and lay a foundation for the development of emitter-optomechanical platforms aimed at nonclassical mechanical state generation~\cite{Sollner:2016aa}, quantum transduction~\cite{Schuetz:2015aa}, and phonon‐mediated quantum gates~\cite{Schuetz:2015aa,Lemonde:2018aa}.

Our experiments are performed on a single self-assembled indium arsenide (InAs) quantum dot in a gallium arsenide (GaAs) crystal. The quantum dot is embedded in a \textit{p-i-n} junction that enables deterministic control of its charge state via an applied bias voltage. We operate at a bias voltage where the dot is loaded with a single electron [Fig.~\ref{fig1}(b)], yielding an ideal two-level system with degenerate polarizations~\cite{Bayer:2002aa}. The transition linewidth is measured to be $\Gamma/2\pi=678$ MHz [Fig.~\ref{fig1}(c)], about five times the lifetime-limited value of $\gamma /2\pi=134$ MHz. We use a continuous-wave laser to optically drive the quantum dot from free space via a confocal microscope. At the same time, we drive the quantum dot with a surface acoustic wave launched via an interdigital transducer lithographically defined on the surface of the GaAs crystal [Fig.~\ref{fig1}(d)]. To enhance the coupling between the quantum dot and the surface acoustic wave, we fabricate two planar Bragg reflectors that define a surface acoustic wave cavity with a quality factor $Q=12,562$ at a frequency $\omega_\text{S}/2\pi=3.53$ GHz. The measured absorption spectra in the presence of the surface acoustic wave verify that we are in the resolved-sideband regime [Fig.~\ref{fig1}(e)], from which we extract a quantum-dot single-phonon coupling strength of $g_0/2\pi=6.6$ kHz. We collect the resonance fluorescence from the quantum dot via the same confocal microscope. Supplemental Material~\cite{supplement} Secs. S1-S3 provide details of device fabrication, characterization, and the measurement setup.

\begin{figure*}[t]
\centering
\includegraphics[width=2.0\columnwidth]{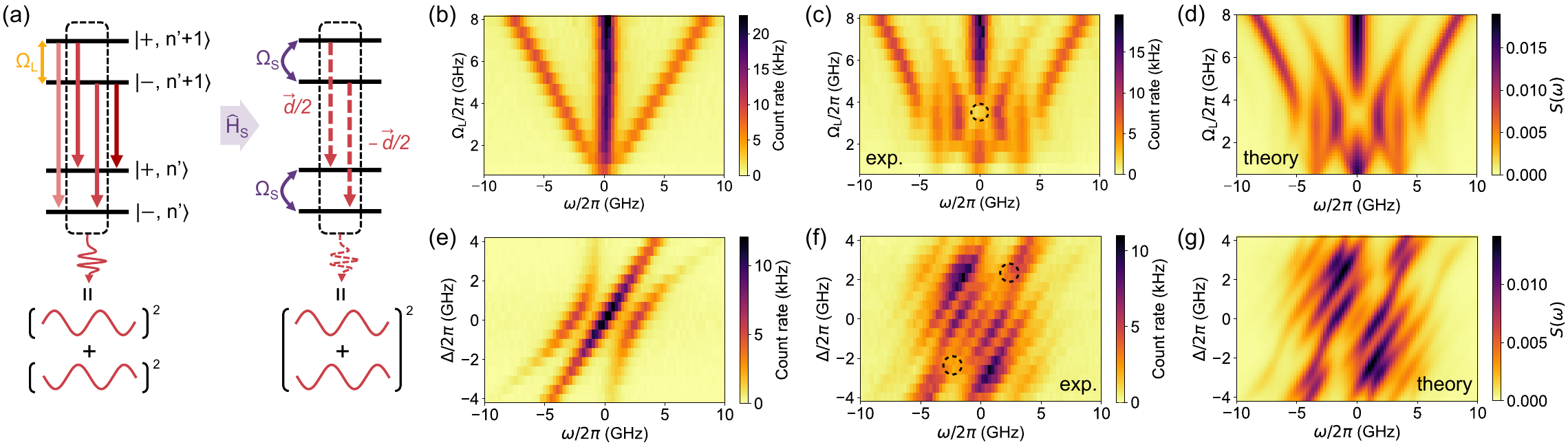}
\caption{(a) Energy levels of the atom-photon dressed states and allowed electric dipole transitions between the dressed states. In the absence of the acoustic drive (left panel), the two transition dipoles that contribute to the central peak of the Mollow triplet oscillate independently, giving rise to a simple intensity summation of their emission at the central frequency. In the presence of the acoustic drive (right panel), these two transition dipoles interfere destructively, resulting in the elimination of central emission peak at the Rabi resonance condition. (b) Experimentally measured resonance fluorescence spectra as a function of the optical Rabi frequency $\Omega_\text{L}$ when the laser is resonant with the quantum dot, showing the standard Mollow triplets. (c) Same measurement as in (b), but with the acoustic drive fixed at $\Omega_\text{S}/2\pi=1.75$ GHz. Black dashed circle marks the dynamical cancellation of emission at the central frequency. (d) Calculated spectra with the same parameters in (c). (e) Measured spectra as a function of the laser detuning $\Delta$ in the absence of the acoustic drive, showing the detuned Mollow triplets. The Rabi frequency is fixed at $\Omega_\text{L}/2\pi=2.625$ GHz. (f) Same measurement as in (e), but with the acoustic drive fixed at $\Omega_\text{S}/2\pi=1.75$ GHz. Black dashed circles mark the dynamical cancellation of emission at the laser frequency. (g) Calculated spectra with the same parameters in (f).}
\label{fig2}
\end{figure*}

We first examine the resonance fluorescence of the quantum dot when driven solely by a laser field. The interaction between the quantum dot and the laser is governed by the Hamiltonian $\hat{H}=-\frac{1}{2}\hbar\Delta\hat{\sigma}_z+\frac{1}{2}\hbar\Omega_\text{L}\hat{\sigma}_x$, where $\Delta$ is the laser frequency detuning from the quantum dot transition, $\Omega_\text{L}$ is the optical Rabi frequency, $\hat{\sigma}_x=\ket{e}\bra{g}+\ket{g}\bra{e}$ and $\hat{\sigma}_z=\ket{e}\bra{e}-\ket{g}\bra{g}$ are the Pauli X and Z operators, and $\ket{g}$ and $\ket{e}$ are the ground and excited states of the quantum dot two-level system. The eigenstates of the Hamiltonian, in the presence of a strong ($\Omega_\text{L}>\Gamma$) and resonant ($\Delta=0$) laser, are the atom-photon dressed states, given by $\ket{\pm,n'}=\frac{1}{\sqrt{2}}\left(\ket{g,n+1}\pm\ket{e,n}\right)$, where $n$ is the number of photons in the bare states, and $n'$ is the number of excitations in the dressed states. Transitions between dressed states with different excitations $n'$ give rise to three distinct emission peaks [Fig.~\ref{fig2}(a), left panel], known as the Mollow triplet~\cite{Mollow:1969aa,Cohen-Tannoudji:1977aa}. Figure~\ref{fig2}(b) shows the experimentally measured resonance fluorescence spectra of the quantum dot as we vary the optical Rabi frequency $\Omega_\text{L}$, exhibiting clear signatures of the Mollow triplet. We note that the central emission peak is twice as strong as each of the Mollow sideband emission, since there are two possible channels that contribute to the central peak, $\ket{+,n'+1}\rightarrow\ket{+,n'}$ and $\ket{-,n'+1}\rightarrow\ket{-,n'}$. These two dipoles are uncorrelated with each other~\cite{Ficek:1999aa}, simply resulting in an intensity summation of their emission [Fig.~\ref{fig2}(a), left panel].

We next apply an acoustic drive to the dressed atom-light system. Figure~\ref{fig2}(c) presents the experimentally measured resonance fluorescence spectra when we fix the acoustic drive to be $\Omega_\text{S}/2\pi=1.75$ GHz but vary the optical Rabi frequency $\Omega_\text{L}$. Here, $\Omega_\text{S}=g_0\sqrt{\bar{m}/2}$ is the acoustic driving strength, where $\bar{m}$ is the average number of phonons in the surface acoustic wave cavity. Compared to Fig.~\ref{fig2}(b), the presence of acoustic drive induces significant modifications to the resonance fluorescence from the strongly driven quantum dot. The measured spectra are well captured by a semiclassical calculation based on Floquet theory and the quantum regression theorem [Fig.~\ref{fig2}(d); see Supplemental Material~\cite{supplement} Sec. S4].

The most notable feature in the resonance fluorescence spectra in Fig.~\ref{fig2}(c) is the dynamical cancellation of the emission at the central peak (marked by the black dashed circle) when the optical Rabi frequency $\Omega_\text{L}$ is tuned in resonance with the surface acoustic wave frequency $\omega_\text{S}$ (referred to as the Rabi resonance condition). This phenomenon arises from destructive quantum interference between the two transition channels $\ket{\pm,n'+1}\rightarrow\ket{\pm,n'}$ that contribute to the central emission line. To understand its origin, we note that a surface acoustic wave introduces a dynamical frequency modulation of the quantum dot, described by $\hat{H}_\text{S}=\hbar\Omega_\text{S}\cos(\omega_\text{S}t)\hat{\sigma}_z$. Under the basis of the atom-photon dressed states, this Hamiltonian can be written as $\hat{H}_\text{S}=-\hbar\Omega_\text{S}\cos(\omega_\text{S}t)\sum_{n'}(\ket{+,n'}\bra{-,n'}+\ket{-,n'}\bra{+,n'})$. Therefore, the surface acoustic wave drives a coherent Rabi oscillation between the dressed states of the same energy ladder, $\ket{\pm,n'}$, by which it synchronizes the oscillations of the two transition dipoles $\ket{\pm,n'+1}\rightarrow\ket{\pm,n'}$ that otherwise oscillate independently. Due to the symmetric and anti-symmetric nature of the superpositions forming the dressed states, the two transition dipoles are $\pi$ out of phase and interfere destructively, resulting in zero emission intensity at the central spectral line [Fig.~\ref{fig2}(a), right panel]. 

A similar dynamical cancellation of resonant emission was previously observed under bichromatic optical drives~\cite{He:2015aa,Gustin:2021aa}, effectively implementing a single amplitude modulated transverse drive. The interaction mechanism in our experiment is qualitatively different, as we employ both a transverse ($\hat{\sigma}_x$) optical and a longitudinal ($\hat{\sigma}_z$) acoustic drive. The latter provides a parametric modulation of the emitter’s transition frequency rather than a second dipole drive. This distinction is analogous to the difference between a direct periodic force applied to a pendulum (transverse) and a periodic modulation of its length (longitudinal). Our results show that a parametric acoustic modulation of the emitter frequency can also lead to the cancellation of resonant emission.

We now extend our study to a detuned laser drive. Figure~\ref{fig2}(e) shows the resonance fluorescence spectra of the quantum dot as we vary the laser frequency. In this measurement, we keep the acoustic drive off and fix the optical Rabi frequency to be $\Omega_\text{L}/2\pi=2.625$ GHz. We observe the standard detuned Mollow triplet with each side peak separated by the generalized Rabi frequency $\Omega_\text{R}=\sqrt{\Omega_\text{L}^2+\Delta^2}$ from the central one. Figure~\ref{fig2}(f) shows the resonance fluorescence spectra of the quantum dot under the same measurement condition, but with an acoustic drive of $\Omega_\text{S}/2\pi=1.75$ GHz. We observe the dynamical cancellation of the central emission peak at two laser detunings, $\Delta/2\pi=\pm 2.36$ GHz [marked by black dashed circles in Fig.~\ref{fig2}(f)], corresponding to the generalized Rabi resonance condition $\omega_\text{S}=\Omega_\text{R}$. This cancellation arises from the same physical mechanism as in the resonant case. The observed emission cancellation is not perfect due to a combination of spectral overlap with other sidebands, residual laser background, and higher-order couplings (see Supplemental Material~\cite{supplement} Sec. S5). The measured spectra in Fig.~\ref{fig2}(f) are well captured by the semiclassical calculation [Fig.~\ref{fig2}(g)].

\begin{figure}[t]
\centering
\includegraphics[width=1.0\columnwidth]{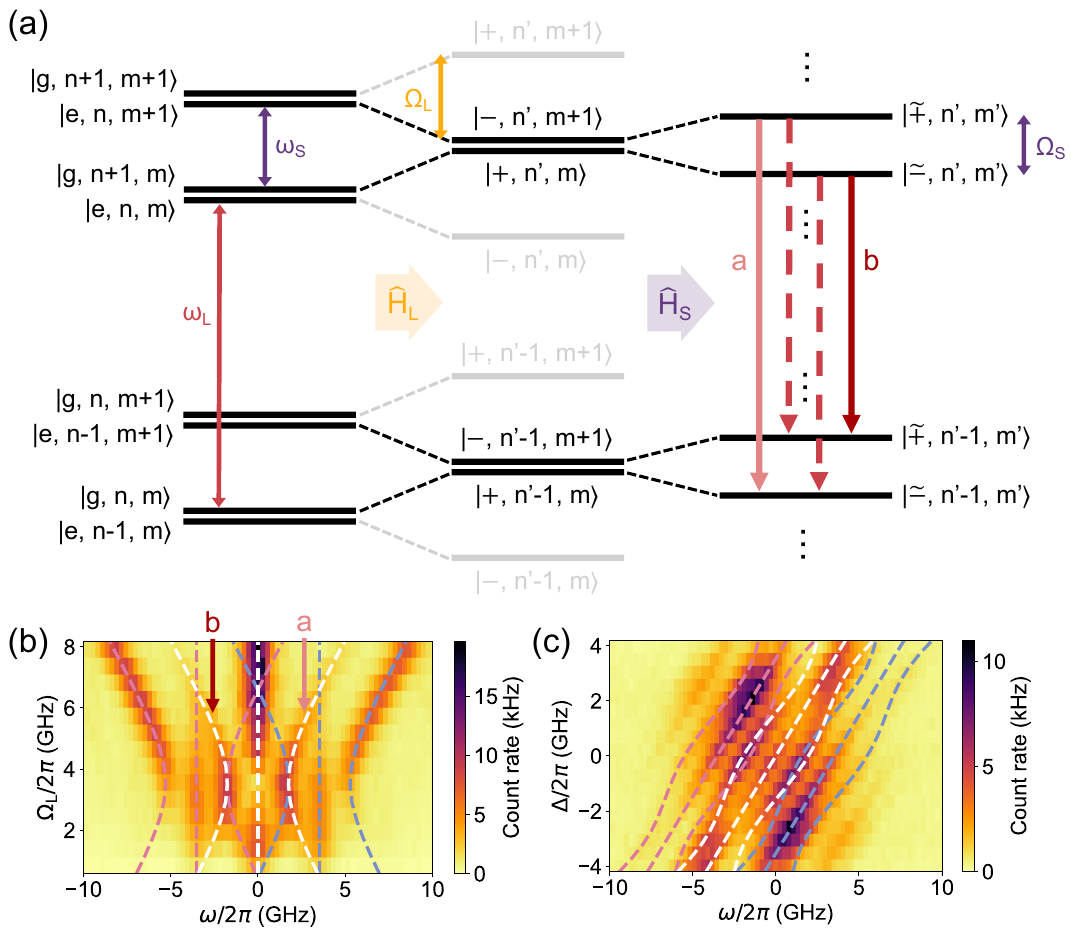}
\caption{(a) Energy levels of the atom-photon-phonon system at Rabi resonance. The optical coupling $\hat{H}_\text{L}$ creates the atom-photon dressed states $\ket{\pm,n',m}=\frac{1}{\sqrt{2}}\left(\ket{g,n+1,m}\pm\ket{e,n,m}\right)$ from the bare states, separated by the optical Rabi frequency $\Omega_\text{L}$. The acoustic coupling $\hat{H}_\text{S}$ further lifts the degeneracy between states $\ket{-,n',m+1}$ and $\ket{+,n',m}$, creating the doubly dressed states $\ket{\widetilde{\pm},n',m'}=\frac{1}{\sqrt{2}}\left(\ket{+,n',m}\pm\ket{-,n',m+1}\right)$, separated by the acoustic driving strength $\Omega_\text{S}$. The dipole matrix elements of the transitions $\ket{\widetilde{\pm},n',m'}\to\ket{\widetilde{\pm},n'-1,m'}$ are zero (red dashed arrows), explaining the elimination of central spectral line at the Rabi resonance condition. The other two transitions, labeled ``a'' and ``b'' (red solid arrows), give rise to two spectral lines that anti-cross each other due to the acoustic dressing. (b)(c) Predicted transition frequencies of the resonance fluorescence spectra by the doubly dressed state picture (dashed lines), overlaid with the experimentally measured spectra shown in Figs.~\ref{fig2}(c) and~\ref{fig2}(f), respectively.}
\label{fig3}
\end{figure}

The resonance fluorescence spectra exhibited in Fig.~\ref{fig2} can be intuitively explained with a doubly dressed state picture illustrated in Fig.~\ref{fig3}(a). Here, we express all quantum states under the new basis $\ket{a,n,m}$, where $a\in\{g,e\}$ is the state of the two-level system, $n$ and $m$ are the photon and phonon numbers, respectively. The optical drive dresses the bare states $\ket{g,n+1,m}$ and $\ket{e,n,m}$ and creates the atom-photon dressed states $\ket{\pm,n',m}$. In the optical resonance condition ($\Delta=0$), the atom-photon dressed states are given by $\ket{\pm,n',m}=\frac{1}{\sqrt{2}}\left(\ket{g,n+1,m}\pm\ket{e,n,m}\right)$, and they are separated by a frequency $\Omega_\text{L}$. When the acoustic drive is close to the Rabi resonance condition $\omega_\text{S}=\Omega_\text{L}$, the symmetric and antisymmetric atom-photon dressed states of different phonon numbers, $\ket{+,n',m}$ and $\ket{-,n',m+1}$, are nearly degenerate in energy. Thus, the acoustic drive dresses these two states and creates the atom-photon-phonon doubly dressed states $\ket{\widetilde{\pm},n',m'}$, given by $\ket{\widetilde{\pm},n',m'}=\frac{1}{\sqrt{2}}\left(\ket{+,n',m}\pm\ket{-,n',m+1}\right)$ at the Rabi resonance, separated by a frequency $\Omega_\text{S}$. The frequencies at which emission peaks reside are the frequency differences between all possible pairs of doubly dressed states that give rise to dipole-allowed transitions, as shown by the dashed lines in Figs.~\ref{fig3}(b) and~\ref{fig3}(c), which agree well with the experimentally measured spectra.

The doubly dressed state picture allows us to intuitively understand the key features in the resonance fluorescence spectra. We first focus on the three emission peaks denoted by the white dashed lines in Fig.~\ref{fig3}(b). This triplet originates from the transitions between $\ket{\widetilde{\pm},n',m'}$ and $\ket{\widetilde{\pm},n'-1,m'}$, as illustrated in Fig.~\ref{fig3}(a). There are two transitions that contribute to the central peak, $\ket{\widetilde{+},n',m'}\rightarrow\ket{\widetilde{+},n'-1,m'}$ and $\ket{\widetilde{-},n',m'}\rightarrow\ket{\widetilde{-},n'-1,m'}$. However, at the Rabi resonance condition, the electric dipole moments of both transitions, $\bra{\widetilde{\pm},n',m'}e\hat{x}\ket{\widetilde{\pm},n'-1,m'}$, are zero, resulting in the elimination of the central emission. The two side peaks of the triplet, labeled ``a" and ``b" in Figs.~\ref{fig3}(a) and~\ref{fig3}(b), originate from the transitions $\ket{\widetilde{+},n',m'}\rightarrow\ket{\widetilde{-},n'-1,m'}$ and $\ket{\widetilde{-},n',m'}\rightarrow\ket{\widetilde{+},n'-1,m'}$, respectively. These two transitions anti-cross each other, with a minimum frequency splitting of $2\Omega_\text{S}$, due to the acoustic dressing of the atom-photon dressed states. Similarly, we find two other groups of triplet emission lines, denoted by the pink and blue dashed lines in Fig.~\ref{fig3}(b). These triplet emission lines originate from the transitions between $\ket{\widetilde{\pm},n',m'}$ and $\ket{\widetilde{\pm},n'-1,m'\pm 1}$ (see Supplemental Material~\cite{supplement} Sec. S5).

Our results mark the first resonance fluorescence spectroscopy of an emitter-optomechanical system in the strong optical driving regime. Unlike cavity optomechanics~\cite{Aspelmeyer:2014aa}, emitter optomechanics possesses intrinsic quantum nonlinearity enabled by the anharmonicity of the two‑level system, a feature crucial for phononic quantum technologies~\cite{Barzanjeh:2022aa}. Emitter-mechanical coupling has been demonstrated experimentally with semiconductor quantum dots~\cite{Metcalfe:2010aa,Yeo:2014aa,Munsch:2017aa,Imany:2022aa,Spinnler:2024aa}, color centers~\cite{Golter:2016aa,Lee:2016aa}, rare earth ions~\cite{Ohta:2021aa}, and 2D materials~\cite{Zalalutdinov:2021aa,Patel:2024aa}. Theoretical studies have shown that a simple two-level emitter can facilitate laser cooling of mechanical motions~\cite{Wilson-Rae:2004aa}, and in the resolved-sideband regime, optimal cooling occurs at the Rabi resonance condition~\cite{Rabl:2010aa}. However, this optimal cooling condition has not been measured directly in experiments, as its observation requires $g_0$ to be large compared with the acoustic cavity decay rate $\gamma_\text{S}$.

Resonance fluorescence spectroscopy provides a direct experimental probe of the phonon cooling rate under various laser driving conditions, as the frequencies and intensities of the emission spectral lines directly reflect the rate of phonon number change in the mechanical mode (see Supplemental Material~\cite{supplement} Sec. S6). Importantly, although the cooling rate strongly depends on the steady-state phonon number in the mechanical mode, the optimal cooling condition does not, even when it is close to ground state cooling (see Supplemental Material~\cite{supplement} Sec. S7). This property allows us to experimentally measure the optimal cooling condition in the presence of the acoustic drive, which enables the resolution of sideband features despite the relatively small $g_0$ of our device compared to $\gamma_\text{S}$.

\begin{figure}[t]
\centering
\includegraphics[width=1.0\columnwidth]{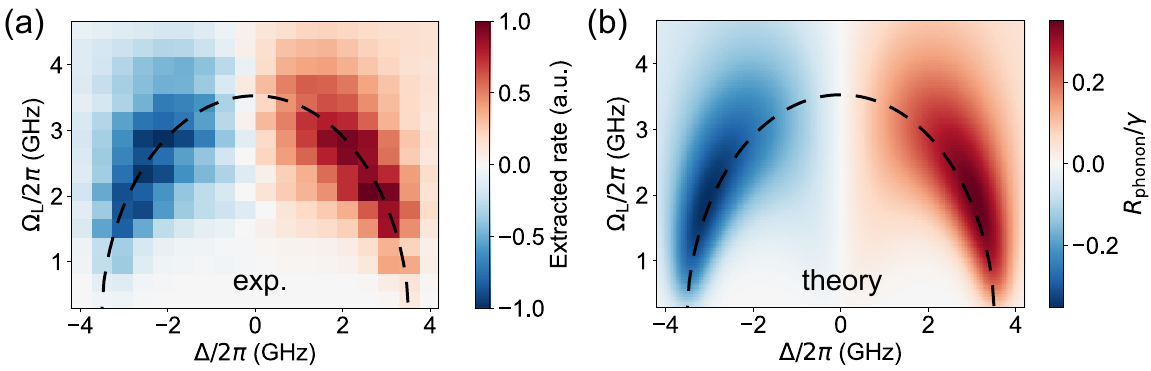}
\caption{(a) Extracted phonon cooling rate from the measured resonance fluorescence spectra as a function of the optical Rabi frequency $\Omega_\text{L}$ and laser detuning $\Delta$. (b) Calculated phonon cooling rate with the same parameters in (a). Black dashed lines in (a) and (b) highlight the Rabi resonance condition, where the generalized Rabi frequency $\Omega_\text{R}$ matches the phonon frequency $\omega_\text{S}$.}
\label{fig4}
\end{figure}

Figure~\ref{fig4}(a) shows the experimentally measured phonon cooling rate as a function of the optical Rabi frequency and laser detuning, with the acoustic drive fixed at $\Omega_\text{S}/2\pi = 1.75$ GHz, corresponding to a steady-state phonon number of $\sim 10^{11}$ in the cavity (see Supplemental Material~\cite{supplement} Sec. S2). As expected, a red-detuned laser removes phonons from the acoustic cavity, while a blue-detuned laser adds phonons. The optimal cooling and heating rates occur when the generalized Rabi frequency, $\Omega_\text{R}=\sqrt{\Omega_\text{L}^2+\Delta^2}$, matches the phonon frequency $\omega_\text{S}$ [black dashed line in Fig.~\ref{fig4}(a)]. These findings are consistent with both previous results~\cite{Rabl:2010aa,Spinnler:2024aa} and our own theoretical calculations [Fig.~\ref{fig4}(b); see Supplemental Material~\cite{supplement} Sec. S6].

We have experimentally demonstrated dynamical control of optical resonance fluorescence from a two-level system via a parametric acoustic drive. The parametric acoustic drive induces quantum interference between competing transition channels among distinct optically dressed states, resulting in the cancellation of spontaneous emission at selected frequencies. The ability to manipulate this interference with a parametric drive will enable Floquet engineering of emission spectra~\cite{Lukin:2020aa}, the development of tunable quantum light sources~\cite{Senellart:2017aa}, and novel coherent control schemes for quantum information processing~\cite{DeCrescent:2024aa}. 

Our findings indicate that even a simple two‑level system can be engineered to support resonances spanning across five orders of magnitude. This property offers new insights for nonlinear frequency mixing~\cite{Weiss:2021aa} in the quantum limit. In addition, the fact that both transverse and longitudinal drives can couple to the same two-level system to induce quantum interference demonstrates a versatile capability for quantum control across disparate physical fields, implemented here between optics and acoustics. To illustrate a specific use case, we have demonstrated experimentally that the optimal condition for optical phonon cooling in emitter optomechanics arises when the generalized optical Rabi frequency equals the mechanical frequency. The same principle may also enable the optimal realization of coherent phonon lasing via two-level systems~\cite{Kabuss:2012aa,Kepesidis:2013aa}. 

The device used in our work is not optimized for record-breaking optomechanical performance, but rather serving for investigating fundamental interactions between optical and acoustic fields via a two-level quantum emitter. Improvements in the quality factor and mode confinement of the mechanical resonator~\cite{DeCrescent:2022aa,Spinnler:2024aa}, combined with the use of a quantum emitter with a larger strain susceptibility~\cite{Tsuchimoto:2022aa}, will enable the cooling of the mechanical motion to the ground state, as well as direct observation of the cooling without an external drive (see Supplemental Material~\cite{supplement} Sec. S8). Integration of the planar surface acoustic wave cavity with vertically tunable open microcavities will further enhance the photon collection efficiency~\cite{Tomm:2021aa,Ding:2025aa}, enabling applications in optomechanical quantum transduction. Further integration of an electron spin to the quantum dot will unlock new avenues towards a spin-phonon quantum interface~\cite{Carter:2018aa,Maity:2020aa} and phonon‐mediated quantum gates~\cite{Schuetz:2015aa,Lemonde:2018aa}.

The work was supported by the National Science Foundation (NSF) (2137953 and 2317149), the Air Force Office of Scientific Research (AFOSR) (FA2386-24-1-4067), and the W. M. Keck Foundation. Y.Z. acknowledges support from NSF QISE-NET award (funded by NSF Grant No. 1747426). S.S. acknowledges support from the Sloan Research Fellowship.

\twocolumngrid
\nocite{*}

\begin{thebibliography}{57}%
\makeatletter
\providecommand \@ifxundefined [1]{%
 \@ifx{#1\undefined}
}%
\providecommand \@ifnum [1]{%
 \ifnum #1\expandafter \@firstoftwo
 \else \expandafter \@secondoftwo
 \fi
}%
\providecommand \@ifx [1]{%
 \ifx #1\expandafter \@firstoftwo
 \else \expandafter \@secondoftwo
 \fi
}%
\providecommand \natexlab [1]{#1}%
\providecommand \enquote  [1]{``#1''}%
\providecommand \bibnamefont  [1]{#1}%
\providecommand \bibfnamefont [1]{#1}%
\providecommand \citenamefont [1]{#1}%
\providecommand \href@noop [0]{\@secondoftwo}%
\providecommand \href [0]{\begingroup \@sanitize@url \@href}%
\providecommand \@href[1]{\@@startlink{#1}\@@href}%
\providecommand \@@href[1]{\endgroup#1\@@endlink}%
\providecommand \@sanitize@url [0]{\catcode `\\12\catcode `\$12\catcode
  `\&12\catcode `\#12\catcode `\^12\catcode `\_12\catcode `\%12\relax}%
\providecommand \@@startlink[1]{}%
\providecommand \@@endlink[0]{}%
\providecommand \url  [0]{\begingroup\@sanitize@url \@url }%
\providecommand \@url [1]{\endgroup\@href {#1}{\urlprefix }}%
\providecommand \urlprefix  [0]{URL }%
\providecommand \Eprint [0]{\href }%
\providecommand \doibase [0]{https://doi.org/}%
\providecommand \selectlanguage [0]{\@gobble}%
\providecommand \bibinfo  [0]{\@secondoftwo}%
\providecommand \bibfield  [0]{\@secondoftwo}%
\providecommand \translation [1]{[#1]}%
\providecommand \BibitemOpen [0]{}%
\providecommand \bibitemStop [0]{}%
\providecommand \bibitemNoStop [0]{.\EOS\space}%
\providecommand \EOS [0]{\spacefactor3000\relax}%
\providecommand \BibitemShut  [1]{\csname bibitem#1\endcsname}%
\let\auto@bib@innerbib\@empty
\bibitem [{\citenamefont {Scully}\ and\ \citenamefont
  {Zubairy}(1997)}]{Scully:1997aa}%
  \BibitemOpen
  \bibfield  {author} {\bibinfo {author} {\bibfnamefont {M.~O.}\ \bibnamefont
  {Scully}}\ and\ \bibinfo {author} {\bibfnamefont {M.~S.}\ \bibnamefont
  {Zubairy}},\ }\href {https://doi.org/DOI: 10.1017/CBO9780511813993} {\emph
  {\bibinfo {title} {Quantum Optics}}}\ (\bibinfo  {publisher} {Cambridge
  University Press},\ \bibinfo {address} {Cambridge},\ \bibinfo {year}
  {1997})\BibitemShut {NoStop}%
\bibitem [{\citenamefont {Mollow}(1969)}]{Mollow:1969aa}%
  \BibitemOpen
  \bibfield  {author} {\bibinfo {author} {\bibfnamefont {B.~R.}\ \bibnamefont
  {Mollow}},\ }\bibfield  {title} {\bibinfo {title} {Power spectrum of light
  scattered by two-level systems},\ }\href
  {https://doi.org/10.1103/PhysRev.188.1969} {\bibfield  {journal} {\bibinfo
  {journal} {Physical Review}\ }\textbf {\bibinfo {volume} {188}},\ \bibinfo
  {pages} {1969} (\bibinfo {year} {1969})}\BibitemShut {NoStop}%
\bibitem [{\citenamefont {Cohen-Tannoudji}\ and\ \citenamefont
  {Reynaud}(1977)}]{Cohen-Tannoudji:1977aa}%
  \BibitemOpen
  \bibfield  {author} {\bibinfo {author} {\bibfnamefont {C.}~\bibnamefont
  {Cohen-Tannoudji}}\ and\ \bibinfo {author} {\bibfnamefont {S.}~\bibnamefont
  {Reynaud}},\ }\bibfield  {title} {\bibinfo {title} {Dressed-atom description
  of resonance fluorescence and absorption spectra of a multi-level atom in an
  intense laser beam},\ }\href {https://doi.org/10.1088/0022-3700/10/3/005}
  {\bibfield  {journal} {\bibinfo  {journal} {Journal of Physics B: Atomic and
  Molecular Physics}\ }\textbf {\bibinfo {volume} {10}},\ \bibinfo {pages}
  {345} (\bibinfo {year} {1977})}\BibitemShut {NoStop}%
\bibitem [{\citenamefont {Aspect}\ \emph {et~al.}(1980)\citenamefont {Aspect},
  \citenamefont {Roger}, \citenamefont {Reynaud}, \citenamefont {Dalibard},\
  and\ \citenamefont {Cohen-Tannoudji}}]{Aspect:1980aa}%
  \BibitemOpen
  \bibfield  {author} {\bibinfo {author} {\bibfnamefont {A.}~\bibnamefont
  {Aspect}}, \bibinfo {author} {\bibfnamefont {G.}~\bibnamefont {Roger}},
  \bibinfo {author} {\bibfnamefont {S.}~\bibnamefont {Reynaud}}, \bibinfo
  {author} {\bibfnamefont {J.}~\bibnamefont {Dalibard}},\ and\ \bibinfo
  {author} {\bibfnamefont {C.}~\bibnamefont {Cohen-Tannoudji}},\ }\bibfield
  {title} {\bibinfo {title} {Time correlations between the two sidebands of the
  resonance fluorescence triplet},\ }\href
  {https://doi.org/10.1103/PhysRevLett.45.617} {\bibfield  {journal} {\bibinfo
  {journal} {Physical Review Letters}\ }\textbf {\bibinfo {volume} {45}},\
  \bibinfo {pages} {617} (\bibinfo {year} {1980})}\BibitemShut {NoStop}%
\bibitem [{\citenamefont {Ates}\ \emph {et~al.}(2009)\citenamefont {Ates},
  \citenamefont {Ulrich}, \citenamefont {Reitzenstein}, \citenamefont
  {L{\"o}ffler}, \citenamefont {Forchel},\ and\ \citenamefont
  {Michler}}]{Ates:2009aa}%
  \BibitemOpen
  \bibfield  {author} {\bibinfo {author} {\bibfnamefont {S.}~\bibnamefont
  {Ates}}, \bibinfo {author} {\bibfnamefont {S.~M.}\ \bibnamefont {Ulrich}},
  \bibinfo {author} {\bibfnamefont {S.}~\bibnamefont {Reitzenstein}}, \bibinfo
  {author} {\bibfnamefont {A.}~\bibnamefont {L{\"o}ffler}}, \bibinfo {author}
  {\bibfnamefont {A.}~\bibnamefont {Forchel}},\ and\ \bibinfo {author}
  {\bibfnamefont {P.}~\bibnamefont {Michler}},\ }\bibfield  {title} {\bibinfo
  {title} {Post-selected indistinguishable photons from the resonance
  fluorescence of a single quantum dot in a microcavity},\ }\href
  {https://doi.org/10.1103/PhysRevLett.103.167402} {\bibfield  {journal}
  {\bibinfo  {journal} {Physical Review Letters}\ }\textbf {\bibinfo {volume}
  {103}},\ \bibinfo {pages} {167402} (\bibinfo {year} {2009})}\BibitemShut
  {NoStop}%
\bibitem [{\citenamefont {Ulhaq}\ \emph {et~al.}(2012)\citenamefont {Ulhaq},
  \citenamefont {Weiler}, \citenamefont {Ulrich}, \citenamefont {Ro{\ss}bach},
  \citenamefont {Jetter},\ and\ \citenamefont {Michler}}]{Ulhaq:2012aa}%
  \BibitemOpen
  \bibfield  {author} {\bibinfo {author} {\bibfnamefont {A.}~\bibnamefont
  {Ulhaq}}, \bibinfo {author} {\bibfnamefont {S.}~\bibnamefont {Weiler}},
  \bibinfo {author} {\bibfnamefont {S.~M.}\ \bibnamefont {Ulrich}}, \bibinfo
  {author} {\bibfnamefont {R.}~\bibnamefont {Ro{\ss}bach}}, \bibinfo {author}
  {\bibfnamefont {M.}~\bibnamefont {Jetter}},\ and\ \bibinfo {author}
  {\bibfnamefont {P.}~\bibnamefont {Michler}},\ }\bibfield  {title} {\bibinfo
  {title} {Cascaded single-photon emission from the mollow triplet sidebands of
  a quantum dot},\ }\href {https://doi.org/10.1038/nphoton.2012.23} {\bibfield
  {journal} {\bibinfo  {journal} {Nature Photonics}\ }\textbf {\bibinfo
  {volume} {6}},\ \bibinfo {pages} {238} (\bibinfo {year} {2012})}\BibitemShut
  {NoStop}%
\bibitem [{\citenamefont {Peiris}\ \emph {et~al.}(2017)\citenamefont {Peiris},
  \citenamefont {Konthasinghe},\ and\ \citenamefont {Muller}}]{Peiris:2017aa}%
  \BibitemOpen
  \bibfield  {author} {\bibinfo {author} {\bibfnamefont {M.}~\bibnamefont
  {Peiris}}, \bibinfo {author} {\bibfnamefont {K.}~\bibnamefont
  {Konthasinghe}},\ and\ \bibinfo {author} {\bibfnamefont {A.}~\bibnamefont
  {Muller}},\ }\bibfield  {title} {\bibinfo {title} {Franson interference
  generated by a two-level system},\ }\href
  {https://doi.org/10.1103/PhysRevLett.118.030501} {\bibfield  {journal}
  {\bibinfo  {journal} {Physical Review Letters}\ }\textbf {\bibinfo {volume}
  {118}},\ \bibinfo {pages} {030501} (\bibinfo {year} {2017})}\BibitemShut
  {NoStop}%
\bibitem [{\citenamefont {Masters}\ \emph {et~al.}(2023)\citenamefont
  {Masters}, \citenamefont {Hu}, \citenamefont {Cordier}, \citenamefont
  {Maron}, \citenamefont {Pache}, \citenamefont {Rauschenbeutel}, \citenamefont
  {Schemmer},\ and\ \citenamefont {Volz}}]{Masters:2023aa}%
  \BibitemOpen
  \bibfield  {author} {\bibinfo {author} {\bibfnamefont {L.}~\bibnamefont
  {Masters}}, \bibinfo {author} {\bibfnamefont {X.-X.}\ \bibnamefont {Hu}},
  \bibinfo {author} {\bibfnamefont {M.}~\bibnamefont {Cordier}}, \bibinfo
  {author} {\bibfnamefont {G.}~\bibnamefont {Maron}}, \bibinfo {author}
  {\bibfnamefont {L.}~\bibnamefont {Pache}}, \bibinfo {author} {\bibfnamefont
  {A.}~\bibnamefont {Rauschenbeutel}}, \bibinfo {author} {\bibfnamefont
  {M.}~\bibnamefont {Schemmer}},\ and\ \bibinfo {author} {\bibfnamefont
  {J.}~\bibnamefont {Volz}},\ }\bibfield  {title} {\bibinfo {title} {On the
  simultaneous scattering of two photons by a single two-level atom},\ }\href
  {https://doi.org/10.1038/s41566-023-01260-7} {\bibfield  {journal} {\bibinfo
  {journal} {Nature Photonics}\ }\textbf {\bibinfo {volume} {17}},\ \bibinfo
  {pages} {972} (\bibinfo {year} {2023})}\BibitemShut {NoStop}%
\bibitem [{\citenamefont {L{\'o}pez~Carre{\~n}o}\ \emph
  {et~al.}(2024)\citenamefont {L{\'o}pez~Carre{\~n}o}, \citenamefont
  {Berm{\'u}dez~Feijoo},\ and\ \citenamefont
  {Stobi{\'n}ska}}]{Lopez-Carreno:2024aa}%
  \BibitemOpen
  \bibfield  {author} {\bibinfo {author} {\bibfnamefont {J.~C.}\ \bibnamefont
  {L{\'o}pez~Carre{\~n}o}}, \bibinfo {author} {\bibfnamefont {S.}~\bibnamefont
  {Berm{\'u}dez~Feijoo}},\ and\ \bibinfo {author} {\bibfnamefont
  {M.}~\bibnamefont {Stobi{\'n}ska}},\ }\bibfield  {title} {\bibinfo {title}
  {Entanglement in resonance fluorescence},\ }\href
  {https://doi.org/10.1038/s44310-023-00001-6} {\bibfield  {journal} {\bibinfo
  {journal} {npj Nanophotonics}\ }\textbf {\bibinfo {volume} {1}},\ \bibinfo
  {pages} {3} (\bibinfo {year} {2024})}\BibitemShut {NoStop}%
\bibitem [{\citenamefont {Mu{\~n}oz}\ \emph {et~al.}(2014)\citenamefont
  {Mu{\~n}oz}, \citenamefont {del Valle}, \citenamefont {Tudela}, \citenamefont
  {M{\"u}ller}, \citenamefont {Lichtmannecker}, \citenamefont {Kaniber},
  \citenamefont {Tejedor}, \citenamefont {Finley},\ and\ \citenamefont
  {Laussy}}]{Munoz:2014aa}%
  \BibitemOpen
  \bibfield  {author} {\bibinfo {author} {\bibfnamefont {C.~S.}\ \bibnamefont
  {Mu{\~n}oz}}, \bibinfo {author} {\bibfnamefont {E.}~\bibnamefont {del
  Valle}}, \bibinfo {author} {\bibfnamefont {A.~G.}\ \bibnamefont {Tudela}},
  \bibinfo {author} {\bibfnamefont {K.}~\bibnamefont {M{\"u}ller}}, \bibinfo
  {author} {\bibfnamefont {S.}~\bibnamefont {Lichtmannecker}}, \bibinfo
  {author} {\bibfnamefont {M.}~\bibnamefont {Kaniber}}, \bibinfo {author}
  {\bibfnamefont {C.}~\bibnamefont {Tejedor}}, \bibinfo {author} {\bibfnamefont
  {J.~J.}\ \bibnamefont {Finley}},\ and\ \bibinfo {author} {\bibfnamefont
  {F.~P.}\ \bibnamefont {Laussy}},\ }\bibfield  {title} {\bibinfo {title}
  {Emitters of n-photon bundles},\ }\href
  {https://doi.org/10.1038/nphoton.2014.114} {\bibfield  {journal} {\bibinfo
  {journal} {Nature Photonics}\ }\textbf {\bibinfo {volume} {8}},\ \bibinfo
  {pages} {550} (\bibinfo {year} {2014})}\BibitemShut {NoStop}%
\bibitem [{\citenamefont {Nick~Vamivakas}\ \emph {et~al.}(2009)\citenamefont
  {Nick~Vamivakas}, \citenamefont {Zhao}, \citenamefont {Lu},\ and\
  \citenamefont {Atat{\"u}re}}]{Nick-Vamivakas:2009aa}%
  \BibitemOpen
  \bibfield  {author} {\bibinfo {author} {\bibfnamefont {A.}~\bibnamefont
  {Nick~Vamivakas}}, \bibinfo {author} {\bibfnamefont {Y.}~\bibnamefont
  {Zhao}}, \bibinfo {author} {\bibfnamefont {C.-Y.}\ \bibnamefont {Lu}},\ and\
  \bibinfo {author} {\bibfnamefont {M.}~\bibnamefont {Atat{\"u}re}},\
  }\bibfield  {title} {\bibinfo {title} {Spin-resolved quantum-dot resonance
  fluorescence},\ }\href {https://doi.org/10.1038/nphys1182} {\bibfield
  {journal} {\bibinfo  {journal} {Nature Physics}\ }\textbf {\bibinfo {volume}
  {5}},\ \bibinfo {pages} {198} (\bibinfo {year} {2009})}\BibitemShut {NoStop}%
\bibitem [{\citenamefont {Zakrzewski}\ \emph {et~al.}(1991)\citenamefont
  {Zakrzewski}, \citenamefont {Lewenstein},\ and\ \citenamefont
  {Mossberg}}]{Zakrzewski:1991aa}%
  \BibitemOpen
  \bibfield  {author} {\bibinfo {author} {\bibfnamefont {J.}~\bibnamefont
  {Zakrzewski}}, \bibinfo {author} {\bibfnamefont {M.}~\bibnamefont
  {Lewenstein}},\ and\ \bibinfo {author} {\bibfnamefont {T.~W.}\ \bibnamefont
  {Mossberg}},\ }\bibfield  {title} {\bibinfo {title} {Theory of dressed-state
  lasers. i. effective hamiltonians and stability properties},\ }\href
  {https://doi.org/10.1103/PhysRevA.44.7717} {\bibfield  {journal} {\bibinfo
  {journal} {Physical Review A}\ }\textbf {\bibinfo {volume} {44}},\ \bibinfo
  {pages} {7717} (\bibinfo {year} {1991})}\BibitemShut {NoStop}%
\bibitem [{\citenamefont {Quang}\ and\ \citenamefont
  {Freedhoff}(1993)}]{Quang:1993aa}%
  \BibitemOpen
  \bibfield  {author} {\bibinfo {author} {\bibfnamefont {T.}~\bibnamefont
  {Quang}}\ and\ \bibinfo {author} {\bibfnamefont {H.}~\bibnamefont
  {Freedhoff}},\ }\bibfield  {title} {\bibinfo {title} {Atomic population
  inversion and enhancement of resonance fluorescence in a cavity},\ }\href
  {https://doi.org/10.1103/PhysRevA.47.2285} {\bibfield  {journal} {\bibinfo
  {journal} {Physical Review A}\ }\textbf {\bibinfo {volume} {47}},\ \bibinfo
  {pages} {2285} (\bibinfo {year} {1993})}\BibitemShut {NoStop}%
\bibitem [{\citenamefont {Ficek}\ and\ \citenamefont
  {Rudolph}(1999)}]{Ficek:1999aa}%
  \BibitemOpen
  \bibfield  {author} {\bibinfo {author} {\bibfnamefont {Z.}~\bibnamefont
  {Ficek}}\ and\ \bibinfo {author} {\bibfnamefont {T.}~\bibnamefont
  {Rudolph}},\ }\bibfield  {title} {\bibinfo {title} {Quantum interference in a
  driven two-level atom},\ }\href {https://doi.org/10.1103/PhysRevA.60.R4245}
  {\bibfield  {journal} {\bibinfo  {journal} {Physical Review A}\ }\textbf
  {\bibinfo {volume} {60}},\ \bibinfo {pages} {R4245} (\bibinfo {year}
  {1999})}\BibitemShut {NoStop}%
\bibitem [{\citenamefont {He}\ \emph {et~al.}(2015)\citenamefont {He},
  \citenamefont {He}, \citenamefont {Liu}, \citenamefont {Wei}, \citenamefont
  {Ram{\'\i}rez}, \citenamefont {Atat{\"u}re}, \citenamefont {Schneider},
  \citenamefont {Kamp}, \citenamefont {H{\"o}fling}, \citenamefont {Lu},\ and\
  \citenamefont {Pan}}]{He:2015aa}%
  \BibitemOpen
  \bibfield  {author} {\bibinfo {author} {\bibfnamefont {Y.}~\bibnamefont
  {He}}, \bibinfo {author} {\bibfnamefont {Y.~M.}\ \bibnamefont {He}}, \bibinfo
  {author} {\bibfnamefont {J.}~\bibnamefont {Liu}}, \bibinfo {author}
  {\bibfnamefont {Y.~J.}\ \bibnamefont {Wei}}, \bibinfo {author} {\bibfnamefont
  {H.~Y.}\ \bibnamefont {Ram{\'\i}rez}}, \bibinfo {author} {\bibfnamefont
  {M.}~\bibnamefont {Atat{\"u}re}}, \bibinfo {author} {\bibfnamefont
  {C.}~\bibnamefont {Schneider}}, \bibinfo {author} {\bibfnamefont
  {M.}~\bibnamefont {Kamp}}, \bibinfo {author} {\bibfnamefont {S.}~\bibnamefont
  {H{\"o}fling}}, \bibinfo {author} {\bibfnamefont {C.~Y.}\ \bibnamefont
  {Lu}},\ and\ \bibinfo {author} {\bibfnamefont {J.~W.}\ \bibnamefont {Pan}},\
  }\bibfield  {title} {\bibinfo {title} {Dynamically controlled resonance
  fluorescence spectra from a doubly dressed single ingaas quantum dot},\
  }\href {https://doi.org/10.1103/PhysRevLett.114.097402} {\bibfield  {journal}
  {\bibinfo  {journal} {Physical Review Letters}\ }\textbf {\bibinfo {volume}
  {114}},\ \bibinfo {pages} {097402} (\bibinfo {year} {2015})}\BibitemShut
  {NoStop}%
\bibitem [{\citenamefont {Gustin}\ \emph {et~al.}(2021)\citenamefont {Gustin},
  \citenamefont {Hanschke}, \citenamefont {Boos}, \citenamefont {M{\"u}ller},
  \citenamefont {Kremser}, \citenamefont {Finley}, \citenamefont {Hughes},\
  and\ \citenamefont {M{\"u}ller}}]{Gustin:2021aa}%
  \BibitemOpen
  \bibfield  {author} {\bibinfo {author} {\bibfnamefont {C.}~\bibnamefont
  {Gustin}}, \bibinfo {author} {\bibfnamefont {L.}~\bibnamefont {Hanschke}},
  \bibinfo {author} {\bibfnamefont {K.}~\bibnamefont {Boos}}, \bibinfo {author}
  {\bibfnamefont {J.~R.~A.}\ \bibnamefont {M{\"u}ller}}, \bibinfo {author}
  {\bibfnamefont {M.}~\bibnamefont {Kremser}}, \bibinfo {author} {\bibfnamefont
  {J.~J.}\ \bibnamefont {Finley}}, \bibinfo {author} {\bibfnamefont
  {S.}~\bibnamefont {Hughes}},\ and\ \bibinfo {author} {\bibfnamefont
  {K.}~\bibnamefont {M{\"u}ller}},\ }\bibfield  {title} {\bibinfo {title}
  {High-resolution spectroscopy of a quantum dot driven bichromatically by two
  strong coherent fields},\ }\href
  {https://doi.org/10.1103/PhysRevResearch.3.013044} {\bibfield  {journal}
  {\bibinfo  {journal} {Physical Review Research}\ }\textbf {\bibinfo {volume}
  {3}},\ \bibinfo {pages} {013044} (\bibinfo {year} {2021})}\BibitemShut
  {NoStop}%
\bibitem [{\citenamefont {Lukin}\ \emph {et~al.}(2020)\citenamefont {Lukin},
  \citenamefont {White}, \citenamefont {Trivedi}, \citenamefont {Guidry},
  \citenamefont {Morioka}, \citenamefont {Babin}, \citenamefont {Soykal},
  \citenamefont {Ul-Hassan}, \citenamefont {Son}, \citenamefont {Ohshima},
  \citenamefont {Vasireddy}, \citenamefont {Nasr}, \citenamefont {Sun},
  \citenamefont {MacLean}, \citenamefont {Dory}, \citenamefont {Nanni},
  \citenamefont {Wrachtrup}, \citenamefont {Kaiser},\ and\ \citenamefont {Vu{\v
  c}kovi{\'c}}}]{Lukin:2020aa}%
  \BibitemOpen
  \bibfield  {author} {\bibinfo {author} {\bibfnamefont {D.~M.}\ \bibnamefont
  {Lukin}}, \bibinfo {author} {\bibfnamefont {A.~D.}\ \bibnamefont {White}},
  \bibinfo {author} {\bibfnamefont {R.}~\bibnamefont {Trivedi}}, \bibinfo
  {author} {\bibfnamefont {M.~A.}\ \bibnamefont {Guidry}}, \bibinfo {author}
  {\bibfnamefont {N.}~\bibnamefont {Morioka}}, \bibinfo {author} {\bibfnamefont
  {C.}~\bibnamefont {Babin}}, \bibinfo {author} {\bibfnamefont {{\"O}.~O.}\
  \bibnamefont {Soykal}}, \bibinfo {author} {\bibfnamefont {J.}~\bibnamefont
  {Ul-Hassan}}, \bibinfo {author} {\bibfnamefont {N.~T.}\ \bibnamefont {Son}},
  \bibinfo {author} {\bibfnamefont {T.}~\bibnamefont {Ohshima}}, \bibinfo
  {author} {\bibfnamefont {P.~K.}\ \bibnamefont {Vasireddy}}, \bibinfo {author}
  {\bibfnamefont {M.~H.}\ \bibnamefont {Nasr}}, \bibinfo {author}
  {\bibfnamefont {S.}~\bibnamefont {Sun}}, \bibinfo {author} {\bibfnamefont
  {J.-P.~W.}\ \bibnamefont {MacLean}}, \bibinfo {author} {\bibfnamefont
  {C.}~\bibnamefont {Dory}}, \bibinfo {author} {\bibfnamefont {E.~A.}\
  \bibnamefont {Nanni}}, \bibinfo {author} {\bibfnamefont {J.}~\bibnamefont
  {Wrachtrup}}, \bibinfo {author} {\bibfnamefont {F.}~\bibnamefont {Kaiser}},\
  and\ \bibinfo {author} {\bibfnamefont {J.}~\bibnamefont {Vu{\v
  c}kovi{\'c}}},\ }\bibfield  {title} {\bibinfo {title} {Spectrally
  reconfigurable quantum emitters enabled by optimized fast modulation},\
  }\href {https://doi.org/10.1038/s41534-020-00310-0} {\bibfield  {journal}
  {\bibinfo  {journal} {npj Quantum Information}\ }\textbf {\bibinfo {volume}
  {6}},\ \bibinfo {pages} {80} (\bibinfo {year} {2020})}\BibitemShut {NoStop}%
\bibitem [{\citenamefont {Groll}\ \emph {et~al.}(2026)\citenamefont {Groll},
  \citenamefont {Wigger}, \citenamefont {Wei{\ss}}, \citenamefont {Yuan},
  \citenamefont {Kuznetsov}, \citenamefont {Hern{\'a}ndez-M{\'\i}nguez},
  \citenamefont {Krenner}, \citenamefont {Kuhn},\ and\ \citenamefont
  {Machnikowski}}]{Groll:2026aa}%
  \BibitemOpen
  \bibfield  {author} {\bibinfo {author} {\bibfnamefont {D.}~\bibnamefont
  {Groll}}, \bibinfo {author} {\bibfnamefont {D.}~\bibnamefont {Wigger}},
  \bibinfo {author} {\bibfnamefont {M.}~\bibnamefont {Wei{\ss}}}, \bibinfo
  {author} {\bibfnamefont {M.}~\bibnamefont {Yuan}}, \bibinfo {author}
  {\bibfnamefont {A.}~\bibnamefont {Kuznetsov}}, \bibinfo {author}
  {\bibfnamefont {A.}~\bibnamefont {Hern{\'a}ndez-M{\'\i}nguez}}, \bibinfo
  {author} {\bibfnamefont {H.~J.}\ \bibnamefont {Krenner}}, \bibinfo {author}
  {\bibfnamefont {T.}~\bibnamefont {Kuhn}},\ and\ \bibinfo {author}
  {\bibfnamefont {P.}~\bibnamefont {Machnikowski}},\ }\bibfield  {title}
  {\bibinfo {title} {Topical review on acousto-optical floquet engineering of
  single-photon emitters},\ }\href {https://doi.org/10.1088/2515-7647/ae4f43}
  {\bibfield  {journal} {\bibinfo  {journal} {Journal of Physics: Photonics}\
  }\textbf {\bibinfo {volume} {8}},\ \bibinfo {pages} {012008} (\bibinfo {year}
  {2026})}\BibitemShut {NoStop}%
\bibitem [{\citenamefont {Ant{\'o}n}\ \emph {et~al.}(2017)\citenamefont
  {Ant{\'o}n}, \citenamefont {Maede-Razavi}, \citenamefont {Carre{\~n}o},
  \citenamefont {Thanopulos},\ and\ \citenamefont {Paspalakis}}]{Anton:2017aa}%
  \BibitemOpen
  \bibfield  {author} {\bibinfo {author} {\bibfnamefont {M.~A.}\ \bibnamefont
  {Ant{\'o}n}}, \bibinfo {author} {\bibfnamefont {S.}~\bibnamefont
  {Maede-Razavi}}, \bibinfo {author} {\bibfnamefont {F.}~\bibnamefont
  {Carre{\~n}o}}, \bibinfo {author} {\bibfnamefont {I.}~\bibnamefont
  {Thanopulos}},\ and\ \bibinfo {author} {\bibfnamefont {E.}~\bibnamefont
  {Paspalakis}},\ }\bibfield  {title} {\bibinfo {title} {Optical and microwave
  control of resonance fluorescence and squeezing spectra in a polar
  molecule},\ }\href {https://doi.org/10.1103/PhysRevA.96.063812} {\bibfield
  {journal} {\bibinfo  {journal} {Physical Review A}\ }\textbf {\bibinfo
  {volume} {96}},\ \bibinfo {pages} {063812} (\bibinfo {year}
  {2017})}\BibitemShut {NoStop}%
\bibitem [{\citenamefont {Munsch}\ \emph {et~al.}(2017)\citenamefont {Munsch},
  \citenamefont {Kuhlmann}, \citenamefont {Cadeddu}, \citenamefont
  {G{\'e}rard}, \citenamefont {Claudon}, \citenamefont {Poggio},\ and\
  \citenamefont {Warburton}}]{Munsch:2017aa}%
  \BibitemOpen
  \bibfield  {author} {\bibinfo {author} {\bibfnamefont {M.}~\bibnamefont
  {Munsch}}, \bibinfo {author} {\bibfnamefont {A.~V.}\ \bibnamefont
  {Kuhlmann}}, \bibinfo {author} {\bibfnamefont {D.}~\bibnamefont {Cadeddu}},
  \bibinfo {author} {\bibfnamefont {J.-M.}\ \bibnamefont {G{\'e}rard}},
  \bibinfo {author} {\bibfnamefont {J.}~\bibnamefont {Claudon}}, \bibinfo
  {author} {\bibfnamefont {M.}~\bibnamefont {Poggio}},\ and\ \bibinfo {author}
  {\bibfnamefont {R.~J.}\ \bibnamefont {Warburton}},\ }\bibfield  {title}
  {\bibinfo {title} {Resonant driving of a single photon emitter embedded in a
  mechanical oscillator},\ }\href {https://doi.org/10.1038/s41467-017-00097-3}
  {\bibfield  {journal} {\bibinfo  {journal} {Nature Communications}\ }\textbf
  {\bibinfo {volume} {8}},\ \bibinfo {pages} {76} (\bibinfo {year}
  {2017})}\BibitemShut {NoStop}%
\bibitem [{\citenamefont {Spinnler}\ \emph {et~al.}(2024)\citenamefont
  {Spinnler}, \citenamefont {Nguyen}, \citenamefont {Wang}, \citenamefont
  {Zhai}, \citenamefont {Javadi}, \citenamefont {Erbe}, \citenamefont {Scholz},
  \citenamefont {Wieck}, \citenamefont {Ludwig}, \citenamefont {Lodahl},
  \citenamefont {Midolo},\ and\ \citenamefont {Warburton}}]{Spinnler:2024aa}%
  \BibitemOpen
  \bibfield  {author} {\bibinfo {author} {\bibfnamefont {C.}~\bibnamefont
  {Spinnler}}, \bibinfo {author} {\bibfnamefont {G.~N.}\ \bibnamefont
  {Nguyen}}, \bibinfo {author} {\bibfnamefont {Y.}~\bibnamefont {Wang}},
  \bibinfo {author} {\bibfnamefont {L.}~\bibnamefont {Zhai}}, \bibinfo {author}
  {\bibfnamefont {A.}~\bibnamefont {Javadi}}, \bibinfo {author} {\bibfnamefont
  {M.}~\bibnamefont {Erbe}}, \bibinfo {author} {\bibfnamefont {S.}~\bibnamefont
  {Scholz}}, \bibinfo {author} {\bibfnamefont {A.~D.}\ \bibnamefont {Wieck}},
  \bibinfo {author} {\bibfnamefont {A.}~\bibnamefont {Ludwig}}, \bibinfo
  {author} {\bibfnamefont {P.}~\bibnamefont {Lodahl}}, \bibinfo {author}
  {\bibfnamefont {L.}~\bibnamefont {Midolo}},\ and\ \bibinfo {author}
  {\bibfnamefont {R.~J.}\ \bibnamefont {Warburton}},\ }\bibfield  {title}
  {\bibinfo {title} {A single-photon emitter coupled to a phononic-crystal
  resonator in the resolved-sideband regime},\ }\href
  {https://doi.org/10.1038/s41467-024-53882-2} {\bibfield  {journal} {\bibinfo
  {journal} {Nature Communications}\ }\textbf {\bibinfo {volume} {15}},\
  \bibinfo {pages} {9509} (\bibinfo {year} {2024})}\BibitemShut {NoStop}%
\bibitem [{\citenamefont {Wilson-Rae}\ \emph {et~al.}(2004)\citenamefont
  {Wilson-Rae}, \citenamefont {Zoller},\ and\ \citenamefont
  {Imamo{\=g}lu}}]{Wilson-Rae:2004aa}%
  \BibitemOpen
  \bibfield  {author} {\bibinfo {author} {\bibfnamefont {I.}~\bibnamefont
  {Wilson-Rae}}, \bibinfo {author} {\bibfnamefont {P.}~\bibnamefont {Zoller}},\
  and\ \bibinfo {author} {\bibfnamefont {A.}~\bibnamefont {Imamo{\=g}lu}},\
  }\bibfield  {title} {\bibinfo {title} {Laser cooling of a nanomechanical
  resonator mode to its quantum ground state},\ }\href
  {https://doi.org/10.1103/PhysRevLett.92.075507} {\bibfield  {journal}
  {\bibinfo  {journal} {Physical Review Letters}\ }\textbf {\bibinfo {volume}
  {92}},\ \bibinfo {pages} {075507} (\bibinfo {year} {2004})}\BibitemShut
  {NoStop}%
\bibitem [{\citenamefont {Rabl}(2010)}]{Rabl:2010aa}%
  \BibitemOpen
  \bibfield  {author} {\bibinfo {author} {\bibfnamefont {P.}~\bibnamefont
  {Rabl}},\ }\bibfield  {title} {\bibinfo {title} {Cooling of mechanical motion
  with a two-level system: The high-temperature regime},\ }\href
  {https://doi.org/10.1103/PhysRevB.82.165320} {\bibfield  {journal} {\bibinfo
  {journal} {Physical Review B}\ }\textbf {\bibinfo {volume} {82}},\ \bibinfo
  {pages} {165320} (\bibinfo {year} {2010})}\BibitemShut {NoStop}%
\bibitem [{\citenamefont {Schuetz}\ \emph {et~al.}(2015)\citenamefont
  {Schuetz}, \citenamefont {Kessler}, \citenamefont {Giedke}, \citenamefont
  {Vandersypen}, \citenamefont {Lukin},\ and\ \citenamefont
  {Cirac}}]{Schuetz:2015aa}%
  \BibitemOpen
  \bibfield  {author} {\bibinfo {author} {\bibfnamefont {M.~J.~A.}\
  \bibnamefont {Schuetz}}, \bibinfo {author} {\bibfnamefont {E.~M.}\
  \bibnamefont {Kessler}}, \bibinfo {author} {\bibfnamefont {G.}~\bibnamefont
  {Giedke}}, \bibinfo {author} {\bibfnamefont {L.~M.~K.}\ \bibnamefont
  {Vandersypen}}, \bibinfo {author} {\bibfnamefont {M.~D.}\ \bibnamefont
  {Lukin}},\ and\ \bibinfo {author} {\bibfnamefont {J.~I.}\ \bibnamefont
  {Cirac}},\ }\bibfield  {title} {\bibinfo {title} {Universal quantum
  transducers based on surface acoustic waves},\ }\href
  {https://doi.org/10.1103/PhysRevX.5.031031} {\bibfield  {journal} {\bibinfo
  {journal} {Physical Review X}\ }\textbf {\bibinfo {volume} {5}},\ \bibinfo
  {pages} {031031} (\bibinfo {year} {2015})}\BibitemShut {NoStop}%
\bibitem [{\citenamefont {Yan}\ \emph {et~al.}(2016)\citenamefont {Yan},
  \citenamefont {L{\"u}}, \citenamefont {Zheng},\ and\ \citenamefont
  {Zhao}}]{Yan:2016aa}%
  \BibitemOpen
  \bibfield  {author} {\bibinfo {author} {\bibfnamefont {Y.}~\bibnamefont
  {Yan}}, \bibinfo {author} {\bibfnamefont {Z.}~\bibnamefont {L{\"u}}},
  \bibinfo {author} {\bibfnamefont {H.}~\bibnamefont {Zheng}},\ and\ \bibinfo
  {author} {\bibfnamefont {Y.}~\bibnamefont {Zhao}},\ }\bibfield  {title}
  {\bibinfo {title} {Exotic fluorescence spectrum of a superconducting qubit
  driven simultaneously by longitudinal and transversal fields},\ }\href
  {https://doi.org/10.1103/PhysRevA.93.033812} {\bibfield  {journal} {\bibinfo
  {journal} {Physical Review A}\ }\textbf {\bibinfo {volume} {93}},\ \bibinfo
  {pages} {033812} (\bibinfo {year} {2016})}\BibitemShut {NoStop}%
\bibitem [{\citenamefont {Yan}\ \emph {et~al.}(2019)\citenamefont {Yan},
  \citenamefont {L{\"u}}, \citenamefont {Luo},\ and\ \citenamefont
  {Zheng}}]{Yan:2019aa}%
  \BibitemOpen
  \bibfield  {author} {\bibinfo {author} {\bibfnamefont {Y.}~\bibnamefont
  {Yan}}, \bibinfo {author} {\bibfnamefont {Z.}~\bibnamefont {L{\"u}}},
  \bibinfo {author} {\bibfnamefont {J.}~\bibnamefont {Luo}},\ and\ \bibinfo
  {author} {\bibfnamefont {H.}~\bibnamefont {Zheng}},\ }\bibfield  {title}
  {\bibinfo {title} {Role of generalized parity in the symmetry of the
  fluorescence spectrum from two-level systems under periodic frequency
  modulation},\ }\href {https://doi.org/10.1103/PhysRevA.100.013823} {\bibfield
   {journal} {\bibinfo  {journal} {Physical Review A}\ }\textbf {\bibinfo
  {volume} {100}},\ \bibinfo {pages} {013823} (\bibinfo {year}
  {2019})}\BibitemShut {NoStop}%
\bibitem [{\citenamefont {S{\"o}llner}\ \emph {et~al.}(2016)\citenamefont
  {S{\"o}llner}, \citenamefont {Midolo},\ and\ \citenamefont
  {Lodahl}}]{Sollner:2016aa}%
  \BibitemOpen
  \bibfield  {author} {\bibinfo {author} {\bibfnamefont {I.}~\bibnamefont
  {S{\"o}llner}}, \bibinfo {author} {\bibfnamefont {L.}~\bibnamefont
  {Midolo}},\ and\ \bibinfo {author} {\bibfnamefont {P.}~\bibnamefont
  {Lodahl}},\ }\bibfield  {title} {\bibinfo {title} {Deterministic
  single-phonon source triggered by a single photon},\ }\href
  {https://doi.org/10.1103/PhysRevLett.116.234301} {\bibfield  {journal}
  {\bibinfo  {journal} {Physical Review Letters}\ }\textbf {\bibinfo {volume}
  {116}},\ \bibinfo {pages} {234301} (\bibinfo {year} {2016})}\BibitemShut
  {NoStop}%
\bibitem [{\citenamefont {Lemonde}\ \emph {et~al.}(2018)\citenamefont
  {Lemonde}, \citenamefont {Meesala}, \citenamefont {Sipahigil}, \citenamefont
  {Schuetz}, \citenamefont {Lukin}, \citenamefont {Loncar},\ and\ \citenamefont
  {Rabl}}]{Lemonde:2018aa}%
  \BibitemOpen
  \bibfield  {author} {\bibinfo {author} {\bibfnamefont {M.~A.}\ \bibnamefont
  {Lemonde}}, \bibinfo {author} {\bibfnamefont {S.}~\bibnamefont {Meesala}},
  \bibinfo {author} {\bibfnamefont {A.}~\bibnamefont {Sipahigil}}, \bibinfo
  {author} {\bibfnamefont {M.~J.~A.}\ \bibnamefont {Schuetz}}, \bibinfo
  {author} {\bibfnamefont {M.~D.}\ \bibnamefont {Lukin}}, \bibinfo {author}
  {\bibfnamefont {M.}~\bibnamefont {Loncar}},\ and\ \bibinfo {author}
  {\bibfnamefont {P.}~\bibnamefont {Rabl}},\ }\bibfield  {title} {\bibinfo
  {title} {Phonon networks with silicon-vacancy centers in diamond
  waveguides},\ }\href {https://doi.org/10.1103/PhysRevLett.120.213603}
  {\bibfield  {journal} {\bibinfo  {journal} {Physical Review Letters}\
  }\textbf {\bibinfo {volume} {120}},\ \bibinfo {pages} {213603} (\bibinfo
  {year} {2018})}\BibitemShut {NoStop}%
\bibitem [{\citenamefont {Bayer}\ \emph {et~al.}(2002)\citenamefont {Bayer},
  \citenamefont {Ortner}, \citenamefont {Stern}, \citenamefont {Kuther},
  \citenamefont {Gorbunov}, \citenamefont {Forchel}, \citenamefont {Hawrylak},
  \citenamefont {Fafard}, \citenamefont {Hinzer}, \citenamefont {Reinecke},
  \citenamefont {Walck}, \citenamefont {Reithmaier}, \citenamefont {Klopf},\
  and\ \citenamefont {Sch{\"a}fer}}]{Bayer:2002aa}%
  \BibitemOpen
  \bibfield  {author} {\bibinfo {author} {\bibfnamefont {M.}~\bibnamefont
  {Bayer}}, \bibinfo {author} {\bibfnamefont {G.}~\bibnamefont {Ortner}},
  \bibinfo {author} {\bibfnamefont {O.}~\bibnamefont {Stern}}, \bibinfo
  {author} {\bibfnamefont {A.}~\bibnamefont {Kuther}}, \bibinfo {author}
  {\bibfnamefont {A.~A.}\ \bibnamefont {Gorbunov}}, \bibinfo {author}
  {\bibfnamefont {A.}~\bibnamefont {Forchel}}, \bibinfo {author} {\bibfnamefont
  {P.}~\bibnamefont {Hawrylak}}, \bibinfo {author} {\bibfnamefont
  {S.}~\bibnamefont {Fafard}}, \bibinfo {author} {\bibfnamefont
  {K.}~\bibnamefont {Hinzer}}, \bibinfo {author} {\bibfnamefont {T.~L.}\
  \bibnamefont {Reinecke}}, \bibinfo {author} {\bibfnamefont {S.~N.}\
  \bibnamefont {Walck}}, \bibinfo {author} {\bibfnamefont {J.~P.}\ \bibnamefont
  {Reithmaier}}, \bibinfo {author} {\bibfnamefont {F.}~\bibnamefont {Klopf}},\
  and\ \bibinfo {author} {\bibfnamefont {F.}~\bibnamefont {Sch{\"a}fer}},\
  }\bibfield  {title} {\bibinfo {title} {Fine structure of neutral and charged
  excitons in self-assembled in(ga)as/(al)gaas quantum dots},\ }\href
  {https://doi.org/10.1103/PhysRevB.65.195315} {\bibfield  {journal} {\bibinfo
  {journal} {Physical Review B}\ }\textbf {\bibinfo {volume} {65}},\ \bibinfo
  {pages} {195315} (\bibinfo {year} {2002})}\BibitemShut {NoStop}%
\bibitem [{sup()}]{supplement}%
  \BibitemOpen
  \href@noop {} {}\bibinfo {note} {See Supplemental Material at [URL will be
  inserted by publisher] for details of sample preparation and device
  characterization, measurement methods, theoretical calculation of resonance
  fluorescence spectrum, the doubly dressed state picture, and phonon cooling
  analysis, which includes
  Refs.~\cite{Wang:2024aa,Metcalfe:2010aa,Warburton:2002aa,Hendriks:1997aa,Lax:1966aa,Yan:2016aa,Groll:2026aa,DeCrescent:2022aa,Tsuchimoto:2022aa,Spinnler:2024aa,Kettler:2021aa,Teufel:2008aa}}\BibitemShut
  {NoStop}%
\bibitem [{\citenamefont {Aspelmeyer}\ \emph {et~al.}(2014)\citenamefont
  {Aspelmeyer}, \citenamefont {Kippenberg},\ and\ \citenamefont
  {Marquardt}}]{Aspelmeyer:2014aa}%
  \BibitemOpen
  \bibfield  {author} {\bibinfo {author} {\bibfnamefont {M.}~\bibnamefont
  {Aspelmeyer}}, \bibinfo {author} {\bibfnamefont {T.~J.}\ \bibnamefont
  {Kippenberg}},\ and\ \bibinfo {author} {\bibfnamefont {F.}~\bibnamefont
  {Marquardt}},\ }\bibfield  {title} {\bibinfo {title} {Cavity optomechanics},\
  }\href {https://doi.org/10.1103/RevModPhys.86.1391} {\bibfield  {journal}
  {\bibinfo  {journal} {Reviews of Modern Physics}\ }\textbf {\bibinfo {volume}
  {86}},\ \bibinfo {pages} {1391} (\bibinfo {year} {2014})}\BibitemShut
  {NoStop}%
\bibitem [{\citenamefont {Barzanjeh}\ \emph {et~al.}(2022)\citenamefont
  {Barzanjeh}, \citenamefont {Xuereb}, \citenamefont {Gr{\"o}blacher},
  \citenamefont {Paternostro}, \citenamefont {Regal},\ and\ \citenamefont
  {Weig}}]{Barzanjeh:2022aa}%
  \BibitemOpen
  \bibfield  {author} {\bibinfo {author} {\bibfnamefont {S.}~\bibnamefont
  {Barzanjeh}}, \bibinfo {author} {\bibfnamefont {A.}~\bibnamefont {Xuereb}},
  \bibinfo {author} {\bibfnamefont {S.}~\bibnamefont {Gr{\"o}blacher}},
  \bibinfo {author} {\bibfnamefont {M.}~\bibnamefont {Paternostro}}, \bibinfo
  {author} {\bibfnamefont {C.~A.}\ \bibnamefont {Regal}},\ and\ \bibinfo
  {author} {\bibfnamefont {E.~M.}\ \bibnamefont {Weig}},\ }\bibfield  {title}
  {\bibinfo {title} {Optomechanics for quantum technologies},\ }\href
  {https://doi.org/10.1038/s41567-021-01402-0} {\bibfield  {journal} {\bibinfo
  {journal} {Nature Physics}\ }\textbf {\bibinfo {volume} {18}},\ \bibinfo
  {pages} {15} (\bibinfo {year} {2022})}\BibitemShut {NoStop}%
\bibitem [{\citenamefont {Metcalfe}\ \emph {et~al.}(2010)\citenamefont
  {Metcalfe}, \citenamefont {Carr}, \citenamefont {Muller}, \citenamefont
  {Solomon},\ and\ \citenamefont {Lawall}}]{Metcalfe:2010aa}%
  \BibitemOpen
  \bibfield  {author} {\bibinfo {author} {\bibfnamefont {M.}~\bibnamefont
  {Metcalfe}}, \bibinfo {author} {\bibfnamefont {S.~M.}\ \bibnamefont {Carr}},
  \bibinfo {author} {\bibfnamefont {A.}~\bibnamefont {Muller}}, \bibinfo
  {author} {\bibfnamefont {G.~S.}\ \bibnamefont {Solomon}},\ and\ \bibinfo
  {author} {\bibfnamefont {J.}~\bibnamefont {Lawall}},\ }\bibfield  {title}
  {\bibinfo {title} {Resolved sideband emission of inas/gaas quantum dots
  strained by surface acoustic waves},\ }\href
  {https://doi.org/10.1103/PhysRevLett.105.037401} {\bibfield  {journal}
  {\bibinfo  {journal} {Physical Review Letters}\ }\textbf {\bibinfo {volume}
  {105}},\ \bibinfo {pages} {037401} (\bibinfo {year} {2010})}\BibitemShut
  {NoStop}%
\bibitem [{\citenamefont {Yeo}\ \emph {et~al.}(2014)\citenamefont {Yeo},
  \citenamefont {de~Assis}, \citenamefont {Gloppe}, \citenamefont
  {Dupont-Ferrier}, \citenamefont {Verlot}, \citenamefont {Malik},
  \citenamefont {Dupuy}, \citenamefont {Claudon}, \citenamefont {G{\'e}rard},
  \citenamefont {Auff{\`e}ves}, \citenamefont {Nogues}, \citenamefont
  {Seidelin}, \citenamefont {Poizat}, \citenamefont {Arcizet},\ and\
  \citenamefont {Richard}}]{Yeo:2014aa}%
  \BibitemOpen
  \bibfield  {author} {\bibinfo {author} {\bibfnamefont {I.}~\bibnamefont
  {Yeo}}, \bibinfo {author} {\bibfnamefont {P.-L.}\ \bibnamefont {de~Assis}},
  \bibinfo {author} {\bibfnamefont {A.}~\bibnamefont {Gloppe}}, \bibinfo
  {author} {\bibfnamefont {E.}~\bibnamefont {Dupont-Ferrier}}, \bibinfo
  {author} {\bibfnamefont {P.}~\bibnamefont {Verlot}}, \bibinfo {author}
  {\bibfnamefont {N.~S.}\ \bibnamefont {Malik}}, \bibinfo {author}
  {\bibfnamefont {E.}~\bibnamefont {Dupuy}}, \bibinfo {author} {\bibfnamefont
  {J.}~\bibnamefont {Claudon}}, \bibinfo {author} {\bibfnamefont {J.-M.}\
  \bibnamefont {G{\'e}rard}}, \bibinfo {author} {\bibfnamefont
  {A.}~\bibnamefont {Auff{\`e}ves}}, \bibinfo {author} {\bibfnamefont
  {G.}~\bibnamefont {Nogues}}, \bibinfo {author} {\bibfnamefont
  {S.}~\bibnamefont {Seidelin}}, \bibinfo {author} {\bibfnamefont {J.-P.}\
  \bibnamefont {Poizat}}, \bibinfo {author} {\bibfnamefont {O.}~\bibnamefont
  {Arcizet}},\ and\ \bibinfo {author} {\bibfnamefont {M.}~\bibnamefont
  {Richard}},\ }\bibfield  {title} {\bibinfo {title} {Strain-mediated coupling
  in a quantum dot--mechanical oscillator hybrid system},\ }\href
  {https://doi.org/10.1038/nnano.2013.274} {\bibfield  {journal} {\bibinfo
  {journal} {Nature Nanotechnology}\ }\textbf {\bibinfo {volume} {9}},\
  \bibinfo {pages} {106} (\bibinfo {year} {2014})}\BibitemShut {NoStop}%
\bibitem [{\citenamefont {Imany}\ \emph {et~al.}(2022)\citenamefont {Imany},
  \citenamefont {Wang}, \citenamefont {DeCrescent}, \citenamefont {Boutelle},
  \citenamefont {McDonald}, \citenamefont {Autry}, \citenamefont {Berweger},
  \citenamefont {Kabos}, \citenamefont {Nam}, \citenamefont {Mirin},\ and\
  \citenamefont {Silverman}}]{Imany:2022aa}%
  \BibitemOpen
  \bibfield  {author} {\bibinfo {author} {\bibfnamefont {P.}~\bibnamefont
  {Imany}}, \bibinfo {author} {\bibfnamefont {Z.}~\bibnamefont {Wang}},
  \bibinfo {author} {\bibfnamefont {R.~A.}\ \bibnamefont {DeCrescent}},
  \bibinfo {author} {\bibfnamefont {R.~C.}\ \bibnamefont {Boutelle}}, \bibinfo
  {author} {\bibfnamefont {C.~A.}\ \bibnamefont {McDonald}}, \bibinfo {author}
  {\bibfnamefont {T.}~\bibnamefont {Autry}}, \bibinfo {author} {\bibfnamefont
  {S.}~\bibnamefont {Berweger}}, \bibinfo {author} {\bibfnamefont
  {P.}~\bibnamefont {Kabos}}, \bibinfo {author} {\bibfnamefont {S.~W.}\
  \bibnamefont {Nam}}, \bibinfo {author} {\bibfnamefont {R.~P.}\ \bibnamefont
  {Mirin}},\ and\ \bibinfo {author} {\bibfnamefont {K.~L.}\ \bibnamefont
  {Silverman}},\ }\bibfield  {title} {\bibinfo {title} {Quantum phase
  modulation with acoustic cavities and quantum dots},\ }\href
  {https://doi.org/10.1364/OPTICA.451418} {\bibfield  {journal} {\bibinfo
  {journal} {Optica}\ }\textbf {\bibinfo {volume} {9}},\ \bibinfo {pages} {501}
  (\bibinfo {year} {2022})}\BibitemShut {NoStop}%
\bibitem [{\citenamefont {Golter}\ \emph {et~al.}(2016)\citenamefont {Golter},
  \citenamefont {Oo}, \citenamefont {Amezcua}, \citenamefont {Stewart},\ and\
  \citenamefont {Wang}}]{Golter:2016aa}%
  \BibitemOpen
  \bibfield  {author} {\bibinfo {author} {\bibfnamefont {D.~A.}\ \bibnamefont
  {Golter}}, \bibinfo {author} {\bibfnamefont {T.}~\bibnamefont {Oo}}, \bibinfo
  {author} {\bibfnamefont {M.}~\bibnamefont {Amezcua}}, \bibinfo {author}
  {\bibfnamefont {K.~A.}\ \bibnamefont {Stewart}},\ and\ \bibinfo {author}
  {\bibfnamefont {H.}~\bibnamefont {Wang}},\ }\bibfield  {title} {\bibinfo
  {title} {Optomechanical quantum control of a nitrogen-vacancy center in
  diamond},\ }\href {https://doi.org/10.1103/PhysRevLett.116.143602} {\bibfield
   {journal} {\bibinfo  {journal} {Physical Review Letters}\ }\textbf {\bibinfo
  {volume} {116}},\ \bibinfo {pages} {143602} (\bibinfo {year}
  {2016})}\BibitemShut {NoStop}%
\bibitem [{\citenamefont {Lee}\ \emph {et~al.}(2016)\citenamefont {Lee},
  \citenamefont {Lee}, \citenamefont {Ovartchaiyapong}, \citenamefont
  {Minguzzi}, \citenamefont {Maze},\ and\ \citenamefont
  {Bleszynski~Jayich}}]{Lee:2016aa}%
  \BibitemOpen
  \bibfield  {author} {\bibinfo {author} {\bibfnamefont {K.~W.}\ \bibnamefont
  {Lee}}, \bibinfo {author} {\bibfnamefont {D.}~\bibnamefont {Lee}}, \bibinfo
  {author} {\bibfnamefont {P.}~\bibnamefont {Ovartchaiyapong}}, \bibinfo
  {author} {\bibfnamefont {J.}~\bibnamefont {Minguzzi}}, \bibinfo {author}
  {\bibfnamefont {J.~R.}\ \bibnamefont {Maze}},\ and\ \bibinfo {author}
  {\bibfnamefont {A.~C.}\ \bibnamefont {Bleszynski~Jayich}},\ }\bibfield
  {title} {\bibinfo {title} {Strain coupling of a mechanical resonator to a
  single quantum emitter in diamond},\ }\href
  {https://doi.org/10.1103/PhysRevApplied.6.034005} {\bibfield  {journal}
  {\bibinfo  {journal} {Physical Review Applied}\ }\textbf {\bibinfo {volume}
  {6}},\ \bibinfo {pages} {034005} (\bibinfo {year} {2016})}\BibitemShut
  {NoStop}%
\bibitem [{\citenamefont {Ohta}\ \emph {et~al.}(2021)\citenamefont {Ohta},
  \citenamefont {Herpin}, \citenamefont {Bastidas}, \citenamefont {Tawara},
  \citenamefont {Yamaguchi},\ and\ \citenamefont {Okamoto}}]{Ohta:2021aa}%
  \BibitemOpen
  \bibfield  {author} {\bibinfo {author} {\bibfnamefont {R.}~\bibnamefont
  {Ohta}}, \bibinfo {author} {\bibfnamefont {L.}~\bibnamefont {Herpin}},
  \bibinfo {author} {\bibfnamefont {V.~M.}\ \bibnamefont {Bastidas}}, \bibinfo
  {author} {\bibfnamefont {T.}~\bibnamefont {Tawara}}, \bibinfo {author}
  {\bibfnamefont {H.}~\bibnamefont {Yamaguchi}},\ and\ \bibinfo {author}
  {\bibfnamefont {H.}~\bibnamefont {Okamoto}},\ }\bibfield  {title} {\bibinfo
  {title} {Rare-earth-mediated optomechanical system in the reversed
  dissipation regime},\ }\href {https://doi.org/10.1103/PhysRevLett.126.047404}
  {\bibfield  {journal} {\bibinfo  {journal} {Physical Review Letters}\
  }\textbf {\bibinfo {volume} {126}},\ \bibinfo {pages} {047404} (\bibinfo
  {year} {2021})}\BibitemShut {NoStop}%
\bibitem [{\citenamefont {Zalalutdinov}\ \emph {et~al.}(2021)\citenamefont
  {Zalalutdinov}, \citenamefont {Robinson}, \citenamefont {Fonseca},
  \citenamefont {LaGasse}, \citenamefont {Pandey}, \citenamefont {Lindsay},
  \citenamefont {Reinecke}, \citenamefont {Photiadis}, \citenamefont
  {Culbertson}, \citenamefont {Cress},\ and\ \citenamefont
  {Houston}}]{Zalalutdinov:2021aa}%
  \BibitemOpen
  \bibfield  {author} {\bibinfo {author} {\bibfnamefont {M.~K.}\ \bibnamefont
  {Zalalutdinov}}, \bibinfo {author} {\bibfnamefont {J.~T.}\ \bibnamefont
  {Robinson}}, \bibinfo {author} {\bibfnamefont {J.~J.}\ \bibnamefont
  {Fonseca}}, \bibinfo {author} {\bibfnamefont {S.~W.}\ \bibnamefont
  {LaGasse}}, \bibinfo {author} {\bibfnamefont {T.}~\bibnamefont {Pandey}},
  \bibinfo {author} {\bibfnamefont {L.~R.}\ \bibnamefont {Lindsay}}, \bibinfo
  {author} {\bibfnamefont {T.~L.}\ \bibnamefont {Reinecke}}, \bibinfo {author}
  {\bibfnamefont {D.~M.}\ \bibnamefont {Photiadis}}, \bibinfo {author}
  {\bibfnamefont {J.~C.}\ \bibnamefont {Culbertson}}, \bibinfo {author}
  {\bibfnamefont {C.~D.}\ \bibnamefont {Cress}},\ and\ \bibinfo {author}
  {\bibfnamefont {B.~H.}\ \bibnamefont {Houston}},\ }\bibfield  {title}
  {\bibinfo {title} {Acoustic cavities in 2d heterostructures},\ }\href
  {https://doi.org/10.1038/s41467-021-23359-7} {\bibfield  {journal} {\bibinfo
  {journal} {Nature Communications}\ }\textbf {\bibinfo {volume} {12}},\
  \bibinfo {pages} {3267} (\bibinfo {year} {2021})}\BibitemShut {NoStop}%
\bibitem [{\citenamefont {Patel}\ \emph {et~al.}(2024)\citenamefont {Patel},
  \citenamefont {Parto}, \citenamefont {Choquer}, \citenamefont {Lewis},
  \citenamefont {Umezawa}, \citenamefont {Hellman}, \citenamefont
  {Polishchuk},\ and\ \citenamefont {Moody}}]{Patel:2024aa}%
  \BibitemOpen
  \bibfield  {author} {\bibinfo {author} {\bibfnamefont {S.~D.}\ \bibnamefont
  {Patel}}, \bibinfo {author} {\bibfnamefont {K.}~\bibnamefont {Parto}},
  \bibinfo {author} {\bibfnamefont {M.}~\bibnamefont {Choquer}}, \bibinfo
  {author} {\bibfnamefont {N.}~\bibnamefont {Lewis}}, \bibinfo {author}
  {\bibfnamefont {S.}~\bibnamefont {Umezawa}}, \bibinfo {author} {\bibfnamefont
  {L.}~\bibnamefont {Hellman}}, \bibinfo {author} {\bibfnamefont
  {D.}~\bibnamefont {Polishchuk}},\ and\ \bibinfo {author} {\bibfnamefont
  {G.}~\bibnamefont {Moody}},\ }\bibfield  {title} {\bibinfo {title} {Surface
  acoustic wave cavity optomechanics with atomically thin {\$}h{\$}-bn and
  {\$}{\{}{$\backslash$}mathrm{\{}wse{\}}{\}}{\_}{\{}2{\}}{\$} single-photon
  emitters},\ }\href {https://doi.org/10.1103/PRXQuantum.5.010330} {\bibfield
  {journal} {\bibinfo  {journal} {PRX Quantum}\ }\textbf {\bibinfo {volume}
  {5}},\ \bibinfo {pages} {010330} (\bibinfo {year} {2024})}\BibitemShut
  {NoStop}%
\bibitem [{\citenamefont {Senellart}\ \emph {et~al.}(2017)\citenamefont
  {Senellart}, \citenamefont {Solomon},\ and\ \citenamefont
  {White}}]{Senellart:2017aa}%
  \BibitemOpen
  \bibfield  {author} {\bibinfo {author} {\bibfnamefont {P.}~\bibnamefont
  {Senellart}}, \bibinfo {author} {\bibfnamefont {G.}~\bibnamefont {Solomon}},\
  and\ \bibinfo {author} {\bibfnamefont {A.}~\bibnamefont {White}},\ }\bibfield
   {title} {\bibinfo {title} {High-performance semiconductor quantum-dot
  single-photon sources},\ }\href {https://doi.org/10.1038/nnano.2017.218}
  {\bibfield  {journal} {\bibinfo  {journal} {Nature Nanotechnology}\ }\textbf
  {\bibinfo {volume} {12}},\ \bibinfo {pages} {1026} (\bibinfo {year}
  {2017})}\BibitemShut {NoStop}%
\bibitem [{\citenamefont {DeCrescent}\ \emph {et~al.}(2024)\citenamefont
  {DeCrescent}, \citenamefont {Wang}, \citenamefont {Bush}, \citenamefont
  {Imany}, \citenamefont {Kwiatkowski}, \citenamefont {Reddy}, \citenamefont
  {Nam}, \citenamefont {Mirin},\ and\ \citenamefont
  {Silverman}}]{DeCrescent:2024aa}%
  \BibitemOpen
  \bibfield  {author} {\bibinfo {author} {\bibfnamefont {R.~A.}\ \bibnamefont
  {DeCrescent}}, \bibinfo {author} {\bibfnamefont {Z.}~\bibnamefont {Wang}},
  \bibinfo {author} {\bibfnamefont {J.~T.}\ \bibnamefont {Bush}}, \bibinfo
  {author} {\bibfnamefont {P.}~\bibnamefont {Imany}}, \bibinfo {author}
  {\bibfnamefont {A.}~\bibnamefont {Kwiatkowski}}, \bibinfo {author}
  {\bibfnamefont {D.~V.}\ \bibnamefont {Reddy}}, \bibinfo {author}
  {\bibfnamefont {S.~W.}\ \bibnamefont {Nam}}, \bibinfo {author} {\bibfnamefont
  {R.~P.}\ \bibnamefont {Mirin}},\ and\ \bibinfo {author} {\bibfnamefont
  {K.~L.}\ \bibnamefont {Silverman}},\ }\bibfield  {title} {\bibinfo {title}
  {Coherent dynamics in an optical quantum dot with phonons and photons},\
  }\href {https://doi.org/10.1364/OPTICA.537726} {\bibfield  {journal}
  {\bibinfo  {journal} {Optica}\ }\textbf {\bibinfo {volume} {11}},\ \bibinfo
  {pages} {1526} (\bibinfo {year} {2024})}\BibitemShut {NoStop}%
\bibitem [{\citenamefont {Wei{\ss}}\ \emph {et~al.}(2021)\citenamefont
  {Wei{\ss}}, \citenamefont {Wigger}, \citenamefont {N{\"a}gele}, \citenamefont
  {M{\"u}ller}, \citenamefont {Finley}, \citenamefont {Kuhn}, \citenamefont
  {Machnikowski},\ and\ \citenamefont {Krenner}}]{Weiss:2021aa}%
  \BibitemOpen
  \bibfield  {author} {\bibinfo {author} {\bibfnamefont {M.}~\bibnamefont
  {Wei{\ss}}}, \bibinfo {author} {\bibfnamefont {D.}~\bibnamefont {Wigger}},
  \bibinfo {author} {\bibfnamefont {M.}~\bibnamefont {N{\"a}gele}}, \bibinfo
  {author} {\bibfnamefont {K.}~\bibnamefont {M{\"u}ller}}, \bibinfo {author}
  {\bibfnamefont {J.~J.}\ \bibnamefont {Finley}}, \bibinfo {author}
  {\bibfnamefont {T.}~\bibnamefont {Kuhn}}, \bibinfo {author} {\bibfnamefont
  {P.}~\bibnamefont {Machnikowski}},\ and\ \bibinfo {author} {\bibfnamefont
  {H.~J.}\ \bibnamefont {Krenner}},\ }\bibfield  {title} {\bibinfo {title}
  {Optomechanical wave mixing by a single quantum dot},\ }\href
  {https://doi.org/10.1364/OPTICA.412201} {\bibfield  {journal} {\bibinfo
  {journal} {Optica}\ }\textbf {\bibinfo {volume} {8}},\ \bibinfo {pages} {291}
  (\bibinfo {year} {2021})}\BibitemShut {NoStop}%
\bibitem [{\citenamefont {Kabuss}\ \emph {et~al.}(2012)\citenamefont {Kabuss},
  \citenamefont {Carmele}, \citenamefont {Brandes},\ and\ \citenamefont
  {Knorr}}]{Kabuss:2012aa}%
  \BibitemOpen
  \bibfield  {author} {\bibinfo {author} {\bibfnamefont {J.}~\bibnamefont
  {Kabuss}}, \bibinfo {author} {\bibfnamefont {A.}~\bibnamefont {Carmele}},
  \bibinfo {author} {\bibfnamefont {T.}~\bibnamefont {Brandes}},\ and\ \bibinfo
  {author} {\bibfnamefont {A.}~\bibnamefont {Knorr}},\ }\bibfield  {title}
  {\bibinfo {title} {Optically driven quantum dots as source of coherent cavity
  phonons: A proposal for a phonon laser scheme},\ }\href
  {https://doi.org/10.1103/PhysRevLett.109.054301} {\bibfield  {journal}
  {\bibinfo  {journal} {Physical Review Letters}\ }\textbf {\bibinfo {volume}
  {109}},\ \bibinfo {pages} {054301} (\bibinfo {year} {2012})}\BibitemShut
  {NoStop}%
\bibitem [{\citenamefont {Kepesidis}\ \emph {et~al.}(2013)\citenamefont
  {Kepesidis}, \citenamefont {Bennett}, \citenamefont {Portolan}, \citenamefont
  {Lukin},\ and\ \citenamefont {Rabl}}]{Kepesidis:2013aa}%
  \BibitemOpen
  \bibfield  {author} {\bibinfo {author} {\bibfnamefont {K.~V.}\ \bibnamefont
  {Kepesidis}}, \bibinfo {author} {\bibfnamefont {S.~D.}\ \bibnamefont
  {Bennett}}, \bibinfo {author} {\bibfnamefont {S.}~\bibnamefont {Portolan}},
  \bibinfo {author} {\bibfnamefont {M.~D.}\ \bibnamefont {Lukin}},\ and\
  \bibinfo {author} {\bibfnamefont {P.}~\bibnamefont {Rabl}},\ }\bibfield
  {title} {\bibinfo {title} {Phonon cooling and lasing with nitrogen-vacancy
  centers in diamond},\ }\href {https://doi.org/10.1103/PhysRevB.88.064105}
  {\bibfield  {journal} {\bibinfo  {journal} {Physical Review B}\ }\textbf
  {\bibinfo {volume} {88}},\ \bibinfo {pages} {064105} (\bibinfo {year}
  {2013})}\BibitemShut {NoStop}%
\bibitem [{\citenamefont {DeCrescent}\ \emph {et~al.}(2022)\citenamefont
  {DeCrescent}, \citenamefont {Wang}, \citenamefont {Imany}, \citenamefont
  {Boutelle}, \citenamefont {McDonald}, \citenamefont {Autry}, \citenamefont
  {Teufel}, \citenamefont {Nam}, \citenamefont {Mirin},\ and\ \citenamefont
  {Silverman}}]{DeCrescent:2022aa}%
  \BibitemOpen
  \bibfield  {author} {\bibinfo {author} {\bibfnamefont {R.~A.}\ \bibnamefont
  {DeCrescent}}, \bibinfo {author} {\bibfnamefont {Z.}~\bibnamefont {Wang}},
  \bibinfo {author} {\bibfnamefont {P.}~\bibnamefont {Imany}}, \bibinfo
  {author} {\bibfnamefont {R.~C.}\ \bibnamefont {Boutelle}}, \bibinfo {author}
  {\bibfnamefont {C.~A.}\ \bibnamefont {McDonald}}, \bibinfo {author}
  {\bibfnamefont {T.}~\bibnamefont {Autry}}, \bibinfo {author} {\bibfnamefont
  {J.~D.}\ \bibnamefont {Teufel}}, \bibinfo {author} {\bibfnamefont {S.~W.}\
  \bibnamefont {Nam}}, \bibinfo {author} {\bibfnamefont {R.~P.}\ \bibnamefont
  {Mirin}},\ and\ \bibinfo {author} {\bibfnamefont {K.~L.}\ \bibnamefont
  {Silverman}},\ }\bibfield  {title} {\bibinfo {title} {Large single-phonon
  optomechanical coupling between quantum dots and tightly confined surface
  acoustic waves in the quantum regime},\ }\href
  {https://doi.org/10.1103/PhysRevApplied.18.034067} {\bibfield  {journal}
  {\bibinfo  {journal} {Physical Review Applied}\ }\textbf {\bibinfo {volume}
  {18}},\ \bibinfo {pages} {034067} (\bibinfo {year} {2022})}\BibitemShut
  {NoStop}%
\bibitem [{\citenamefont {Tsuchimoto}\ \emph {et~al.}(2022)\citenamefont
  {Tsuchimoto}, \citenamefont {Sun}, \citenamefont {Togan}, \citenamefont
  {F{\"a}lt}, \citenamefont {Wegscheider}, \citenamefont {Wallraff},
  \citenamefont {Ensslin}, \citenamefont {{\.I}mamo{\u g}lu},\ and\
  \citenamefont {Kroner}}]{Tsuchimoto:2022aa}%
  \BibitemOpen
  \bibfield  {author} {\bibinfo {author} {\bibfnamefont {Y.}~\bibnamefont
  {Tsuchimoto}}, \bibinfo {author} {\bibfnamefont {Z.}~\bibnamefont {Sun}},
  \bibinfo {author} {\bibfnamefont {E.}~\bibnamefont {Togan}}, \bibinfo
  {author} {\bibfnamefont {S.}~\bibnamefont {F{\"a}lt}}, \bibinfo {author}
  {\bibfnamefont {W.}~\bibnamefont {Wegscheider}}, \bibinfo {author}
  {\bibfnamefont {A.}~\bibnamefont {Wallraff}}, \bibinfo {author}
  {\bibfnamefont {K.}~\bibnamefont {Ensslin}}, \bibinfo {author} {\bibfnamefont
  {A.}~\bibnamefont {{\.I}mamo{\u g}lu}},\ and\ \bibinfo {author}
  {\bibfnamefont {M.}~\bibnamefont {Kroner}},\ }\bibfield  {title} {\bibinfo
  {title} {Large-bandwidth transduction between an optical single quantum dot
  molecule and a superconducting resonator},\ }\href
  {https://doi.org/10.1103/PRXQuantum.3.030336} {\bibfield  {journal} {\bibinfo
   {journal} {PRX Quantum}\ }\textbf {\bibinfo {volume} {3}},\ \bibinfo {pages}
  {030336} (\bibinfo {year} {2022})}\BibitemShut {NoStop}%
\bibitem [{\citenamefont {Tomm}\ \emph {et~al.}(2021)\citenamefont {Tomm},
  \citenamefont {Javadi}, \citenamefont {Antoniadis}, \citenamefont {Najer},
  \citenamefont {L{\"o}bl}, \citenamefont {Korsch}, \citenamefont {Schott},
  \citenamefont {Valentin}, \citenamefont {Wieck}, \citenamefont {Ludwig},\
  and\ \citenamefont {Warburton}}]{Tomm:2021aa}%
  \BibitemOpen
  \bibfield  {author} {\bibinfo {author} {\bibfnamefont {N.}~\bibnamefont
  {Tomm}}, \bibinfo {author} {\bibfnamefont {A.}~\bibnamefont {Javadi}},
  \bibinfo {author} {\bibfnamefont {N.~O.}\ \bibnamefont {Antoniadis}},
  \bibinfo {author} {\bibfnamefont {D.}~\bibnamefont {Najer}}, \bibinfo
  {author} {\bibfnamefont {M.~C.}\ \bibnamefont {L{\"o}bl}}, \bibinfo {author}
  {\bibfnamefont {A.~R.}\ \bibnamefont {Korsch}}, \bibinfo {author}
  {\bibfnamefont {R.}~\bibnamefont {Schott}}, \bibinfo {author} {\bibfnamefont
  {S.~R.}\ \bibnamefont {Valentin}}, \bibinfo {author} {\bibfnamefont {A.~D.}\
  \bibnamefont {Wieck}}, \bibinfo {author} {\bibfnamefont {A.}~\bibnamefont
  {Ludwig}},\ and\ \bibinfo {author} {\bibfnamefont {R.~J.}\ \bibnamefont
  {Warburton}},\ }\bibfield  {title} {\bibinfo {title} {A bright and fast
  source of coherent single photons},\ }\href
  {https://doi.org/10.1038/s41565-020-00831-x} {\bibfield  {journal} {\bibinfo
  {journal} {Nature Nanotechnology}\ }\textbf {\bibinfo {volume} {16}},\
  \bibinfo {pages} {399} (\bibinfo {year} {2021})}\BibitemShut {NoStop}%
\bibitem [{\citenamefont {Ding}\ \emph {et~al.}(2025)\citenamefont {Ding},
  \citenamefont {Guo}, \citenamefont {Xu}, \citenamefont {Liu}, \citenamefont
  {Zou}, \citenamefont {Zhao}, \citenamefont {Ge}, \citenamefont {Zhang},
  \citenamefont {Liu}, \citenamefont {Wang}, \citenamefont {Chen},
  \citenamefont {Wang}, \citenamefont {He}, \citenamefont {Huo}, \citenamefont
  {Lu},\ and\ \citenamefont {Pan}}]{Ding:2025aa}%
  \BibitemOpen
  \bibfield  {author} {\bibinfo {author} {\bibfnamefont {X.}~\bibnamefont
  {Ding}}, \bibinfo {author} {\bibfnamefont {Y.-P.}\ \bibnamefont {Guo}},
  \bibinfo {author} {\bibfnamefont {M.-C.}\ \bibnamefont {Xu}}, \bibinfo
  {author} {\bibfnamefont {R.-Z.}\ \bibnamefont {Liu}}, \bibinfo {author}
  {\bibfnamefont {G.-Y.}\ \bibnamefont {Zou}}, \bibinfo {author} {\bibfnamefont
  {J.-Y.}\ \bibnamefont {Zhao}}, \bibinfo {author} {\bibfnamefont {Z.-X.}\
  \bibnamefont {Ge}}, \bibinfo {author} {\bibfnamefont {Q.-H.}\ \bibnamefont
  {Zhang}}, \bibinfo {author} {\bibfnamefont {H.-L.}\ \bibnamefont {Liu}},
  \bibinfo {author} {\bibfnamefont {L.-J.}\ \bibnamefont {Wang}}, \bibinfo
  {author} {\bibfnamefont {M.-C.}\ \bibnamefont {Chen}}, \bibinfo {author}
  {\bibfnamefont {H.}~\bibnamefont {Wang}}, \bibinfo {author} {\bibfnamefont
  {Y.-M.}\ \bibnamefont {He}}, \bibinfo {author} {\bibfnamefont {Y.-H.}\
  \bibnamefont {Huo}}, \bibinfo {author} {\bibfnamefont {C.-Y.}\ \bibnamefont
  {Lu}},\ and\ \bibinfo {author} {\bibfnamefont {J.-W.}\ \bibnamefont {Pan}},\
  }\bibfield  {title} {\bibinfo {title} {High-efficiency single-photon source
  above the loss-tolerant threshold for efficient linear optical quantum
  computing},\ }\href {https://doi.org/10.1038/s41566-025-01639-8} {\bibfield
  {journal} {\bibinfo  {journal} {Nature Photonics}\ }\textbf {\bibinfo
  {volume} {19}},\ \bibinfo {pages} {387} (\bibinfo {year} {2025})}\BibitemShut
  {NoStop}%
\bibitem [{\citenamefont {Carter}\ \emph {et~al.}(2018)\citenamefont {Carter},
  \citenamefont {Bracker}, \citenamefont {Bryant}, \citenamefont {Kim},
  \citenamefont {Kim}, \citenamefont {Zalalutdinov}, \citenamefont {Yakes},
  \citenamefont {Czarnocki}, \citenamefont {Casara}, \citenamefont
  {Scheibner},\ and\ \citenamefont {Gammon}}]{Carter:2018aa}%
  \BibitemOpen
  \bibfield  {author} {\bibinfo {author} {\bibfnamefont {S.~G.}\ \bibnamefont
  {Carter}}, \bibinfo {author} {\bibfnamefont {A.~S.}\ \bibnamefont {Bracker}},
  \bibinfo {author} {\bibfnamefont {G.~W.}\ \bibnamefont {Bryant}}, \bibinfo
  {author} {\bibfnamefont {M.}~\bibnamefont {Kim}}, \bibinfo {author}
  {\bibfnamefont {C.~S.}\ \bibnamefont {Kim}}, \bibinfo {author} {\bibfnamefont
  {M.~K.}\ \bibnamefont {Zalalutdinov}}, \bibinfo {author} {\bibfnamefont
  {M.~K.}\ \bibnamefont {Yakes}}, \bibinfo {author} {\bibfnamefont
  {C.}~\bibnamefont {Czarnocki}}, \bibinfo {author} {\bibfnamefont
  {J.}~\bibnamefont {Casara}}, \bibinfo {author} {\bibfnamefont
  {M.}~\bibnamefont {Scheibner}},\ and\ \bibinfo {author} {\bibfnamefont
  {D.}~\bibnamefont {Gammon}},\ }\bibfield  {title} {\bibinfo {title}
  {Spin-mechanical coupling of an inas quantum dot embedded in a mechanical
  resonator},\ }\href {https://doi.org/10.1103/PhysRevLett.121.246801}
  {\bibfield  {journal} {\bibinfo  {journal} {Physical Review Letters}\
  }\textbf {\bibinfo {volume} {121}},\ \bibinfo {pages} {246801} (\bibinfo
  {year} {2018})}\BibitemShut {NoStop}%
\bibitem [{\citenamefont {Maity}\ \emph {et~al.}(2020)\citenamefont {Maity},
  \citenamefont {Shao}, \citenamefont {Bogdanovi{\'c}}, \citenamefont
  {Meesala}, \citenamefont {Sohn}, \citenamefont {Sinclair}, \citenamefont
  {Pingault}, \citenamefont {Chalupnik}, \citenamefont {Chia}, \citenamefont
  {Zheng}, \citenamefont {Lai},\ and\ \citenamefont {Lon{\v
  c}ar}}]{Maity:2020aa}%
  \BibitemOpen
  \bibfield  {author} {\bibinfo {author} {\bibfnamefont {S.}~\bibnamefont
  {Maity}}, \bibinfo {author} {\bibfnamefont {L.}~\bibnamefont {Shao}},
  \bibinfo {author} {\bibfnamefont {S.}~\bibnamefont {Bogdanovi{\'c}}},
  \bibinfo {author} {\bibfnamefont {S.}~\bibnamefont {Meesala}}, \bibinfo
  {author} {\bibfnamefont {Y.-I.}\ \bibnamefont {Sohn}}, \bibinfo {author}
  {\bibfnamefont {N.}~\bibnamefont {Sinclair}}, \bibinfo {author}
  {\bibfnamefont {B.}~\bibnamefont {Pingault}}, \bibinfo {author}
  {\bibfnamefont {M.}~\bibnamefont {Chalupnik}}, \bibinfo {author}
  {\bibfnamefont {C.}~\bibnamefont {Chia}}, \bibinfo {author} {\bibfnamefont
  {L.}~\bibnamefont {Zheng}}, \bibinfo {author} {\bibfnamefont
  {K.}~\bibnamefont {Lai}},\ and\ \bibinfo {author} {\bibfnamefont
  {M.}~\bibnamefont {Lon{\v c}ar}},\ }\bibfield  {title} {\bibinfo {title}
  {Coherent acoustic control of a single silicon vacancy spin in diamond},\
  }\href {https://doi.org/10.1038/s41467-019-13822-x} {\bibfield  {journal}
  {\bibinfo  {journal} {Nature Communications}\ }\textbf {\bibinfo {volume}
  {11}},\ \bibinfo {pages} {193} (\bibinfo {year} {2020})}\BibitemShut
  {NoStop}%
\bibitem [{\citenamefont {Teufel}\ \emph {et~al.}(2008)\citenamefont {Teufel},
  \citenamefont {Harlow}, \citenamefont {Regal},\ and\ \citenamefont
  {Lehnert}}]{Teufel:2008aa}%
  \BibitemOpen
  \bibfield  {author} {\bibinfo {author} {\bibfnamefont {J.~D.}\ \bibnamefont
  {Teufel}}, \bibinfo {author} {\bibfnamefont {J.~W.}\ \bibnamefont {Harlow}},
  \bibinfo {author} {\bibfnamefont {C.~A.}\ \bibnamefont {Regal}},\ and\
  \bibinfo {author} {\bibfnamefont {K.~W.}\ \bibnamefont {Lehnert}},\
  }\bibfield  {title} {\bibinfo {title} {Dynamical backaction of microwave
  fields on a nanomechanical oscillator},\ }\href
  {https://doi.org/10.1103/PhysRevLett.101.197203} {\bibfield  {journal}
  {\bibinfo  {journal} {Physical Review Letters}\ }\textbf {\bibinfo {volume}
  {101}},\ \bibinfo {pages} {197203} (\bibinfo {year} {2008})}\BibitemShut
  {NoStop}%
\bibitem [{\citenamefont {Kettler}\ \emph {et~al.}(2021)\citenamefont
  {Kettler}, \citenamefont {Vaish}, \citenamefont {de~L{\'e}pinay},
  \citenamefont {Besga}, \citenamefont {de~Assis}, \citenamefont {Bourgeois},
  \citenamefont {Auff{\`e}ves}, \citenamefont {Richard}, \citenamefont
  {Claudon}, \citenamefont {G{\'e}rard}, \citenamefont {Pigeau}, \citenamefont
  {Arcizet}, \citenamefont {Verlot},\ and\ \citenamefont
  {Poizat}}]{Kettler:2021aa}%
  \BibitemOpen
  \bibfield  {author} {\bibinfo {author} {\bibfnamefont {J.}~\bibnamefont
  {Kettler}}, \bibinfo {author} {\bibfnamefont {N.}~\bibnamefont {Vaish}},
  \bibinfo {author} {\bibfnamefont {L.~M.}\ \bibnamefont {de~L{\'e}pinay}},
  \bibinfo {author} {\bibfnamefont {B.}~\bibnamefont {Besga}}, \bibinfo
  {author} {\bibfnamefont {P.-L.}\ \bibnamefont {de~Assis}}, \bibinfo {author}
  {\bibfnamefont {O.}~\bibnamefont {Bourgeois}}, \bibinfo {author}
  {\bibfnamefont {A.}~\bibnamefont {Auff{\`e}ves}}, \bibinfo {author}
  {\bibfnamefont {M.}~\bibnamefont {Richard}}, \bibinfo {author} {\bibfnamefont
  {J.}~\bibnamefont {Claudon}}, \bibinfo {author} {\bibfnamefont {J.-M.}\
  \bibnamefont {G{\'e}rard}}, \bibinfo {author} {\bibfnamefont
  {B.}~\bibnamefont {Pigeau}}, \bibinfo {author} {\bibfnamefont
  {O.}~\bibnamefont {Arcizet}}, \bibinfo {author} {\bibfnamefont
  {P.}~\bibnamefont {Verlot}},\ and\ \bibinfo {author} {\bibfnamefont {J.-P.}\
  \bibnamefont {Poizat}},\ }\bibfield  {title} {\bibinfo {title} {Inducing
  micromechanical motion by optical excitation of a single quantum dot},\
  }\href {https://doi.org/10.1038/s41565-020-00814-y} {\bibfield  {journal}
  {\bibinfo  {journal} {Nature Nanotechnology}\ }\textbf {\bibinfo {volume}
  {16}},\ \bibinfo {pages} {283} (\bibinfo {year} {2021})}\BibitemShut
  {NoStop}%
\bibitem [{\citenamefont {Lax}(1966)}]{Lax:1966aa}%
  \BibitemOpen
  \bibfield  {author} {\bibinfo {author} {\bibfnamefont {M.}~\bibnamefont
  {Lax}},\ }\bibfield  {title} {\bibinfo {title} {Quantum noise. iv. quantum
  theory of noise sources},\ }\href {https://doi.org/10.1103/PhysRev.145.110}
  {\bibfield  {journal} {\bibinfo  {journal} {Physical Review}\ }\textbf
  {\bibinfo {volume} {145}},\ \bibinfo {pages} {110} (\bibinfo {year}
  {1966})}\BibitemShut {NoStop}%
\bibitem [{\citenamefont {Warburton}\ \emph {et~al.}(2002)\citenamefont
  {Warburton}, \citenamefont {Schulhauser}, \citenamefont {Haft}, \citenamefont
  {Sch{\"a}flein}, \citenamefont {Karrai}, \citenamefont {Garcia},
  \citenamefont {Schoenfeld},\ and\ \citenamefont
  {Petroff}}]{Warburton:2002aa}%
  \BibitemOpen
  \bibfield  {author} {\bibinfo {author} {\bibfnamefont {R.~J.}\ \bibnamefont
  {Warburton}}, \bibinfo {author} {\bibfnamefont {C.}~\bibnamefont
  {Schulhauser}}, \bibinfo {author} {\bibfnamefont {D.}~\bibnamefont {Haft}},
  \bibinfo {author} {\bibfnamefont {C.}~\bibnamefont {Sch{\"a}flein}}, \bibinfo
  {author} {\bibfnamefont {K.}~\bibnamefont {Karrai}}, \bibinfo {author}
  {\bibfnamefont {J.~M.}\ \bibnamefont {Garcia}}, \bibinfo {author}
  {\bibfnamefont {W.}~\bibnamefont {Schoenfeld}},\ and\ \bibinfo {author}
  {\bibfnamefont {P.~M.}\ \bibnamefont {Petroff}},\ }\bibfield  {title}
  {\bibinfo {title} {Giant permanent dipole moments of excitons in
  semiconductor nanostructures},\ }\href
  {https://doi.org/10.1103/PhysRevB.65.113303} {\bibfield  {journal} {\bibinfo
  {journal} {Physical Review B}\ }\textbf {\bibinfo {volume} {65}},\ \bibinfo
  {pages} {113303} (\bibinfo {year} {2002})}\BibitemShut {NoStop}%
\bibitem [{\citenamefont {Hendriks}\ \emph {et~al.}(1997)\citenamefont
  {Hendriks}, \citenamefont {van Exter}, \citenamefont {Woerdman},
  \citenamefont {van Geelen}, \citenamefont {Weegels}, \citenamefont {Gulden},\
  and\ \citenamefont {Moser}}]{Hendriks:1997aa}%
  \BibitemOpen
  \bibfield  {author} {\bibinfo {author} {\bibfnamefont {R.~F.~M.}\
  \bibnamefont {Hendriks}}, \bibinfo {author} {\bibfnamefont {M.~P.}\
  \bibnamefont {van Exter}}, \bibinfo {author} {\bibfnamefont {J.~P.}\
  \bibnamefont {Woerdman}}, \bibinfo {author} {\bibfnamefont {A.}~\bibnamefont
  {van Geelen}}, \bibinfo {author} {\bibfnamefont {L.}~\bibnamefont {Weegels}},
  \bibinfo {author} {\bibfnamefont {K.~H.}\ \bibnamefont {Gulden}},\ and\
  \bibinfo {author} {\bibfnamefont {M.}~\bibnamefont {Moser}},\ }\bibfield
  {title} {\bibinfo {title} {Electro-optic birefringence in semiconductor
  vertical-cavity lasers},\ }\href {https://doi.org/10.1063/1.119340}
  {\bibfield  {journal} {\bibinfo  {journal} {Applied Physics Letters}\
  }\textbf {\bibinfo {volume} {71}},\ \bibinfo {pages} {2599} (\bibinfo {year}
  {1997})}\BibitemShut {NoStop}%
\bibitem [{\citenamefont {Wang}\ \emph {et~al.}(2024)\citenamefont {Wang},
  \citenamefont {DeCrescent}, \citenamefont {Imany}, \citenamefont {Bush},
  \citenamefont {Reddy}, \citenamefont {Woo~Nam}, \citenamefont {Mirin},\ and\
  \citenamefont {Silverman}}]{Wang:2024aa}%
  \BibitemOpen
  \bibfield  {author} {\bibinfo {author} {\bibfnamefont {Z.}~\bibnamefont
  {Wang}}, \bibinfo {author} {\bibfnamefont {R.~A.}\ \bibnamefont
  {DeCrescent}}, \bibinfo {author} {\bibfnamefont {P.}~\bibnamefont {Imany}},
  \bibinfo {author} {\bibfnamefont {J.~T.}\ \bibnamefont {Bush}}, \bibinfo
  {author} {\bibfnamefont {D.~V.}\ \bibnamefont {Reddy}}, \bibinfo {author}
  {\bibfnamefont {S.}~\bibnamefont {Woo~Nam}}, \bibinfo {author} {\bibfnamefont
  {R.~P.}\ \bibnamefont {Mirin}},\ and\ \bibinfo {author} {\bibfnamefont
  {K.~L.}\ \bibnamefont {Silverman}},\ }\bibfield  {title} {\bibinfo {title}
  {Gated inas quantum dots embedded in surface acoustic wave cavities for
  low-noise optomechanics},\ }\href {https://doi.org/10.1364/OE.538480}
  {\bibfield  {journal} {\bibinfo  {journal} {Optics Express}\ }\textbf
  {\bibinfo {volume} {32}},\ \bibinfo {pages} {38384} (\bibinfo {year}
  {2024})}\BibitemShut {NoStop}%
\end{thebibliography}
\providecommand{\noopsort}[1]{}\providecommand{\singleletter}[1]{#1}%

\end{document}


\preprint{APS/123-QED}

\title{Supplemental Material: Hybrid Acousto-Optical Double Dressing of a Two-Level System}
\author{Yuan Zhan}
\affiliation{JILA, University of Colorado, Boulder, Colorado 80309, USA}
\affiliation{Department of Physics, University of Colorado, Boulder, Colorado 80309, USA}

\author{Zixuan Wang}
\affiliation{Department of Physics, University of Colorado, Boulder, Colorado 80309, USA}
\affiliation{National Institute of Standards and Technology, Boulder, Colorado 80305, USA}

\author{Richard P. Mirin}
\affiliation{National Institute of Standards and Technology, Boulder, Colorado 80305, USA}

\author{Kevin L. Silverman}
\affiliation{National Institute of Standards and Technology, Boulder, Colorado 80305, USA}

\author{Shuo Sun}
\email{shuosun@colorado.edu}
\affiliation{JILA, University of Colorado, Boulder, Colorado 80309, USA}
\affiliation{Department of Physics, University of Colorado, Boulder, Colorado 80309, USA}

\date{\today}

\maketitle

\onecolumngrid

\section{Material synthesis and device fabrication}
\label{fab}

\begin{figure}[b!]
\centering
\includegraphics[width=0.4\columnwidth]{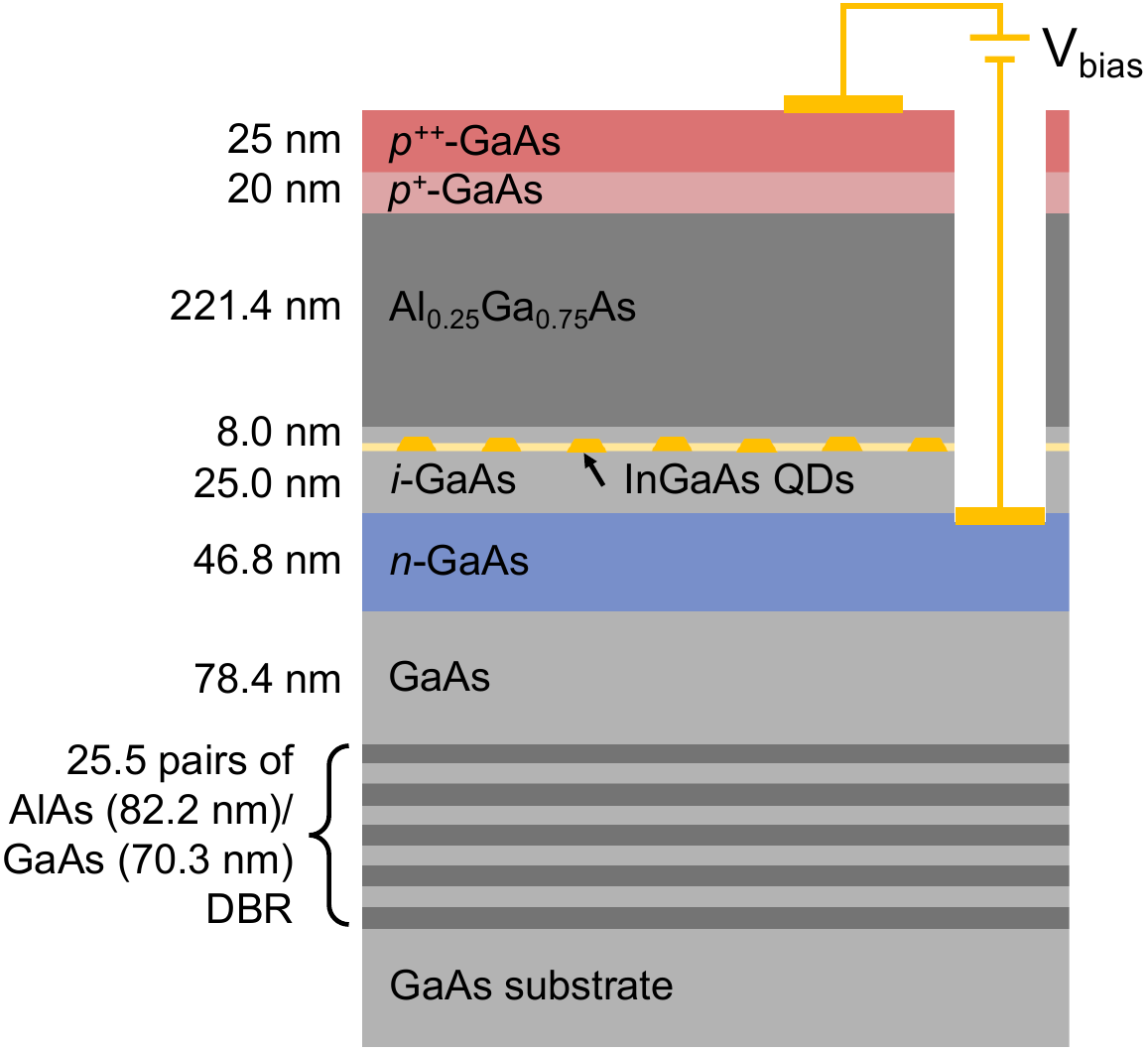}
\caption{Material stack structure of the sample containing the quantum dots.}
\label{sfig1}
\end{figure}

Figure~\ref{sfig1} shows the material stack structure of the sample containing the quantum dots. The sample is grown by molecular beam epitaxy on a GaAs wafer. We first grow 25.5 pairs of alternating AlAs (82.2~nm) and GaAs (70.3~nm) layers to form the bottom distributed Bragg reflector, followed by a 78.4-nm-thick GaAs buffer layer. We then grow the \textit{n}-doped GaAs with silicon doping ($\sim 2\times 10^{18}$~cm$^{-3}$ concentration) to a thickness of 46.8~nm. After that we grow a 25-nm-thick intrinsic GaAs layer, a monolayer of InGaAs quantum dots, and another 8-nm-thick intrinsic GaAs capping layer. We then grow a 221.4-nm-thick Al$_{0.25}$Ga$_{0.75}$As layer for current blockage. Finally, we grow the \textit{p}-doped GaAs layer with carbon doping at the top, which consists of a 20-nm-thick \textit{p}$^+$-doped layer ($\sim 2\times 10^{18}$~cm$^{-3}$ concentration) and a 25-nm-thick \textit{p}$^{++}$-doped layer ($\sim 1\times 10^{19}$~cm$^{-3}$ concentration). The quantum dot density is estimated to be $\sim 2$~$\mu$m$^{-2}$ for the device we use in the experiments.

After growth, devices are fabricated to combine charge control of the quantum dots with electrical control of the surface acoustic wave. The fabrication process starts with creating mirror trenches for the surface acoustic wave cavity via electron beam lithography. Next, electrical contacts to the \textit{n}- and \textit{p}-doped layers are fabricated. Electrical contacts to the \textit{n}-doped layer are created by first etching vias around the surface acoustic wave cavities down to approximately 80 nm above the \textit{n}-layer, and then depositing 400-nm-thick Ni$/$AuGe$/$Ni$/$Au via electron beam evaporation. Ohmic contacts to the \textit{p}-doped layer are created by depositing 200-nm-thick Pt$/$Au to the surface. Both metals are then annealed at 430 $^\circ$C to ensure good ohmic contacts. Following the electrical contacts for the diode, the \textit{n}-doped layer outside the surface acoustic wave cavity region is etched away to avoid cross talk between the DC bias and the coplanar waveguides to be defined. The coplanar waveguides are then formed by 400-nm-thick Ti$/$Au with their pattern defined by photolithography and metal liftoff. Finally, interdigital transducers made of 20-nm-thick aluminum are defined by electron beam lithography and metal liftoff. More details on device fabrication can be found in Ref.~\cite{Wang:2024aa}.

\section{Characterization of device and driving parameters}
\label{saw}
\begin{figure}[b!]
\centering
\includegraphics[width=0.92\columnwidth]{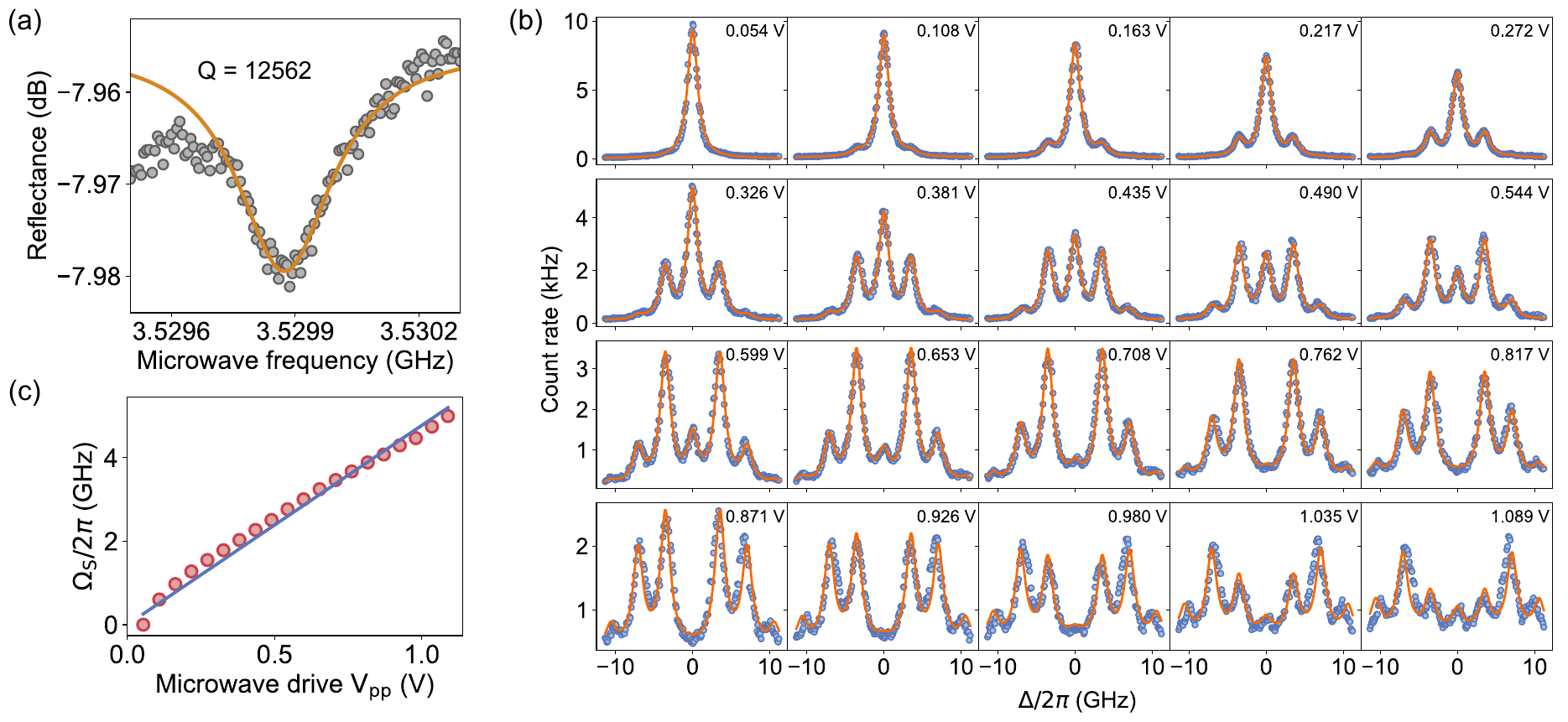}
\caption{(a) Microwave reflection spectrum of the surface acoustic wave cavity. The gray circles are the measured data and the orange solid line is the numerical fit to a Lorentzian function. (b) Quantum dot absorption spectra at varied acoustic driving strength. The blue circles are the measured data and the orange solid lines are the numerical fits to the analytical function given by Eq.~\ref{seq1}. The peak-to-peak voltage of the microwave drive (from a signal generator) used in each measurement is labeled in the right top corner of each panel. (c) Numerically extracted acoustic driving strength $\Omega_\text{S}$ as a function of the microwave drive amplitude $\text{V}_\text{pp}$. The red circles are the measured data and the blue solid line is a linear fit.}
\label{sfig2}
\end{figure}

We characterize the surface acoustic wave cavity at 5 K by measuring the microwave reflection spectrum with a vector network analyzer. By fitting the measured data [gray circles in Fig.~\ref{sfig2}(a)] to a Lorentzian function [orange solid line in Fig.~\ref{sfig2}(a)], we extract the cavity quality factor to be $Q=12,562$ and the resonance frequency to be $\omega_\text{S}/2\pi=3.5299$ GHz.

To extract the acoustic driving strength $\Omega_\text{S}$ used in our experiments, we measure the quantum dot absorption spectra as we vary the amplitude of the microwave drive used for launching the surface acoustic wave [Fig.~\ref{sfig2}(b)]. We fit the measured spectra to the following analytical function~\cite{Metcalfe:2010aa},
\begin{equation}
\begin{aligned}
P(\Delta)=\sum_{n=-\infty}^{\infty}\frac{J_n^2(\frac{2\Omega_\text{S}}{\omega_\text{S}})}{(\Delta-n\omega_\text{S})^2+\left(\frac{\Gamma}{2}\right)^2},
\end{aligned}
\label{seq1}
\end{equation}
where $\Delta$ is the frequency detuning between the laser and the bare quantum dot transition, $\Gamma$ is the optical linewidth of the quantum dot, $\omega_\text{S}$ is the resonance frequency of the surface acoustic wave cavity, and $J_n$ is the $n$-th order Bessel function of the first kind. In the fit, we fix the surface acoustic wave cavity resonance at its independently measured value from Fig.~\ref{sfig2}(a), $\omega_\text{S}/2\pi=3.5299$ GHz, and leave the acoustic driving strength $\Omega_\text{S}$ as an independent fit parameter for each spectrum.

The red circles in Fig.~\ref{sfig2}(c) show the extracted $\Omega_\text{S}$ from each measured spectrum, plotted as a function of the microwave drive amplitude $\text{V}_\text{pp}$ (the peak-to-peak voltage). As expected, the acoustic driving strength $\Omega_\text{S}$ scales linearly with $\text{V}_\text{pp}$, with a slope of $\frac{\Omega_\text{S}}{2\pi\text{V}_\text{pp}}=4.77$ GHz/V from a linear fit [blue solid line in Fig.~\ref{sfig2}(c)]. This coefficient provides a direct mapping from the experimentally controlled microwave drive amplitude to the acoustic driving strength $\Omega_\text{S}$ quoted in the main text.

The single-phonon coupling strength between the quantum dot and the surface acoustic wave cavity, $g_0$, is related to the acoustic driving strength $\Omega_\text{S}$ via $g_0=\sqrt{2}\Omega_\text{S}/\sqrt{\bar{m}}$. Here, $\bar{m}$ represents the average steady-state phonon occupancy in the cavity, defined as
\begin{equation}
\begin{aligned}
\bar{m}=\frac{P_\text{mw}}{\hbar\omega_\text{S}}\eta_\text{IDT}\tau,
\end{aligned}
\label{seq1-2}
\end{equation}
where $P_\text{mw}$ is the microwave power applied to the interdigital transducer (IDT), $\eta_\text{IDT}$ is the IDT electromechanical efficiency, and $\tau=Q/\omega_\text{S}=567$~ns is the phonon lifetime. We determine $P_\text{mw}$ using the microwave drive amplitude at the source ($\text{V}_\text{pp}$) and a calibrated line attenuation of $\sim 4.6$~dB. The efficiency $\eta_\text{IDT}\approx 0.485\%$ is extracted from the contrast of the reflectance dip in the microwave reflection spectrum [Fig.~\ref{sfig2}(a)]. Based on these parameters, we calculate a coupling strength of $g_0/2\pi=6.6$ kHz for the device studied in the main text. Furthermore, we estimate $\bar{m}\approx 1.4\times 10^{11}$ for the specific microwave power used in Figs.~2(c),~2(f), and~4.

\section{Measurement setup and rejection of the laser surface reflection}
\label{differentialDetection}
The sample is mounted in an optically accessible closed-cycle cryostat and cooled down to 5 K. Optical excitation and collection are performed with a confocal microscope using an objective lens with a numerical aperture of 0.7. Photoluminescence spectra [Fig.~1(b) of the main text] are obtained by exciting the quantum dot with a laser at 780 nm and measuring the signal with a grating spectrometer that has a resolution of 30 GHz. Absorption spectra [Figs.~1(c) and~1(e) of the main text] are obtained by exciting the quantum dot with a frequency tunable continuous-wave laser and recording the fluorescence intensity with a single-photon counting module and a pulse counter. To reject the spurious surface reflection of the laser, we collect signals in the orthogonal polarization with respect to the excitation. The signal is then coupled into a single-mode fiber, which spatially filters out any higher-order reflections transmitted through the cross polarizers. With this cross-polarization detection, we achieve an extinction ratio of $\sim 7\times 10^6$.

\begin{figure}[t!]
\centering
\includegraphics[width=0.85\columnwidth]{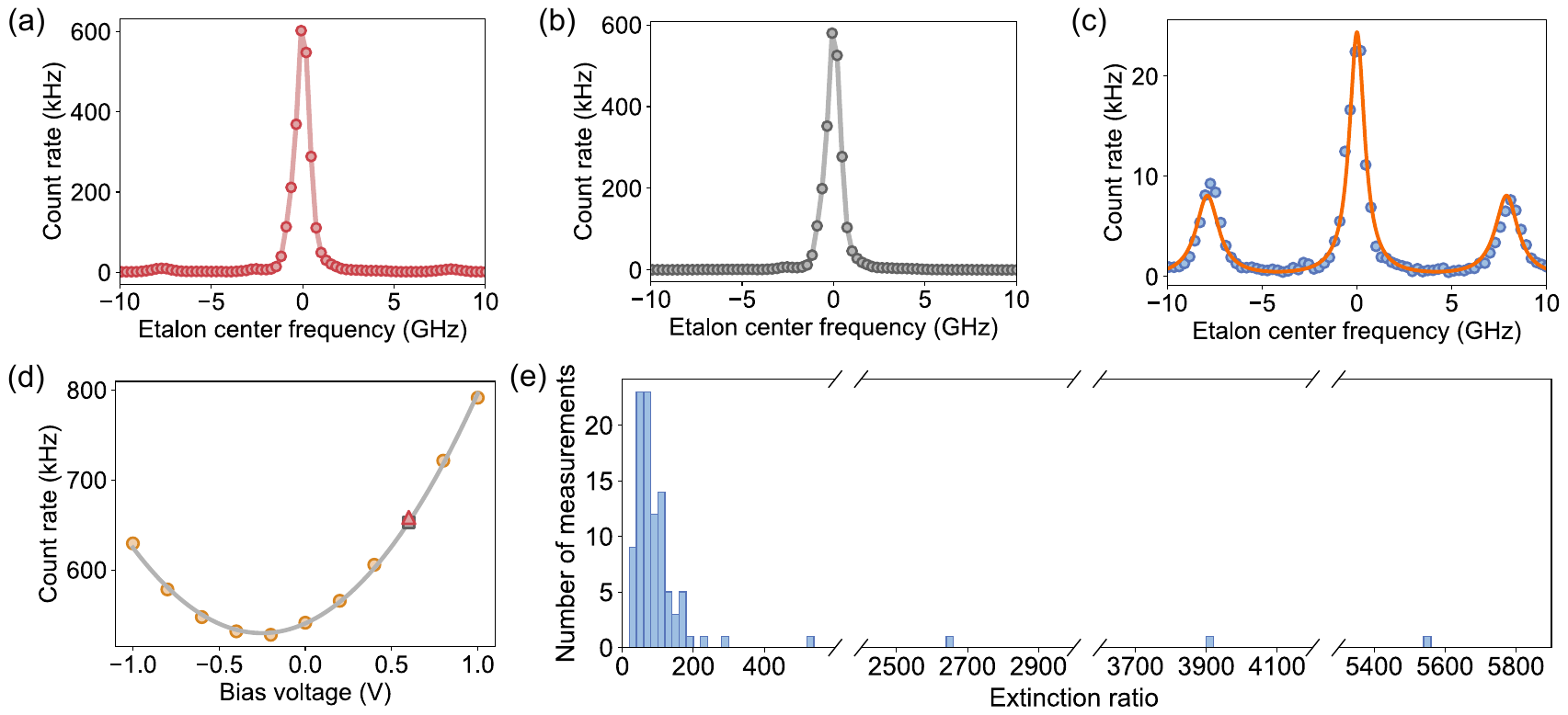}
\caption{(a) Resonance fluorescence spectrum of a quantum dot driven by a strong resonant laser, measured with only cross-polarization detection and without differential detection. The optical Rabi frequency is set at $\Omega_\text{L}/2\pi=7.9$ GHz. (b) Extracted laser reflection intensity for the spectrum shown in (a). (c) Differential signals between (a) and (b), showing the expected Mollow triplet. The blue circles are the measured data and the orange solid line is the fit to the theoretical Mollow triplet lineshape. (d) Recorded laser reflection intensity under cross-polarization detection as a function of the bias voltage. The 10 data points at off-resonant bias voltages (yellow circles) are numerically fit to a quadratic function (gray solid line) to extract the background count rate at the bias voltage of 0.6 V (gray square). The red triangle shows the truly measured laser reflection intensity at 0.6 V. (e) Histogram of the extinction ratio of the differential detection technique extracted from 100 repeated measurements as shown in (d).}
\label{sfig3}
\end{figure}

Spectrally resolved resonance fluorescence (Fig.~2 of the main text) at each laser power and detuning is obtained by recording the fluorescence intensity of the quantum dot as a function of the center frequency of a scanning etalon that has a linewidth of 525 MHz and a free spectral range of 20 GHz. Under strong laser excitation, the extinction ratio achieved with cross-polarization detection alone is still insufficient to isolate the quantum dot signal at the laser frequency. As an example, Fig.~\ref{sfig3}(a) shows the measured Mollow triplet spectrum at an optical Rabi frequency of $\Omega_\text{L}/2\pi=7.9$ GHz using cross-polarization detection alone, where the central peak is dominated by the residue laser reflection, giving incorrect intensity ratio between the central peak and the two side peaks.

To further isolate the weak quantum dot signal from the strong laser reflection, we employ a combination of cross-polarization rejection and differential detection. The differential detection operates as follows in principle. For each spectrum, we record two measurements, one with the quantum dot tuned to its target transition frequency and one with the quantum dot far detuned. The tuning is accomplished by modulating the bias voltage on the \textit{p-i-n} diode via the quantum-confined Stark effect~\cite{Warburton:2002aa}. The on-resonance measurement contains both the quantum dot fluorescence and the residual laser reflection, while the off-resonance measurement captures only the reflection background. Subtracting the off-resonance spectrum from the on-resonance spectrum then yields the pure fluorescence signal.

In practice, we find that the residual laser reflection under cross-polarization detection varies as we change the applied bias voltage. The yellow circles in Fig.~\ref{sfig3}(d) show the laser reflection intensity under cross-polarization detection as a function of the applied bias voltage. The laser reflection intensity depends quadratically on the bias voltage, shown as the gray solid line in Fig.~\ref{sfig3}(d), which we attribute to the GaAs birefringence induced by the electro-optic effect~\cite{Hendriks:1997aa}. This phenomenon complicates our differential detection scheme, as the laser reflection measured with quantum dot far detuned becomes different from the actual reflection level when the quantum dot is at its target frequency. To correct for this effect, for each data point in the spectra presented in the main text, we take 11 measurements by sweeping the bias voltage from $-1.0$ V to 1.0 V with steps of 0.2 V at a frequency of 181.8 Hz. Only one out of the 11 measurements contains the quantum dot fluorescence, and the rest only captures the bias-dependent laser reflection. We fit the bias-dependent laser reflection in these 10 measurements to a quadratic function and obtain the actual reflection background by extrapolating the fit to the bias voltage when the quantum dot is at its target frequency. Figure~\ref{sfig3}(b) shows the extracted background count rate as a function of the etalon center frequency when applying this technique to the Mollow triplet measurement in Fig.~\ref{sfig3}(a). The blue circles in Fig.~\ref{sfig3}(c) show the differential signals between the measured count rates [Fig.~\ref{sfig3}(a)] and the extracted background count rates [Fig.~\ref{sfig3}(b)]. The measured data agree very well with the theoretical spectrum of the Mollow triplet [orange solid line in Fig.~\ref{sfig3}(c)], demonstrating the effectiveness of the differential detection technique.

To quantify the extinction ratio achieved with the differential detection technique, we move the laser away from the quantum dot such that we can directly measure the laser reflection intensity even at the bias voltage of 0.6 V where the quantum dot is at its target transition frequency. The extinction ratio is then given by $\eta=I_\text{meas}/|I_\text{meas}-I_\text{ext}|$, where $I_\text{meas}$ is the measured laser reflection intensity at the bias voltage of 0.6 V, and $I_\text{ext}$ is the extrapolated laser reflection intensity at 0.6 V using measurements performed at bias voltages other than 0.6 V. Figure~\ref{sfig3}(d) shows the measurement result with a laser power of 83 $\mu$W at the objective lens. The red triangle is the measured laser reflection intensity at 0.6 V, whereas the gray square is the extrapolated laser reflection intensity at 0.6 V using measurements taken at all other bias voltages (yellow circles). From these results, we calculate an extinction ratio of 127. Figure~\ref{sfig3}(e) shows the histogram of the extinction ratio obtained by repeating the measurements in Fig.~\ref{sfig3}(d) for 100 times. The extinction ratio distribution follows closely to the Poisson distribution with a mean value of 76. Together, the combined cross-polarization detection and differential detection provide an overall extinction ratio of $\sim 5\times 10^8$.

\section{Theoretical calculation of the resonance fluorescence spectrum}
\label{spectrum}
The resonance fluorescence spectra in Figs.~2(d) and 2(g) of the main text are calculated from the Fourier transform of the first-order correlation function in the steady-state limit, given by
\begin{equation}
\begin{aligned}
S(\omega)=\frac{1}{\pi}\text{Re}\int_0^\infty\lim_{t\to\infty}\langle\hat{\sigma}_+(t)\hat{\sigma}_-(t+\tau)\rangle e^{i(\omega-\omega_\text{L})\tau}d\tau,
\end{aligned}
\label{seq2}
\end{equation}
where $\omega_\text{L}$ is the frequency of the excitation laser, and $\hat{\sigma}_+(t)$ and $\hat{\sigma}_-(t+\tau)$ are the Heisenberg-picture operators for raising and lowering the two-level system at times $t$ and $t+\tau$, respectively. The two-time correlation function $\langle\hat{\sigma}_+(t)\hat{\sigma}_-(t+\tau)\rangle$ can be evaluated using the quantum regression theorem~\cite{Lax:1966aa}, which states that its value is given by $\langle\hat{\sigma}_+(t)\hat{\sigma}_-(t+\tau)\rangle=\text{Tr}\left[\hat{\sigma}_-\tilde{\rho}(t+\tau)\right]$, where $\tilde{\rho}(t+\tau)$ is the effective density matrix of the system at time $t+\tau$ when the system state is set to be $\tilde{\rho}(t)=\rho(t)\hat{\sigma}_+$ at time $t$, and $\rho(t)$ is the true density matrix of the system at time $t$. Therefore, evaluating the correlation function $\langle\hat{\sigma}_+(t)\hat{\sigma}_-(t+\tau)\rangle$ reduces to solving for the time evolution of $\langle\hat{\sigma}_-(t+\tau) \rangle$ with respect to $\tau$.

For a two-level system, the time evolution of $\langle\hat{\sigma}_-(t)\rangle$ is described by the optical Bloch equations, given by
\begin{equation}
\begin{aligned}
&\frac{d}{dt}\langle\hat{\sigma}_+(t)\rangle=\left(-i\Delta-\frac{\gamma}{2}\right)\langle\hat{\sigma}_+(t)\rangle-i\frac{\Omega_\text{L}}{2}\langle\hat{\sigma}_z(t)\rangle, \\
&\frac{d}{dt}\langle\hat{\sigma}_-(t)\rangle=\left(i\Delta-\frac{\gamma}{2}\right)\langle\hat{\sigma}_-(t)\rangle+i\frac{\Omega_\text{L}}{2}\langle\hat{\sigma}_z(t)\rangle, \\
&\frac{d}{dt}\langle\hat{\sigma}_z(t)\rangle=-i\Omega_\text{L}\left(\langle\hat{\sigma}_+(t)\rangle-\langle\hat{\sigma}_-(t)\rangle\right)-\gamma\left(\langle\hat{\sigma}_z(t)\rangle+1\right). \\
\end{aligned}
\label{seq3}
\end{equation}
In Eq.~\ref{seq3}, $\Delta$ is the frequency detuning between the laser and the two-level system, $\Omega_\text{L}$ is the optical Rabi frequency, $\gamma$ is the spontaneous emission rate, and $\hat{\sigma}_z$ is the Pauli Z operator of the two-level system. In our experiments, the frequency of the two-level system is periodically modulated by the surface acoustic wave. Therefore, we can simply describe the dynamics using the same optical Bloch equations but replacing the frequency detuning $\Delta$ by $\Delta-2\Omega_\text{S}\cos{(\omega_\text{S}t)}$, where $\Omega_\text{S}$ is the acoustic driving strength, and $\omega_\text{S}$ is the resonance frequency of the surface acoustic wave cavity. We can then write Eq.~\ref{seq3} into the matrix form, given by
\begin{equation}
\begin{aligned}
\frac{d}{dt}\left(
    \begin{matrix}
    \langle\hat{\sigma}_+(t)\rangle \\ 
    \langle\hat{\sigma}_-(t)\rangle \\
    \langle\hat{\sigma}_z(t)\rangle
    \end{matrix}
    \right)=M(t)\left(
    \begin{matrix}
    \langle\hat{\sigma}_+(t)\rangle \\ 
    \langle\hat{\sigma}_-(t)\rangle \\
    \langle\hat{\sigma}_z(t)\rangle
    \end{matrix}
    \right)+\left(
    \begin{matrix}
    0 \\
    0 \\
    -\gamma
    \end{matrix}
    \right),
\end{aligned}
\label{seq4}
\end{equation}
where
\begin{equation}
\begin{aligned}
\quad M(t)=\left(
    \begin{matrix}
    -i\Delta+2i\Omega_\text{S}\cos{(\omega_\text{S}t)}-\frac{\gamma}{2} &0 &-i\frac{\Omega_\text{L}}{2} \\
    0 &i\Delta-2i\Omega_\text{S}\cos{(\omega_\text{S}t)}-\frac{\gamma}{2} &i\frac{\Omega_\text{L}}{2} \\
    -i\Omega_\text{L} &i\Omega_\text{L} &-\gamma
    \end{matrix}
    \right),
\end{aligned}
\label{seq5}
\end{equation}
satisfying $M(t)=M(t+\frac{2\pi}{\omega_\text{S}})$. Since $M(t)$ is periodic, we can solve Eq.~\ref{seq4} using the Floquet theory. We refer the readers to Ref.~\cite{Yan:2016aa} for details of the derivation.

In our calculations, we set $\gamma/2\pi=134$ MHz, obtained from an independent measurement of the quantum dot radiative lifetime. To account for the spectral diffusion of the quantum dot, the calculated spectra are averaged over a Gaussian distribution of the laser detuning $\Delta$, with a full width at half maximum equal to the measured optical linewidth $\Gamma/2\pi=678$ MHz [Fig.~1(c) of the main text]. We further convolve the calculated spectra with the transmission function of the etalon with a linewidth of 525 MHz to obtain the theoretically predicted spectra shown in Figs.~2(d) and 2(g) of the main text.

\section{Doubly dressed state picture}
\label{doublyDressedStatePic}
In this section, we present the doubly dressed state picture we use to predict the emission frequency of each resonance fluorescence peak from the quantum dot under various driving conditions. We consider a Hamiltonian of a quantum dot interacting with a quantized laser field and acoustic field, $\hat{H}=\hat{H}_0+\hat{H}_\text{L}+\hat{H}_\text{S}$, where $\hat{H}_0$ describes the bare system consisting of the quantum dot, optical photons, and acoustic phonons, $\hat{H}_\text{L}$ describes the interaction between the quantum dot and the optical photons, and $\hat{H}_\text{S}$ describes the interaction between the quantum dot and the acoustic phonons. The three components of the Hamiltonian are given by
\begin{equation}
\begin{aligned}
&\hat{H}_0=\hbar\frac{\omega_0}{2}\hat{\sigma}_z+\hbar\omega_\text{L}\hat{a}^\dagger\hat{a}+\hbar\omega_\text{S}\hat{b}^\dagger\hat{b}, \\
&\hat{H}_\text{L}=\hbar\frac{g_\text{L}}{2}(\hat{a}^\dagger\hat{\sigma}_-+\hat{a}\hat{\sigma}_+), \\
&\hat{H}_\text{S}=\hbar\frac{g_0}{2}\hat{\sigma}_z(\hat{b}+\hat{b}^\dagger).
\end{aligned}
\label{seq6}
\end{equation}
In Eq.~(\ref{seq6}), $\hat{\sigma}_z$ is the Pauli Z operator of the quantum dot two-level system, $\hat{\sigma}_+$ and $\hat{\sigma}_-$ are the raising and lowering operators of the quantum dot two-level system, $\hat{a}^\dagger$ and $\hat{a}$ are the creation and annihilation operators of the optical photons, $\hat{b}^\dagger$ and $\hat{b}$ are the creation and annihilation operators of the acoustic phonons, $\omega_0$ is the quantum dot transition frequency, $\omega_\text{L}$ is the frequency of the laser field, $\omega_\text{S}$ is the frequency of the acoustic field, $g_\text{L}$ is the coupling strength between the quantum dot and a single photon, and $g_0$ is the coupling strength between the quantum dot and a single phonon. Diagonalizing $\hat{H}$ yields the doubly dressed eigenstates, and each emission peak in the resonance fluorescence spectrum corresponds to a dipole-allowed transition between two of these eigenstates.

\begin{figure}[t!]
\centering
\includegraphics[width=1.0\columnwidth]{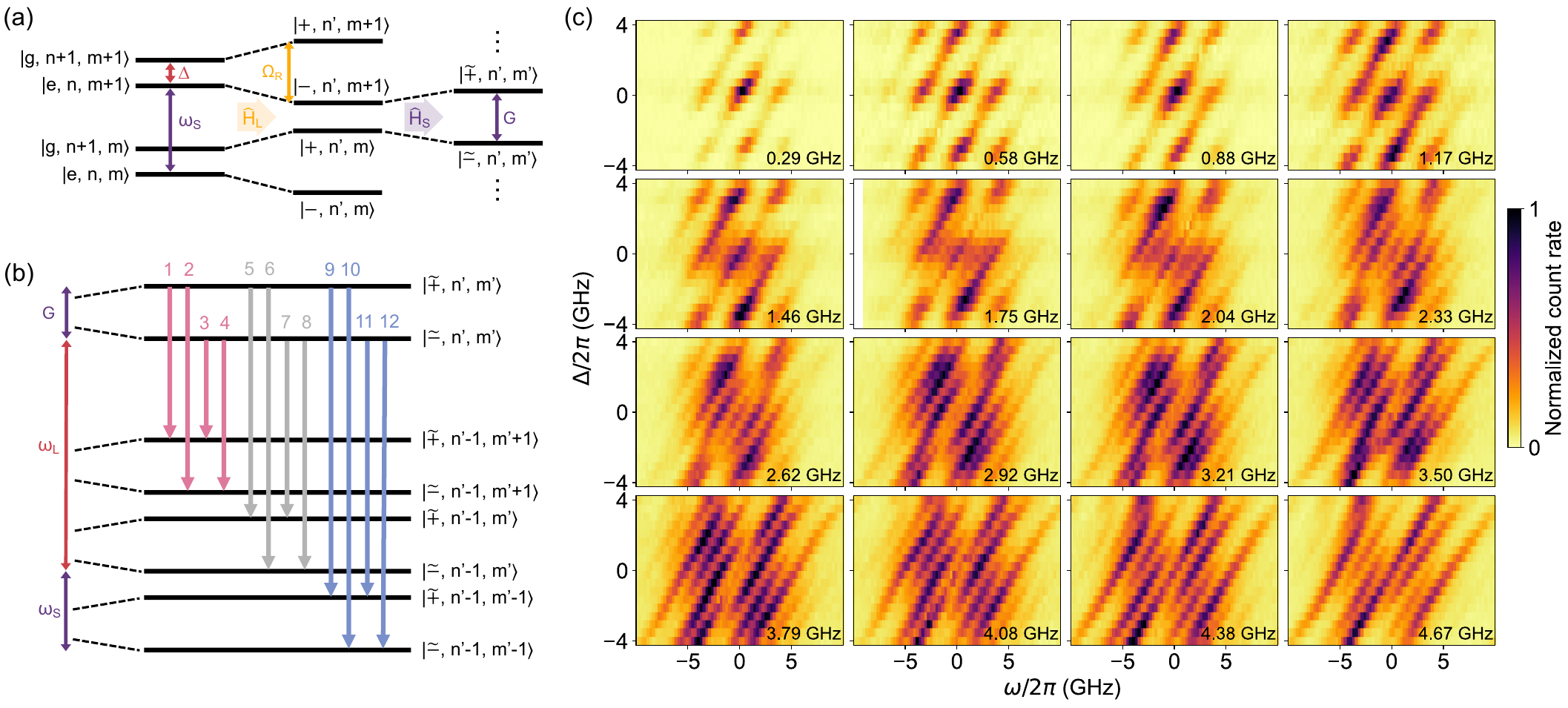}
\caption{(a) Energy-level diagram of the atom-photon-phonon system when the laser is detuned from the atomic transition and the optical Rabi frequency is detuned from the acoustic frequency. (b) Dipole-allowed transitions between doubly dressed states. The pink, gray, and blue groups of transition lines correspond to the pink, white, and blue groups of spectral components shown in Figs.~3(b) and 3(c) of the main text, respectively. (c) Measured spectra across a range of laser detuning $\Delta$ and optical Rabi frequency $\Omega_\text{L}$. The value of $\Omega_\text{L}/2\pi$ is indicated in the lower-right corner of each panel. The acoustic driving strength is fixed at $\Omega_\text{S}/2\pi= 1.75$ GHz.}
\label{sfig4}
\end{figure}

We express all the states under the eigenbasis of $\hat{H}_0$, given by $\ket{a,n,m}$, where $a\in\{g,e\}$ is the atomic state of the quantum dot, and $n$ and $m$ are photon and phonon numbers, respectively [Fig.~\ref{sfig4}(a), left panel]. The coupling term $\hat{H}_\text{L}$ creates the atom-photon dressed states [Fig.~\ref{sfig4}(a), central panel], given by
\begin{equation}
\begin{aligned}
&\ket{+,n',m}=\cos{\theta_\text{L}}\ket{g,n+1,m}+\sin{\theta_\text{L}}\ket{e,n,m},\text{ with an eigenfrequency }\omega_{+,n',m}=\left(n'-\frac{1}{2}\right)\omega_\text{L}+m\omega_\text{S}+\frac{\Omega_\text{R}}{2}, \\
&\ket{-,n',m}=\sin{\theta_\text{L}}\ket{g,n+1,m}-\cos{\theta_\text{L}}\ket{e,n,m},\text{ with an eigenfrequency }\omega_{-,n',m}=\left(n'-\frac{1}{2}\right)\omega_\text{L}+m\omega_\text{S}-\frac{\Omega_\text{R}}{2}.
\end{aligned}
\label{seq7}
\end{equation}
In Eq.~\ref{seq7}, $\theta_\text{L}=\frac{1}{2}\tan^{-1}\left(\frac{g_\text{L}\sqrt{n}}{\Delta}\right)$, $\Delta=\omega_\text{L}-\omega_0$ is the frequency detuning between the laser and the quantum dot transition, and $\Omega_\text{R}=\sqrt{g_\text{L}^2n+\Delta^2}$ is the generalized optical Rabi frequency. The coupling term $\hat{H}_\text{S}$ further dresses the atom-photon dressed states, yielding the atom-photon-phonon doubly dressed states as the eigenstates of the system [Fig.~\ref{sfig4}(a), right panel]. In the regime where $g_0\sqrt{m}\lesssim\omega_\text{S}$, the doubly dressed states can be approximated by
\begin{equation}
\begin{aligned}
\ket{\widetilde{+},n',m'}=&\cos{\theta_\text{S}}\ket{-,n',m+1}+\sin{\theta_\text{S}}\ket{+,n',m}, \\
&\text{ with an eigenfrequency }\omega_{\widetilde{+},n',m'}=\left(n'-\frac{1}{2}\right)\omega_\text{L}+\left(m'+\frac{1}{2}\right)\omega_\text{S}+\frac{G}{2}, \\
\ket{\widetilde{-},n',m'}=&\sin{\theta_\text{S}}\ket{-,n',m+1}-\cos{\theta_\text{S}}\ket{+,n',m}, \\
&\text{ with an eigenfrequency }\omega_{\widetilde{-},n',m'}=\left(n'-\frac{1}{2}\right)\omega_\text{L}+\left(m'+\frac{1}{2}\right)\omega_\text{S}-\frac{G}{2}.
\end{aligned}
\label{seq8}
\end{equation}
In Eq.~\ref{seq8}, $\theta_\text{S}=\frac{1}{2}\tan^{-1}\left(\frac{g_0\sqrt{m}\sin{2\theta_\text{L}}}{\omega_\text{S}-\Omega_\text{R}}\right)$, and $G=\sqrt{\left(\omega_\text{S}-\Omega_\text{R}\right)^2+\left(g_0\sqrt{m}\sin{2\theta_\text{L}}\right)^2}$.

Each emission peak in the resonance fluorescence spectrum corresponds to a dipole-allowed transition between two of the eigenstates given by Eq.~\ref{seq8}. Without loss of generality, we consider transitions starting from eigenstates $\ket{\widetilde{\pm},n',m'}$. Figure~\ref{sfig4}(b) shows all the 12 dipole-allowed transitions initiated from the eigenstates $\ket{\widetilde{\pm},n',m'}$. The emission frequency of each transition can be calculated from the eigenfrequency difference between the initial and final states, as summarized in Tab.~\ref{stab1}. The predicted transition frequencies of all the emission peaks are plotted as dashed lines in Figs.~3(b) and 3(c) of the main text and agree well with the experimentally measured spectra.

The 12 dipole-allowed transitions can be divided into three groups, shown as the pink, gray, and blue solid arrows in Fig.~\ref{sfig4}(b), corresponding to the pink, white, and blue dashed lines in Figs.~3(b) and 3(c) of the main text. In each group, two transitions share the same emission frequency and form the central peak of a triplet, while the remaining two form the side peaks. Consequently, the spectrum consists of three distinct triplets, as clearly observed in Fig.~3(b) of the main text. The two side peaks of each triplet anti-cross each other in Fig.~3(b) of the main text. This is due to the acoustic dressing of the atom-photon dressed states, resulting in a splitting of $2G$ between the two side peaks of each triplet, which has a minimum value of $2\Omega_\text{S}$. We also note that there are several emission lines observed in Fig.~3(c) of the main text that are not predicted by the doubly dressed state picture. These emission lines originate from higher-order dressing of the atom-photon-phonon states, which is not taken into account in our analysis.

Besides the transition frequency, the doubly dressed state picture also allows us to predict the emission intensity of each transition via the corresponding transition dipole moment, defined as $\bra{f}\hat{d}\ket{i}  =d\bra{f}\hat{\sigma}_x\ket{i}$, where $\hat{d}$ is the electric dipole operator, $\ket{i}$ and $\ket{f}$ are the initial and final states of the transition, respectively, and $d\equiv\bra{g}\hat{d}\ket{e}$ is the oscillator strength of the bare quantum dot. Table~\ref{stab1} presents the calculated values of $|\bra{f}\hat{\sigma}_x\ket{i}|^2$ for all the 12 transitions. We note that the dipole moments for both transitions 5 and 8 (the two transitions with frequencies at the laser frequency $\omega_\text{L}$) vanish under the Rabi resonance condition ($\cos{\theta_\text{S}}=\sin{\theta_\text{S}}=\frac{1}{\sqrt{2}}$), resulting in the cancellation of emission from the central peak of the gray triplet. The central peaks of the other two triplets are always nonzero and thus always show up. These predictions agree well with our experimental observations in Figs.~2(c) and~2(f) of the main text.

We further note that photon emission at the predicted Rabi resonance condition in Fig.~2(f) of the main text (black dashed circles) is not completely suppressed. This arises from three physical and experimental factors. First, the two side peaks of the central triplet are closely spaced in frequency and have finite linewidths broadened by both the spectral diffusion and convolution with the etalon response. Their spectral overlap at the central frequency therefore produces non-zero photon counts. Second, the high laser power used in this measurement results in residual leakage from spurious surface reflection at the exact frequency where dynamical cancellation is expected, despite the combined use of cross-polarization and differential detection techniques. Third, our doubly-dressed-state model neglects far-off-resonant couplings between neighboring ladders, i.e., acoustic coupling between $\ket{\pm,n',m}$ and all levels in $\ket{\pm,n',m\pm 1}$ manifolds. This approximation is only strictly valid when $\Omega_\text{S}\gg\omega_\text{S}$, and becomes imperfect under our experimental condition where $\frac{\Omega_\text{S}}{\omega_\text{S}}\sim 0.5$. These higher-order couplings can slightly shift the Rabi resonance condition away from the ideal prediction, an effect known as the Bloch-Siegert shift~\cite{Groll:2026aa}. A complete study of this ultra-strong driving regime is beyond the scope of this work.

\begin{table}[h]
\centering
\begin{tabular}{@{\extracolsep{1pt}} c|c|c|c} 
\hline
\hline
Transition index &Transition frequency &$\left|\bra{f}\hat{\sigma}_x\ket{i}\right|^2$ & $\delta N_{\text{phonon}}$ \\
\hline
1 &$\omega_\text{L}-\omega_\text{S}$ &$\cos^4{\theta_\text{L}}\cos^2{\theta_\text{S}}\sin^2{\theta_\text{S}}$ &$1$ \\
2 &$\omega_\text{L}-\omega_\text{S}+G$ &$\cos^4{\theta_\text{L}}\cos^4{\theta_\text{S}}$ &$2\sin^2{\theta_\text{S}}$ \\
3 &$\omega_\text{L}-\omega_\text{S}-G$ &$\cos^4{\theta_\text{L}}\sin^4{\theta_\text{S}}$ &$2\cos^2{\theta_\text{S}}$ \\
4 &$\omega_\text{L}-\omega_\text{S}$ &$\cos^4{\theta_\text{L}}\cos^2{\theta_\text{S}}\sin^2{\theta_\text{S}}$ &$1$ \\
\hline
5 &$\omega_\text{L}$ &$\cos^2{\theta_\text{L}}\sin^2{\theta_\text{L}}\left|\sin^2{\theta_\text{S}}-\cos^2{\theta_\text{S}}\right|^2$ &$0$ \\
6 &$\omega_\text{L}+G$ &$4\cos^2{\theta_\text{L}}\sin^2{\theta_\text{L}}\cos^2{\theta_\text{S}}\sin^2{\theta_\text{S}}$ &$\sin^2{\theta_\text{S}}-\cos^2{\theta_\text{S}}$ \\
7 &$\omega_\text{L}-G$ &$4\cos^2{\theta_\text{L}}\sin^2{\theta_\text{L}}\cos^2{\theta_\text{S}}\sin^2{\theta_\text{S}}$ &$\cos^2{\theta_\text{S}}-\sin^2{\theta_\text{S}}$ \\
8 &$\omega_\text{L}$ &$\cos^2{\theta_\text{L}}\sin^2{\theta_\text{L}}\left|\cos^2{\theta_\text{S}}-\sin^2{\theta_\text{S}}\right|^2$ &$0$ \\
\hline
9 &$\omega_\text{L}+\omega_\text{S}$ &$\sin^4{\theta_\text{L}}\cos^2{\theta_\text{S}}\sin^2{\theta_\text{S}}$ &$-1$ \\
10 &$\omega_\text{L}+\omega_\text{S}+G$ &$\sin^4{\theta_\text{L}}\sin^4{\theta_\text{S}}$ &$-2\cos^2{\theta_\text{S}}$ \\
11 &$\omega_\text{L}+\omega_\text{S}-G$ &$\sin^4{\theta_\text{L}}\cos^4{\theta_\text{S}}$ &$-2\sin^2{\theta_\text{S}}$ \\
12 &$\omega_\text{L}+\omega_\text{S}$ &$\sin^4{\theta_\text{L}}\cos^2{\theta_\text{S}}\sin^2{\theta_\text{S}}$ &$-1$ \\
\hline
\hline
\end{tabular}
\caption{Transition frequencies, dipole moments, and phonon number change per emitted photon for the 12 transitions labeled in Fig.~\ref{sfig4}(b).}
\label{stab1}
\end{table}

\section{Extraction of the phonon cooling rate from the resonance fluorescence spectrum}
\label{coolingRate}
To obtain Fig.~4(a) of the main text, we measure the resonance fluorescence spectra from our device under 16 different optical Rabi frequencies and 18 different laser detunings, while keeping the acoustic drive fixed at $\Omega_\text{S}/2\pi=1.75$ GHz. Figure~\ref{sfig4}(c) shows all the recorded spectra. For each Rabi frequency (value of $\Omega_\text{L}/2\pi$ labeled in the bottom right corner of each panel), we scan the laser detuning from $\Delta/2\pi=-4.25$ GHz to 4.25 GHz in steps of 0.5 GHz. As we will discuss below, each measured spectrum allows us to extract the optical phonon cooling rate at its corresponding optical Rabi frequency and laser detuning.

Based on the doubly dressed state picture described in Sec.~\ref{doublyDressedStatePic}, a single photon emitted via transition $\alpha$ ($\alpha=1,2,\dots,12$) changes the phonon number in the acoustic cavity by
\begin{equation}
\begin{aligned}
\delta N_{\text{phonon},\alpha}={}_\alpha\langle f|\hat{b}^\dagger\hat{b}|f\rangle_\alpha-{}_\alpha\langle i|\hat{b}^\dagger\hat{b}|i\rangle_\alpha,
\end{aligned}
\label{seq9}
\end{equation}
where $\hat{b}^\dagger$ and $\hat{b}$ are the creation and annihilation operators of the acoustic phonons, and $\ket{i}_\alpha$ and $\ket{f}_\alpha$ are the initial and final states of transition $\alpha$. The last column of Tab.~\ref{stab1} lists the value of $\delta N_\text{phonon}$ for each of the 12 transitions. The mean phonon number change per emitted photon is thus given by
\begin{equation}
\begin{aligned}
\delta_\text{phonon}=\sum_{\alpha=1}^{12}\delta N_{\text{phonon},\alpha}\left|{}_\alpha\langle f|\hat{\sigma}_x|i\rangle_\alpha\right|^2,
\end{aligned}
\label{seq10}
\end{equation}
where $\left|{}_\alpha\langle f|\hat{\sigma}_x|i\rangle_\alpha\right|^2$ is the probability of photon emission via transition $\alpha$. The phonon cooling rate (i.e. the phonon number change per unit time) can thus be calculated by
\begin{equation}
\begin{aligned}
R_\text{phonon}=\delta_\text{phonon}\cdot R_\text{photon}=\sum_{\alpha=1}^{12}\delta N_{\text{phonon},\alpha}\left|{}_\alpha\langle f|\hat{\sigma}_x|i\rangle_\alpha\right|^2\cdot\gamma\bar{\rho}_\text{ee}\propto\sum_{\alpha=1}^{12} \delta N_{\text{phonon},\alpha}I_\alpha,
\end{aligned}
\label{seq11}
\end{equation}
where $R_\text{photon}=\gamma\bar{\rho}_\text{ee}$ is the number of photons emitted per unit time, $\gamma$ is the spontaneous emission rate of the bare quantum dot, $\bar{\rho}_\text{ee}$ is the steady-state population of the quantum dot excited state, and $I_\alpha$ is the emission intensity of transition $\alpha$ in the measured resonance fluorescence spectrum. Therefore, by extracting the emission intensity of each transition, we can directly calculate the phonon cooling rate $R_\text{phonon}$ following Eq.~\ref{seq11}.

In practice, however, distinct transitions may overlap spectrally under certain conditions, making it difficult to resolve all of them individually. However, one can show that
\begin{equation}
\begin{aligned}
\sum_{\alpha=1}^{12}\delta N_{\text{phonon},\alpha}\left|{}_\alpha\langle f|\hat{\sigma}_x|i\rangle_\alpha\right|^2=2\times\sum_{\alpha=1,4,9,12}\delta N_{\text{phonon},\alpha}\left|{}_\alpha\langle f|\hat{\sigma}_x|i\rangle_\alpha\right|^2.
\end{aligned}
\label{seq12}
\end{equation}
This identity implies that the optical phonon cooling rate is fully determined by the emission intensities of transitions 1, 4, 9 and 12, which lie at only two frequencies, $\omega_\text{L}\pm\omega_\text{S}$, where $\omega_\text{L}$ is the laser frequency and $\omega_\text{S}$ is the resonance frequency of the acoustic cavity. Because those two first-order phonon sideband peaks are always well resolved in our spectra, the intensity difference between them suffices to extract the cooling rate reliably.

To obtain Fig.~4(b) of the main text, we directly apply Eq.~\ref{seq11} to calculate the optical phonon cooling rate at different optical Rabi frequencies and laser detunings. Using the values of $\delta N_{\text{phonon},\alpha}$ and $\left|{}_\alpha\langle f|\hat{\sigma}_x|i\rangle_\alpha\right|^2$ for each transition as listed in Tab.~\ref{stab1}, Eq.~\ref{seq11} simplifies to 
\begin{equation}
R_\text{phonon}=\frac{\Delta}{\Omega_\text{R}}\frac{\Omega_\text{L}^2\Omega_\text{S}^2}{(\omega_\text{S}-\Omega_\text{R})^2\Omega_\text{R}^2+\Omega_\text{L}^2\Omega_\text{S}^2}\cdot \gamma\bar{\rho}_\text{ee}
\label{seq13}
\end{equation}
where $\Omega_\text{R}=\sqrt{\Omega_\text{L}^2+\Delta^2}$ is the generalized optical Rabi frequency. The value of $\bar{\rho}_\text{ee}$ is determined via the steady-state solution of Eq.~\ref{seq4} using the Floquet theory~\cite{Yan:2016aa}. To account for the spectral diffusion, we average the phonon cooling rate over a Gaussian distributed laser detuning with a full width at half maximum of $\Gamma/2\pi=678$ MHz.

\begin{figure}[t]
\centering
\includegraphics[width=0.9\columnwidth]{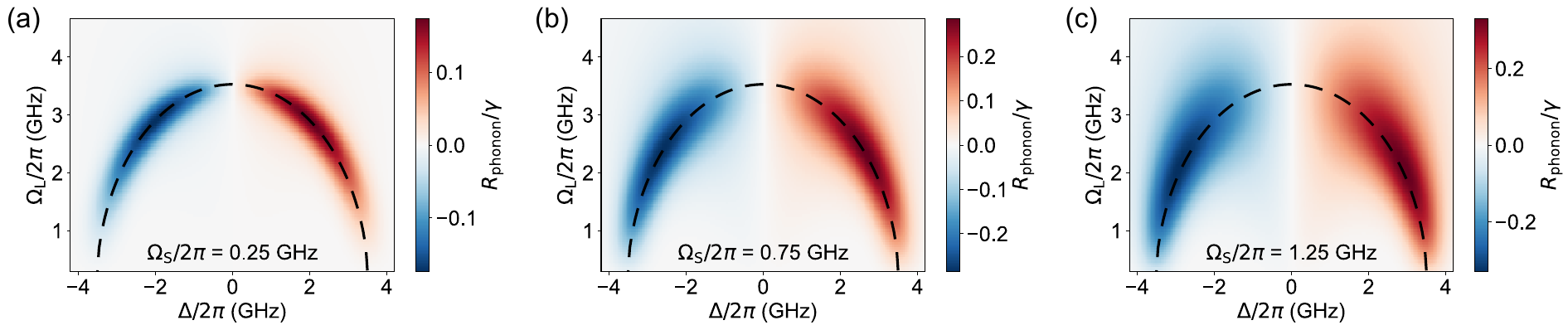}
\caption{Calculated phonon cooling rate $R_\text{phonon}$ as a function of the laser detuning $\Delta$ and the optical Rabi frequency $\Omega_\text{L}$ for acoustic driving strength $\Omega_\text{S}/2\pi=0.25$ GHz (a), 0.75 GHz (b), and 1.25 GHz (c), respectively. The simulation parameters are identical to those used in Fig.~4(b) of the main text. Black dashed lines indicate the Rabi resonance condition.}
\label{sfig5}
\end{figure}

Figure~\ref{sfig5} shows the same calculation as Fig.~4(b) of the main text, but at different values of $\Omega_\text{S}/2\pi=0.25$, 0.75, and 1.25 GHz, respectively. We observe that the optimal phonon cooling rate always occurs at the Rabi resonance condition, indicated as the dashed lines in Fig.~\ref{sfig5}, regardless of the value of $\Omega_\text{S}$. In the next section, we will find that the Rabi resonance condition remains optimal for optical phonon cooling even in the absence of the acoustic drive.

\section{Analysis of optimal phonon cooling in the absence of the external acoustic drive}
\label{groundStateCooling}
In this section, we discuss the optimal phonon cooling condition in the absence of the external acoustic drive. We assume that without laser cooling, the phonon number in the acoustic cavity follows the Boltzmann distribution, with an average phonon number $m_\text{th}=[\exp{(\omega_\text{S}/k_BT)}-1]^{-1}$, where $\omega_\text{S}$ is the resonance frequency of the acoustic cavity, $T$ is the temperature of the phonon bath, and $k_B$ is the Boltzmann constant. Under these conditions, the analytical treatment based on the doubly dressed state picture (Sec.~\ref{coolingRate}) becomes inaccurate, since it neglects both the anharmonic energy spacing at low phonon numbers and the full phonon‐number statistics. We therefore calculate the cooling rates by numerically solving the system’s master equation.

Our goal is to calculate the average phonon number $m_\text{ss}$ of our system in the steady state in the presence of the cooling laser, defined as $m_\text{ss}=\langle b^\dagger b\rangle_\text{ss}$, where $\hat{b}^\dagger$ and $\hat{b}$ are creation and annihilation operators of the acoustic phonons. We characterize the cooling performance using a normalized parameter $C=(m_\text{ss}-m_\text{th})/m_\text{th}$. Under this definition, $C$ takes negative values when the acoustic cavity is cooled while positive values in the case of heating, and its absolute value quantifies the cooling or heating performance.

We calculate $m_\text{ss}$ via the master equation given by
\begin{equation}
\begin{aligned}
\frac{d}{dt}\rho(t)=-\frac{i}{\hbar}[\hat{H}(t),\rho(t)]+\sum_i\frac{1}{2}[2\hat{c}_i\rho(t)\hat{c}_i^\dagger-\rho(t)\hat{c}_i^\dagger\hat{c}_i-\hat{c}_i^\dagger\hat{c}_i\rho(t)],
\end{aligned}
\label{seq14}
\end{equation}
where $\rho(t)$ is the density matrix of the system at time $t$, $\hat{H}$ is the Hamiltonian of the system, and $\hat{c}_i$ are the jump operators. The Hamiltonian of the system is given by
\begin{equation}
\begin{aligned}
\hat{H}(t)=-\hbar\frac{\Delta}{2}\hat{\sigma}_z+\hbar\frac{\Omega_\text{L}}{2}\hat{\sigma}_x+\hbar\omega_\text{S}\hat{b}^\dagger\hat{b}+\hbar\frac{g_0}{2}\hat{\sigma}_z(\hat{b}+\hat{b}^\dagger),
\end{aligned}
\label{seq15}
\end{equation}
where $\Delta$ is the frequency detuning between the laser and the quantum dot transition, $\Omega_\text{L}$ is the optical Rabi frequency, $g_0$ is the single-phonon coupling strength, and $\hat{\sigma}_x$ and $\hat{\sigma}_z$ are the Pauli X and Z operators of the quantum dot two-level system. The jump operators are given by
\begin{equation}
\begin{aligned}
\hat{c}_1&=\sqrt{\gamma}\hat{\sigma}_-, \\
\hat{c}_2&=\sqrt{\gamma_\text{S}m_\text{th}}\hat{b}^\dagger, \\
\hat{c}_3&=\sqrt{\gamma_\text{S}(m_\text{th}+1)}\hat{b}, \\
\end{aligned}
\label{seq16}
\end{equation}
where $\gamma$ is the spontaneous emission rate of the quantum dot, $\gamma_\text{S}=\omega_\text{S}/Q$ is the acoustic cavity dissipation rate, $Q$ is the quality factor of the acoustic cavity, and $\hat{\sigma}_-$ is the lowering operator of the quantum dot two-level system. 

\begin{figure}[t!]
\centering
\includegraphics[width=0.65\columnwidth]{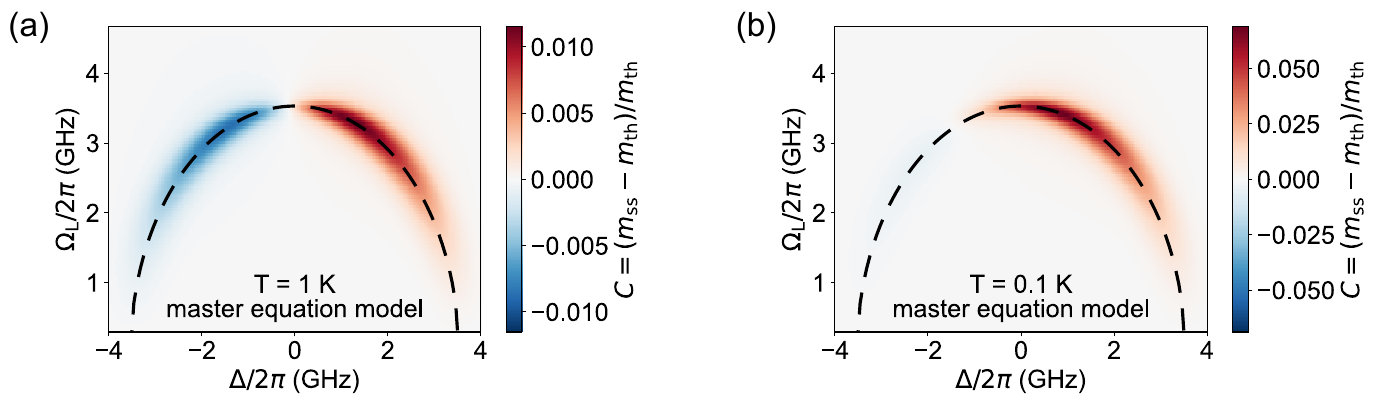}
\caption{Calculated cooling performance $C$ from the master equation simulation at $T=1$ K (a) and $T=0.1$ K (b), respectively. The black dashed lines indicate the Rabi resonance condition.}
\label{sfig6}
\end{figure}

Figures~\ref{sfig6} shows the calculated cooling performance $C$ as a function of the laser detuning $\Delta$ and the optical Rabi frequency $\Omega_\text{L}$. In our calculations, we fix $\omega_\text{S}/2\pi=3.5299$ GHz and $Q=12,562$ as measured from our device [Fig.~\ref{sfig2}(a)], $g_0/2\pi=1.2$ MHz as has been characterized in a similar device in our previous work~\cite{DeCrescent:2022aa}, and $\gamma/2\pi=134$ MHz as obtained from the independently measured quantum dot radiative lifetime. To account for the spectral diffusion, we average the calculated cooling performance over a Gaussian distributed laser detuning with a full width at half maximum of $\Gamma/2\pi=678$ MHz. Figure~\ref{sfig6}(a) shows the cooling performance in the high temperature regime ($T=1$ K) where $m_\text{th} = 5.4 \gg 1$, whereas Fig.~\ref{sfig6}(b) shows the cooling performance in the low temperature regime ($T=0.1$ K) where $m_\text{th} = 0.2 \ll 1$. In both temperature regimes, the optimal cooling condition occurs at the Rabi resonance condition, as indicated by the black dashed lines. The numerical calculations confirm that the Rabi resonance condition remains optimal for optical phonon cooling, even in the limit of ground state cooling.

\section{Feasibility of direct measurements of ground-state cooling and emitter-induced backaction}
\label{directObservation}
\begin{figure}[t!]
\centering
\includegraphics[width=0.75\columnwidth]{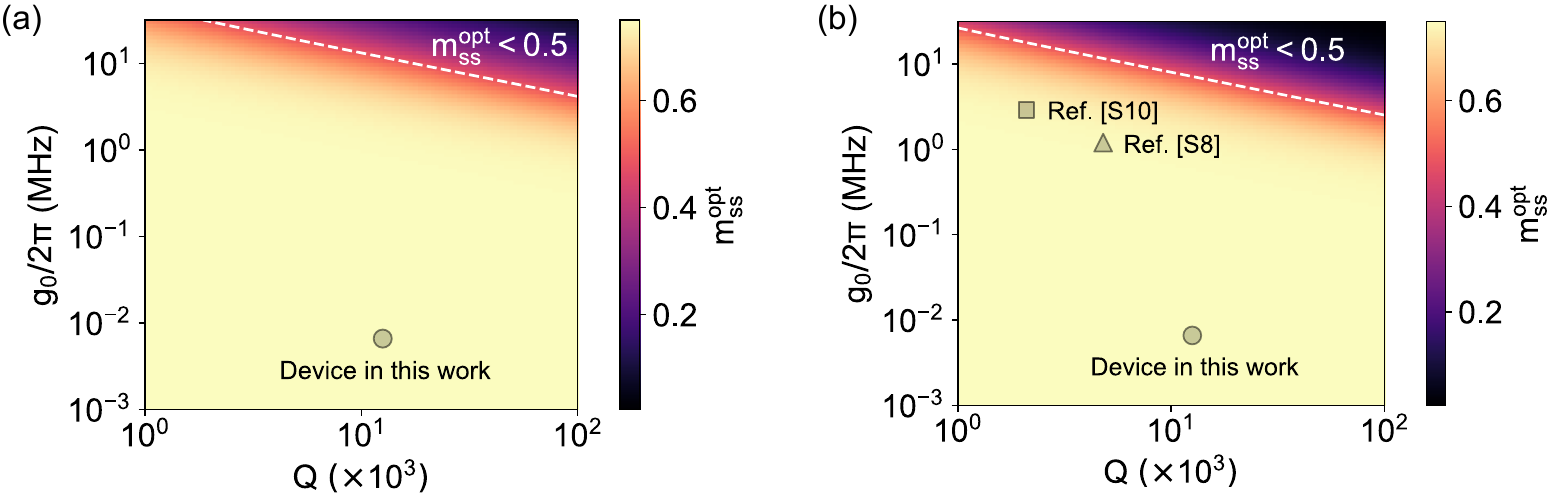}
\caption{Steady-state phonon number under optimal cooling, $m_\text{ss}^\text{opt}$, as a function of the single-phonon coupling strength $g_0$ and the acoustic cavity quality factor $Q$. Panel (a) assumes a quantum-dot linewidth of $\Gamma/2\pi=678$ MHz due to spectral diffusion, whereas panel (b) assumes transform-limited linewidth. The gray circles in (a) and (b) mark the parameters of the current device, whereas the gray triangle and square in (b) mark the parameters reported in Refs.~\cite{DeCrescent:2022aa} and~\cite{Spinnler:2024aa}, respectively. The white dashed lines denote the device parameters required to achieve $m_\text{ss}^\text{opt}=0.5$ at $T=0.2$~K, where $m_\text{th}=0.75$.}
\label{sfig7}
\end{figure}

We first quantitatively analyze the feasibility of reaching ground-state cooling in quantum-dot-based optomechanical devices. Figure~\ref{sfig7} shows the calculated steady-state phonon number in the acoustic cavity under optimal cooling conditions, $m_\text{ss}^\text{opt}$, as a function of the single-phonon coupling strength $g_0$ and the acoustic cavity quality factor $Q$. The calculations are based on the master-equation model described in Sec.~\ref{groundStateCooling}, evaluated at a fixed temperature of $T=0.2$ K where the thermal phonon population is $m_\text{th}=0.75$. In our calculations, we fix the mechanical frequency at $\omega_\text{S}/2\pi=3.5299$ GHz and the spontaneous emission rate of the quantum dot at $\gamma/2\pi=134$ MHz, both the same as our current device. Figure~\ref{sfig7}(a) shows the calculated $m_\text{ss}^\text{opt}$ with a quantum-dot linewidth $\Gamma/2\pi=678$ MHz due to spectral diffusion, whereas Fig.~\ref{sfig7}(b) assumes a transform-limited linewidth.

Direct observation of cooling requires a measurable reduction in the phonon number. Here we qualitatively set $m_\text{ss}^\text{opt}=0.5$ as the target, indicated by the white dashed lines in Fig.~\ref{sfig7}. The gray circles in Fig.~\ref{sfig7} indicate the $g_0$ and $Q$ corresponding to the device used in our present work, which lie well below this target. Therefore, to directly observe ground-state cooling requires substantial improvements in the current device parameters. 

The $g_0$ of our current device is mainly limited by the large mode volume ($\sim 5,000 \lambda^3$) of the planar surface acoustic wave cavity. By using a focusing surface acoustic wave cavity with a mode volume of 6$\lambda^3$, a $g_0$ of $2\pi\times 1.2$ MHz~\cite{DeCrescent:2022aa} has been reported, as indicated by the gray triangle in Fig.~\ref{sfig7}(b). Achieving $m_\text{ss}^\text{opt}=0.5$ at $T=0.2$~K would require the cavity $Q$ in Ref.~\cite{DeCrescent:2022aa} to reach the same level as that of planar devices, along with a further fivefold increase in $g_0$. Alternatively, by employing a quantum dot molecule that possesses an order-of-magnitude larger longitudinal strain susceptibility~\cite{Tsuchimoto:2022aa}, we expect to achieve a $g_0$ of $\sim 2\pi\times 10$ MHz with the focusing surface acoustic wave cavity in Ref.~\cite{DeCrescent:2022aa}, directly enabling the reach of $m_\text{ss}^\text{opt}=0.5$ at $T=0.2$~K without further improvement in $Q$. We also note that the use of a phononic crystal cavity has enabled a higher $g_0$ of $2\pi\times 2.9$ MHz for a single quantum dot~\cite{Spinnler:2024aa} but a lower $Q$ factor, as indicated by the gray square in Fig.~\ref{sfig7}(b). An order of magnitude increase in their $Q$ would also enable ground-state cooling.

An alternative approach for direct observation of cooling is to measure the optomechanical broadening and frequency shift of the acoustic cavity. In our system, optical excitation of the quantum dot creates an exciton that generates a local electric field, which induces stress via the piezoelectric response of GaAs~\cite{Kettler:2021aa}. This piezoelectrically generated stress acts as a backaction force on the mechanical mode, modifying both its resonance frequency and damping rate. These signatures can be directly characterized by probing the mechanical susceptibility via a weak external drive applied through the interdigital transducer.

To explore whether this effect is observable under our current device parameters, we numerically calculate the mechanical resonance shift $\delta\omega_\text{S}$ and damping rate modification $\delta\gamma_\text{S}$ by solving a classical equation of motion for the displacement field $x(t)$ of the mechanical oscillator,
\begin{equation}
\begin{aligned}
m_\text{eff}\frac{d^2x(t)}{dt^2}+m_\text{eff}\gamma_\text{S}\frac{dx(t)}{dt}+m_\text{eff}\omega_\text{S}^2x(t)=F_\text{QD}(t)+F_\text{p}(t),
\end{aligned}
\label{seq18}
\end{equation}
where $m_\text{eff}$ is the effective mass, $F_\text{QD}(t)$ is the backaction force exerted by the quantum dot, and $F_\text{p}(t)$ is the weak probe force. The backaction force is given by
\begin{equation}
\begin{aligned}
F_\text{QD}(t)=-\frac{\partial}{\partial x}\left\langle\hbar\frac{g_0}{2}\hat{\sigma}_z(\hat{b}+\hat{b}^\dagger)\right\rangle=-\hbar G\left(\rho_\text{ee}(t)-\frac{1}{2}\right),
\end{aligned}
\label{seq19}
\end{equation}
where $G=\frac{\partial\omega_0}{\partial x}$ is the displacement susceptibility of the quantum-dot frequency $\omega_0$, and $\rho_\text{ee}(t)=\langle\hat{\sigma}_+\hat{\sigma}_-\rangle$ is the excited-state population of the quantum dot. In obtaining Eq.~\ref{seq19}, we used $x(t)=x_\text{zpf}\langle\hat{b}+\hat{b}^\dagger\rangle$, where $x_\text{zpf}=g_0/G$ is the zero-point motion.

The excited-state population $\rho_\text{ee}(t)$ is calculated semiclassically by treating the displacement field as a classical drive, $x(t)=\frac{2\Omega_\text{S}}{G}\cos(\omega_\text{S}t)$, and numerically solving the master equation, given by
\begin{equation}
\begin{aligned}
\frac{d}{dt}\rho(t)=-\frac{i}{\hbar}[\hat{H}(t),\rho(t)]+\frac{1}{2}[2\hat{c}\rho(t)\hat{c}^\dagger-\rho(t)\hat{c}^\dagger\hat{c}-\hat{c}^\dagger\hat{c}\rho(t)],
\end{aligned}
\label{seq20}
\end{equation}
where $\hat{H}$ is the system Hamiltonian, given by 
\begin{equation}
\begin{aligned}
\hat{H}(t)=-\hbar\frac{1}{2}\left[\Delta-2\Omega_\text{S}\cos(\omega_\text{S}t)\right]\hat{\sigma}_z+\hbar\frac{\Omega_\text{L}}{2}\hat{\sigma}_x,
\end{aligned}
\label{seq21}
\end{equation}
and $\hat{c}=\sqrt{\gamma}\hat{\sigma}_-$ is the jump operator corresponding to the spontaneous emission of the quantum dot. 

\begin{figure}[t!]
\centering
\includegraphics[width=0.65\columnwidth]{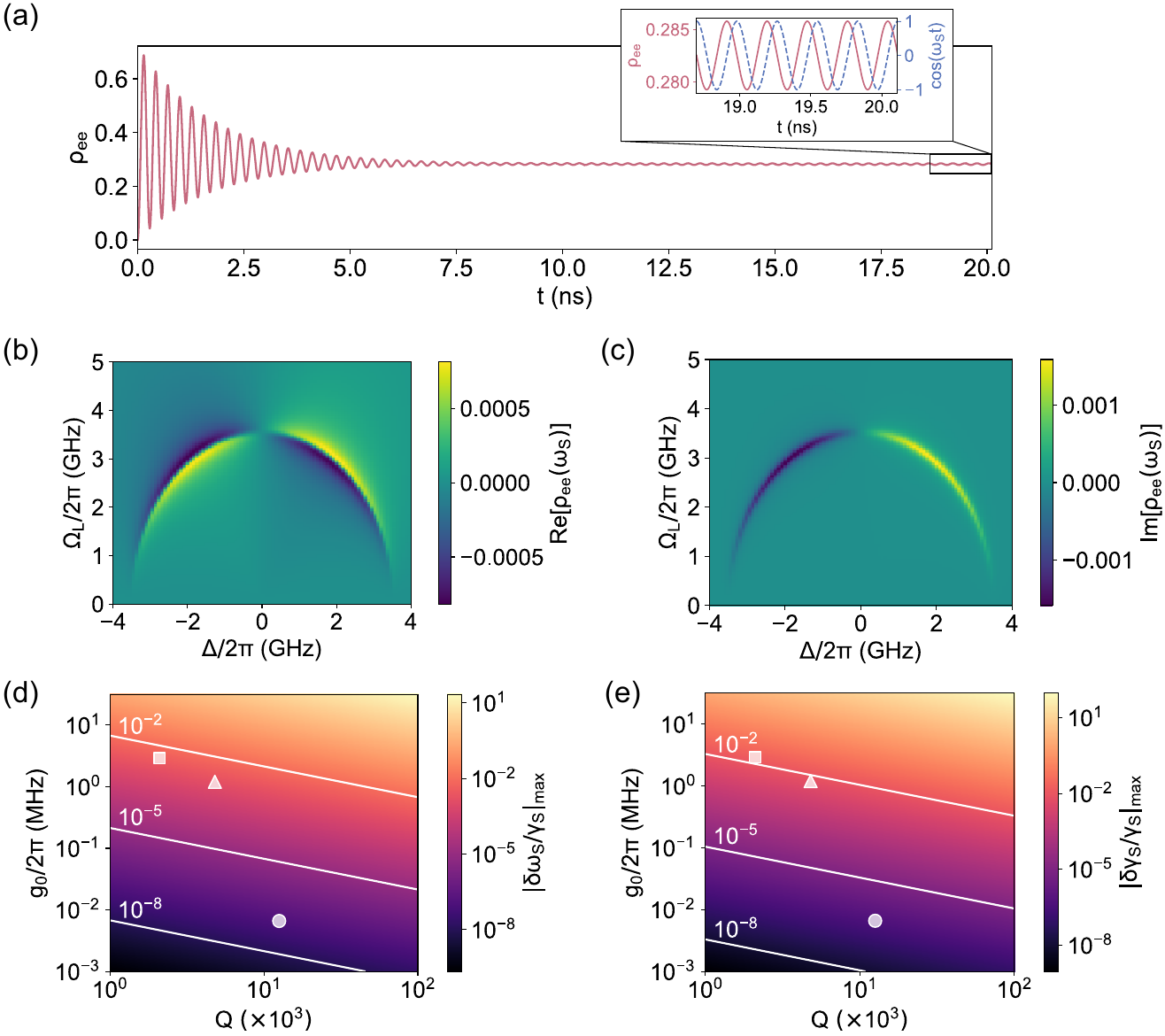}
\caption{(a) Excited-state population $\rho_\text{ee}$ as a function of time under an optical drive satisfying the Rabi resonance condition. Inset: long-time behavior of $\rho_\text{ee}(t)$ (solid pink) and displacement field $x(t)$ (dashed blue). (b)(c) Real (b) and imaginary (c) parts of the Fourier component of $\rho_\text{ee}$ at the mechanical frequency $\omega_\text{S}$, as functions of the laser detuning $\Delta$ and the optical Rabi frequency $\Omega_\text{L}$. The acoustic drive strength is fixed at $\Omega_\text{S}/2\pi=1.0$ MHz in (a)-(c). (d)(e) Maximal mechanical frequency shift (d) and damping rate modification (e) as functions of the single-phonon coupling strength $g_0$ and the acoustic cavity quality factor $Q$. The white contour lines indicate these fractional change levels of $10^{-8}$, $10^{-5}$, and $10^{-2}$. The white circles mark the values of $g_0$ and $Q$ for the current device. The while triangles and squares mark the values of $g_0$ and $Q$ reported in Refs.~\cite{DeCrescent:2022aa} and~\cite{Spinnler:2024aa}, respectively. A transform-limited emitter linewidth is assumed in all simulations.}
\label{sfig8}
\end{figure}

Figure~\ref{sfig8}(a) illustrates the calculated temporal evolution of the excited-state population $\rho_\text{ee}(t)$ under a weak acoustic drive ($\Omega_\text{S}/2\pi = 1.0$ MHz). We model a transform-limited quantum dot subjected to a laser drive satisfying the Rabi resonance condition for optimal cooling, with a Rabi frequency $\Omega_\text{L}/2\pi=3.0$ GHz and a laser detuning $\Delta_\text{L}=-\sqrt{\omega_\text{S}^2-\Omega_\text{L}^2}=-1.86$ GHz. In the long-time limit, $\rho_\text{ee}(t)$ reaches a steady state, oscillating at the mechanical frequency $\omega_\text{S}$ [solid pink trace, Fig.~\ref{sfig8}(a) inset]. Crucially, this population oscillation exhibits a characteristic phase shift relative to the mechanical displacement $x(t)$ [dashed blue trace]~\cite{Spinnler:2024aa}. This phase shift is the fundamental signature of the quantum-dot backaction on the acoustic cavity. The in-phase component of the induced force acts as an additional restoring force (the optical spring effect), shifting the resonance frequency, while the quadrature component modifies the effective damping rate, providing the mechanism for optomechanical cooling.

To quantitatively extract $\delta\omega_\text{S}$ and $\delta\gamma_\text{S}$, we rewrite Eq.~\ref{seq18} in the frequency domain,
\begin{equation}
\begin{aligned}
m_\text{eff}\left[\omega_\text{S}^2-\omega_\text{p}^2\right]x(\omega_\text{p})-im_\text{eff}\omega_\text{p}\gamma_\text{S} x(\omega_\text{p})+\hbar G\rho_\text{ee}(\omega_\text{p})=F_\text{p}(\omega_\text{p}),
\end{aligned}
\label{seq22}
\end{equation}
where $\omega_\text{p}$ is the probe frequency. With our phase convention, $x(\omega_\text{p})$ only takes real values, so the real and imaginary parts of $\rho_\text{ee}(\omega_\text{p})$ directly relate to $\delta\omega_\text{S}$ and $\delta\gamma_\text{S}$, respectively. In the regime $\omega_\text{S}\gg\gamma_\text{S},\delta\omega_\text{S},\delta\gamma_\text{S}$, $\rho_\text{ee}(\omega_\text{p})$ can be evaluated at $\omega_\text{p}=\omega_\text{S}$. From Eq.~\ref{seq22}, $\delta\omega_\text{S}$ and $\delta\gamma_\text{S}$ are given by
\begin{equation}
\begin{aligned}
\frac{\delta\omega_\text{S}}{\gamma_\text{S}}&=\frac{g_0^2Q}{\omega_\text{S}}\frac{\operatorname{Re}\left[\rho_\text{ee}(\omega_\text{S})\right]}{\Omega_\text{S}}, \\
\frac{\delta\gamma_\text{S}}{\gamma_\text{S}}&=-\frac{2g_0^2Q}{\omega_\text{S}}\frac{\operatorname{Im}\left[\rho_\text{ee}(\omega_\text{S})\right]}{\Omega_\text{S}},
\end{aligned}
\label{seq23}
\end{equation}
where $\rho_\text{ee}(\omega_\text{S})$ is the Fourier transform of $\rho_\text{ee}(t)$ at frequency $\omega_\text{S}$ in the long-time limit. 

Figures~\ref{sfig8}(b) and~\ref{sfig8}(c) show the calculated real and imaginary parts of $\rho_\text{ee}(\omega_\text{S})$ as functions of $\Delta$ and $\Omega_\text{L}$, with all other parameters identical to those in Fig.~\ref{sfig8}(a). We observe that the strongest backaction, in terms of both the frequency shift and the damping rate modification, emerges near the Rabi resonance condition, in contrast to the behavior of standard cavity optomechanics~\cite{Teufel:2008aa}. This reinforces the fact that for an emitter-based optomechanical system, the optimal cooling or heating occurs specifically at the Rabi resonance condition.

Figures~\ref{sfig8}(d) and~\ref{sfig8}(e) show the calculated $\left|\frac{\delta\omega_\text{S}}{\gamma_\text{S}}\right|_\text{max}$ and $\left|\frac{\delta\gamma_\text{S}}{\gamma_\text{S}}\right|_\text{max}$ under optimal cooling conditions as functions of the device parameters $g_0$ and $Q$. The white circles indicate the parameters of our current device. Even assuming a transform-limited quantum-dot linewidth, we obtain $\left|\frac{\delta\omega_\text{S}}{\gamma_\text{S}}\right|_\text{max}=1.2\times 10^{-7}$ and $\left|\frac{\delta\gamma_\text{S}}{\gamma_\text{S}}\right|_\text{max}=5.0\times 10^{-7}$. Both are vanishingly small and fall well below the resolution of our instruments. The white triangles and squares mark the $g_0$ and $Q$ values reported in Refs.~\cite{DeCrescent:2022aa} and~\cite{Spinnler:2024aa}, respectively, which achieve much larger $g_0$ values due to the significantly smaller mode volumes of their acoustic cavities. This demonstrates that state-of-the-art parameters for quantum-dot-based optomechanical systems are already at a level where these fractional changes can reach $\sim 1\%$, which is detectable. Consequently, further improvements in $g_0$ and $Q$, even by modest amounts, would enable direct electrical measurement of both the mechanical resonance shift and the damping modification arising from backaction.

\providecommand{\noopsort}[1]{}\providecommand{\singleletter}[1]{#1}%